\documentclass[12pt,letterpaper]{article}
\pdfoutput=1

\usepackage{amsmath,amssymb,calc, amsthm,bbm, epsfig,psfrag, mathtools,comment}
\usepackage{mathrsfs}
\usepackage{needspace}
\usepackage{graphicx, enumerate}
\usepackage{float}
\usepackage{comment}
 \usepackage[all,cmtip]{xy}
\usepackage[numbers,sort&compress]{natbib}
\usepackage[dvipsnames]{xcolor}
\usepackage{tikz}
\usepackage{tikz-cd}
\usepackage{physics}
\usepackage{tensor}
\usepackage{enumitem}
\usepackage[compat=1.1.0]{tikz-feynman}
\usepackage[utf8]{inputenc} 
\definecolor{azure}{rgb}{0.0, 0.5, 1.0}
\definecolor{darkblue}{rgb}{0.15,0.35,0.7}
\definecolor{reddish}{rgb}{0.65, 0.2, 0.2}
\definecolor{brandeisblue}{rgb}{0.0, 0.44, 1.0}
\definecolor{ceruleanblue}{rgb}{0.16, 0.32, 0.75}
\definecolor{indigo(dye)}{rgb}{0.0, 0.25, 0.42}
\usepackage[linktocpage=true]{hyperref}
\hypersetup{
colorlinks=true,
citecolor=ceruleanblue,
linkcolor=ceruleanblue,
urlcolor=ceruleanblue,
pdfauthor={},
pdftitle={},
pdfsubject={}
}

\usepackage{tikz}
\tikzset{sines/.style={
        thick,
        line join=round, 
        draw=black, 
        decorate, 
        decoration={complete sines, amplitude=1mm,
        segment length=2mm}
    }}
    
\tikzset{esines/.style={
        thick,
        line join=round, 
        decorate, 
        decoration={complete sines, amplitude=1mm,
        segment length=2mm}
    }}

\usetikzlibrary{arrows.meta}

\usetikzlibrary{decorations}

\tikzset{/pgf/decoration/.cd,
    number of sines/.initial=10,
    angle step/.initial=20,
}
\newdimen\tmpdimen
\pgfdeclaredecoration{complete sines}{initial}
{
    \state{initial}[
        width=+0pt,
        next state=move,
        persistent precomputation={
            \pgfmathparse{\pgfkeysvalueof{/pgf/decoration/angle step}}%
            \let\anglestep=\pgfmathresult%
            \let\currentangle=\pgfmathresult%
            \pgfmathsetlengthmacro{\pointsperanglestep}%
                {(\pgfdecoratedremainingdistance/\pgfkeysvalueof{/pgf/decoration/number of sines})/360*\anglestep}%
        }] {}
    \state{move}[width=+\pointsperanglestep, next state=draw]{
        \pgfpathmoveto{\pgfpointorigin}
    }
    \state{draw}[width=+\pointsperanglestep, switch if less than=1.25*\pointsperanglestep to final, 
        persistent postcomputation={
        \pgfmathparse{mod(\currentangle+\anglestep, 360)}%
        \let\currentangle=\pgfmathresult%
    }]{%
        \pgfmathsin{+\currentangle}%
        \tmpdimen=\pgfdecorationsegmentamplitude%
        \tmpdimen=\pgfmathresult\tmpdimen%
        \divide\tmpdimen by2\relax%
        \pgfpathlineto{\pgfqpoint{0pt}{\tmpdimen}}%
    }
    \state{final}{
        \ifdim\pgfdecoratedremainingdistance>0pt\relax
            \pgfpathlineto{\pgfpointdecoratedpathlast}
        \fi
   }
}

\definecolor{indigo(dye)}{rgb}{0.0, 0.25, 0.42}

\newcommand{\overbar}[1]{\mkern 1.5mu\overline{\mkern-1.5mu#1\mkern-1.5mu}\mkern 1.5mu}
\newcommand\Tb{\overbar{T}}

\def\TT{{T\overbar{T}}}

\newcommand{\Th}{\widehat{T}}
\newcommand{\Tt}{\widetilde{T}}

\usepackage{amsthm}

\usepackage{cleveref}
\usepackage{bm}
\crefname{lem}{lemma}{lemmas}
\crefname{thm}{theorem}{theorems}
\crefname{cor}{corollary}{corollaries}
\crefname{rem}{remark}{remarks}
\crefname{prop}{proposition}{propositions}

\makeatletter
\renewcommand\section{\@startsection {section}{1}{\z@}%
                               {-3.5ex \@plus -1ex \@minus -.2ex}
                               {2.3ex \@plus.2ex}%
                               {\normalfont\large\bfseries}}
\renewcommand\subsection{\@startsection{subsection}{2}{\z@}%
                                 {-3.25ex\@plus -1ex \@minus -.2ex}%
                                 {1.5ex \@plus .2ex}%
                                 {\normalfont\bfseries}}
\makeatother

\let\non\nonumber

\def\bea#1\eea{\begin{align}#1\end{align}}

\def\bes #1\ees{\begin{split}#1\end{split}}

\newcommand{\be}{\begin{equation}}
\newcommand{\ee}{\end{equation}}

\newfont{\goth}{ygoth.tfm scaled 1200}                   

\renewcommand{\thesection}{\arabic{section}}

\numberwithin{equation}{section}

\newcommand{\ul}{\underline}

\begin{document}
\begin{titlepage}

\begin{center}

\hfill         \phantom{xxx}  

\vskip 2 cm {\Large \bf \scalebox{1}{Root-$\TT$ Deformed Boundary Conditions in Holography}}

\vskip 1.25 cm {\bf Stephen Ebert,$^1$ Christian Ferko$^{2}$ and Zhengdi Sun$^3$}\non\\

\vskip 0.2 cm
{\it $^1$ Mani L. Bhaumik Institute for Theoretical Physics,	
\\ University of California, Los Angeles, CA 90095, USA}

\vskip 0.2 cm
 {\it $^2$ Center for Quantum Mathematics and Physics (QMAP), \\ Department of Physics \& Astronomy,  University of California, Davis, CA 95616, USA}

\vskip 0.2 cm
 {\it $^3$ Department of Physics, University of California, San Diego, CA 92093, USA}
\end{center}
\vskip 1.5 cm

\begin{abstract}

\noindent We develop the holographic dictionary for pure $\mathrm{AdS}_3$ gravity where the Lagrangian of the dual $2d$ conformal field theory has been deformed by an arbitrary function of the energy-momentum tensor. In addition to the $\TT$ deformation, examples of such functions include a class of marginal stress tensor deformations which are special because they leave the generating functional of connected correlators unchanged up to a redefinition of the source and expectation value. Within this marginal class, we identify the unique deformation that commutes with the $\TT$ flow, which is the root-$\TT$ operator, and write down the modified boundary conditions corresponding to this root-$\TT$ deformation. We also identify the unique marginal stress tensor flow for the cylinder spectrum of the dual CFT which commutes with the inviscid Burgers' flow driven by $\TT$, and we propose this unique flow as a candidate root-$\TT$ deformation of the energy levels. We study BTZ black holes in $\mathrm{AdS}_3$ subject to root-$\TT$ deformed boundary conditions, and find that their masses flow in a way which is identical to that of our candidate root-$\TT$ energy flow equation, which offers evidence that this flow is the correct one. Finally, we also obtain the root-$\TT$ deformed boundary conditions for the gauge field in the Chern-Simons formulation of $\mathrm{AdS}_3$ gravity.

\baselineskip=18pt

\end{abstract}

\end{titlepage}

\tableofcontents

\section{Introduction} \label{intro}

A promising strategy for learning more about holography is to begin with a relatively well-understood holographic correspondence and then to deform it in some controlled way. We will focus on the case of an asymptotically $\mathrm{AdS}_3$ bulk which is dual to a two-dimensional conformal field theory. Given such a holographic boundary theory, we can view the $\mathrm{CFT}_2$ as essentially defining the $3d$ gravitational theory. More precisely, the $\mathrm{CFT}_2$ defines the boundary conditions which the fields of the bulk gravity theory should obey at infinity.

To take a concrete example, we recall that every translation-invariant quantum field theory admits a conserved stress tensor operator $T_{\alpha \beta}$. In the holographic dictionary, this boundary stress tensor operator is dual to the asymptotic bulk metric. One way to see this is to vary the action $S$ of the $3d$ gravitational theory, including both the Einstein-Hilbert term and appropriate boundary terms, and put this varied quantity on-shell using the bulk equations of motion. The resulting expression can be written as a boundary integral
\begin{align}\label{3d_bulk_variation}
    \delta S \Big\vert_{\text{on-shell}} = \frac{1}{2} \int_{\partial \mathcal{M}} d^2 x \, \sqrt{\gamma} \, T_{\alpha \beta} \, \delta \gamma^{\alpha \beta} \, ,
\end{align}
where $\mathcal{M}$ is the $3d$ spacetime manifold, $\partial \mathcal{M}$ is its $2d$ boundary, and $\gamma^{\alpha \beta}$ is the metric on $\partial \mathcal{M}$. In order for the on-shell variation of the action to vanish, we require $\delta \gamma^{\alpha \beta} = 0$ on $\partial \mathcal{M}$, which means that we impose Dirichlet boundary conditions on the metric near infinity. The quantity $T_{\alpha \beta}$ which appears in (\ref{3d_bulk_variation}) is then identified with the expectation value of the stress tensor operator of the boundary theory; the procedure described above furnishes an explicit expression for $T_{\alpha \beta}$ in terms of functions appearing in the Fefferman-Graham expansion of the metric near infinity. We interpret this by saying that the asymptotic metric $\gamma^{\alpha \beta}$ is a source for the stress tensor operator of the dual $\mathrm{CFT}_2$.

Now consider a deformation of the boundary conformal field theory. One familiar way to perform such a deformation is to add an integrated local operator to the action defining the $2d$ theory, so that
\begin{align}\label{general_O_deformation}
    S_0 \longrightarrow S_0 + \delta S = S_0 + \mu \int d^2 x \, \sqrt{\gamma} \, \mathcal{O} ( x ) \, , 
\end{align}
where $\mathcal{O} ( x )$ is a local operator and $\mu$ is a real parameter. Because the $\mathrm{CFT}_2$ defines the boundary conditions which the bulk fields obey at infinity, it is natural to expect that such a deformation would change these boundary conditions. This has been shown to be the case for many such multi-trace deformations \cite{Witten:2001ua}, at least subject to the usual caveats that one should restrict attention to the effects on light single-trace operators at large $N$.

For instance, one much-studied example is a double-trace deformation, where the object $\mathcal{O} ( x ) $ appearing in (\ref{general_O_deformation}) is the square of an operator which is dual to a fundamental field in the gravity theory. In this work, we will focus on deformations constructed from the stress-energy tensor $T_{\alpha \beta}$; because this operator is present in \emph{any} translation-invariant quantum field theory, such deformations are in a sense universal. An operator $\mathcal{O} ( x )$ which is constructed from products of components $T_{\alpha \beta}$ is a double-trace operator, by the definition given above, because the stress-energy tensor is dual to the bulk metric, which is a fundamental field of the gravity theory. One particularly nice Lorentz-invariant double-trace combination of components $T_{\alpha \beta}$ is
\begin{align}\label{OTT_def}
    \mathcal{O}_{\TT} = T^{\alpha \beta} T_{\alpha \beta} - \left( \tensor{T}{^\alpha_\alpha} \right)^2 \, .
\end{align}
This combination defines the so-called $\TT$ operator, which has generated considerable research interest in recent years. For the moment, let us focus on the properties of this operator purely as an object in the $2d$ boundary theory (and postponing its bulk interpretation). By infinitesimally adding this operator $\mathcal{O}_{\TT}$ at each step along a flow, one can define a one-parameter family of theories which obeys the differential equation
\begin{align}\label{TT_flow}
    \frac{\partial S^{(\lambda)}}{\partial \lambda} = - \frac{1}{2} \int d^2 x \, \sqrt{\gamma} \, \mathcal{O}^{(\lambda)}_{\TT} ( x ) \, ,
\end{align}
where the superscript $\lambda$ is meant to emphasize that we must re-compute the operator $\mathcal{O}_{\TT}^{(\lambda)}$ using the deformed stress tensor $T_{\alpha \beta}^{(\lambda)}$ at each point along the flow.\footnote{\label{notation_footnote}Throughout this work, we always use the symbol $\lambda$ to denote the parameter of a $\TT$ flow, while we use the symbol $\mu$ either for the parameter of a generic deformation of a boundary field theory, or for the parameter of the root-$\TT$ flow, which we introduce shortly. Note that $\mu$ is never a spacetime index.}

We make three sets of observations.
\begin{enumerate}[label=(\Roman*)]
    \item\label{TT_one} First note that $\mathcal{O}_{\TT}$ is a dimension-four operator, which means that it is irrelevant in the Wilsonian sense. As a consequence the flow equation (\ref{TT_flow}) is quite unusual, from the perspective of the renormalization group. Ordinarily one imagines beginning with a conformal field theory and then adding an integrated \emph{relevant} operator in the spectrum of the theory, which triggers a flow to the infrared.  In a loose sense, the $\TT$ flow is the inverse of this familiar paradigm, as we add an integrated \emph{irrelevant} operator which modifies the definition of the theory in the ultraviolet.
    
    \item\label{TT_two} The quantity $\mathcal{O}_{\TT}$ defined in (\ref{OTT_def}) involves products of stress tensor operators. As products of coincident local operators are generally divergent in quantum field theory, it is far from obvious that the combination $\mathcal{O}_{\TT}$ actually defines a local operator at all. However, it has been shown that one can begin with a point-split quantity 
    \begin{align}
        \mathcal{O}_{\TT} ( x, y ) = T^{\alpha \beta} ( x ) \, T_{\alpha \beta} ( y ) -  \tensor{T}{^\alpha_\alpha} ( x ) \, \tensor{T}{^\beta_\beta} ( y ) \, , 
    \end{align}
    and then take a coincident point-limit $\lim\limits_{y \to x} \mathcal{O}_{\TT} ( x, y )$. Surprisingly, this procedure \emph{does} define a sensible local operator, up to certain total derivative ambiguities which can be ignored \cite{Zamolodchikov:2004ce,Smirnov:2016lqw}.
    
    \item\label{TT_three} This deformation is ``nice'' in the sense that it preserves many desirable properties of the undeformed theory, such as integrability \cite{Smirnov:2016lqw,Cavaglia:2016oda,Conti:2018jho,Chen:2021aid} and supersymmetry \cite{Baggio:2018rpv,Chang:2018dge,Jiang:2019hux,Chang:2019kiu, Ferko:2019oyv,Ferko:2021loo,Ebert:2022xfh}. Relatedly, observables in the deformed theory can often be described with simple closed-form expressions; a few examples include the finite-volume spectrum \cite{Smirnov:2016lqw,Cavaglia:2016oda}, $S$-matrix \cite{Dubovsky:2017cnj}, and torus partition function \cite{Cardy:2018sdv,Datta:2018thy,Aharony:2018bad}.
\end{enumerate}

Because the operator $\mathcal{O}_{\TT}$ appears to be rather special from the field theory perspective, one might suspect that this deformation corresponds to some fairly natural modification of the asymptotic boundary conditions for the bulk fields in the holographic dual. This turns out to be the case \cite{Guica:2019nzm}. To see this, one first defines the $\lambda$-dependent quantities
\begin{align} \label{TT_deformed_gamma_T}\begin{split}
    \gamma^{(\lambda)}_{\alpha \beta} &= \gamma^{(0)}_{\alpha \beta} - 2 \lambda \widehat{T}_{\alpha \beta}^{(0)} + \lambda^2 \widehat{T}^{(0)}_{\alpha \rho} \,  \widehat{T}^{(0)}_{\sigma \beta} \, \gamma^{(0) \rho \sigma} \, , \\
    \widehat{T}^{(\lambda)}_{\alpha \beta} &= \widehat{T}^{(0)}_{\alpha \beta} - \lambda \widehat{T}^{(0)}_{\alpha \rho} \, \widehat{T}^{(0)}_{\sigma \beta} \gamma^{(0) \rho \sigma} \, ,
\end{split}\end{align}
where $\widehat{T}_{\alpha \beta} = T_{\alpha \beta} - \gamma_{\alpha \beta} \tensor{T}{^\rho_\rho}$ is the trace-reversed stress tensor. In terms of these quantities, the boundary action which solves the $\TT$ flow equation (\ref{TT_flow}) has the property that its variation can be written as
\begin{align}\label{deformed_delta_S}
    \delta S = \frac{1}{2} \int d^2 x \, \sqrt{\gamma^{(\lambda)} } \, T_{\alpha \beta}^{(\lambda)} \, \delta \gamma^{(\lambda) \alpha \beta} \, .
\end{align}
This is \emph{exactly} of the same form as the usual on-shell bulk variation, equation (\ref{3d_bulk_variation}), except written in terms of the $\lambda$-dependent metric and stress tensor. In order for the variation of the action to vanish, we now require that $\delta \gamma^{(\lambda) \alpha \beta} = 0$, which means that we impose Dirichlet boundary conditions on the deformed metric $\gamma^{(\lambda) \alpha \beta}$ at infinity. In terms of the original variables, this looks like a certain choice of mixed boundary conditions on the metric at infinity, since we now hold fixed a combination of the original metric $\gamma^{(0)}_{\alpha \beta}$ and its radial derivative, which is related to $T^{(0)}_{\alpha \beta}$.

One might ask whether there are other universal deformations constructed from stress tensors which admit interpretations as particularly simple modified boundary conditions. Another candidate is the recently-proposed root-$\TT$ operator \cite{Ferko:2022cix}, which is defined as
\begin{align}\label{root_TT_def}
    \mathcal{R} = \sqrt{ \frac{1}{2} T^{\alpha \beta} T_{\alpha \beta} - \frac{1}{4} \left( \tensor{T}{^\alpha_\alpha} \right)^2 } \, .
\end{align}
By way of comparison, let us revisit the three points \ref{TT_one} - \ref{TT_three} which we made concerning the $\TT$ operator and consider the analogous statements for root-$\TT$. 

\begin{enumerate}[label = ($\widetilde{\mathrm{\Roman*}}$)]

    \item Whereas $\TT$ is an irrelevant operator, the root-$\TT$ operator is classically \emph{marginal}. For instance, it has been checked in a large class of examples that the stress tensor of a root-$\TT$ deformed CFT still has vanishing trace. As a consequence, the coupling constant $\mu$ parameterizing the root-$\TT$ flow is dimensionless.
    
    \item Although $\TT$ is quantum-mechanically well-defined, it is not known whether the root-$\TT$ operator can be defined at the quantum level by point-splitting. Understanding the quantum properties of this operator remains an important open problem.
    
    \item The root-$\TT$ deformation shares some of the ``niceness'' properties of the ordinary $\TT$ deformation. For instance, flow equations for the root-$\TT$-deformed Lagrangian can often be solved in closed form \cite{Ferko:2022cix}, and the root-$\TT$ deformation preserves classical integrability in many examples \cite{Borsato:2022tmu}. However, formulas for root-$\TT$ deformed spectra, $S$-matrices, and partition functions have not been obtained.
\end{enumerate}

Although much less is known about the root-$\TT$ operator, there are many hints that this deformation might lead to an interesting class of models. One is the relation to the ModMax theory \cite{Bandos:2020jsw,Bandos:2020hgy,Bandos:2021rqy,Lechner:2022qhb} in four dimensions. This theory and its Born-Infeld extension obey $4d$ analogues of the root-$\TT$ and $\TT$ flow equations, respectively \cite{Babaei-Aghbolagh:2022uij,Conti:2018jho}, and both flows can be supersymmetrized \cite{Ferko:2022iru,Ferko:2023ruw}. The root-$\TT$ operator also appears in a flow equation which generates the $3d$ Born-Infeld Lagrangian or its supersymmetric extension \cite{Ferko:2023sps}. Further, the dimensional reduction of the ModMax theory is identical to the theory obtained by root-$\TT$ deforming a collection of $2d$ free scalars \cite{Babaei-Aghbolagh:2022leo, Conti:2022egv}. A $(0+1)$-dimensional version of the root-$\TT$ deformation was studied in \cite{Garcia:2022wad}, which also preserves integrability. This operator has been connected to ultra/non-relativistic limits and the BMS group in three dimensions \cite{Rodriguez:2021tcz,Bagchi:2022nvj}, and to nonlinear automorphisms of the conformal algebra \cite{Tempo:2022ndz}. See also \cite{Hou:2022csf} for an analysis of $\TT$ and root-$\TT$-like deformations using characteristic flows.

Given the interest in the root-$\TT$ operator from the field theory perspective, it is natural to ask whether there are modified boundary conditions for the bulk metric which implement this deformation, as (\ref{TT_deformed_gamma_T}) do in the $\TT$ case. In this work, we will argue that the answer to this question is yes, and the analogous expressions are
\begin{align}\begin{split}\label{root_TT_deformed_bcs}
    \gamma_{\alpha \beta}^{(\mu)} &= \cosh ( \mu ) \gamma_{\alpha \beta}^{(0)} + \frac{\sinh ( \mu )}{\mathcal{R}^{(0)}} \widetilde{T}_{\alpha \beta}^{(0)} \, , \\
    \Tt_{\alpha \beta}^{(\mu)} &= \cosh ( \mu ) \Tt_{\alpha \beta}^{(0)} + \sinh ( \mu ) \mathcal{R}^{(0)} \gamma^{(0)}_{\alpha \beta} \, ,
\end{split}\end{align}
where we have defined $\Tt_{\alpha \beta} = T_{\alpha \beta} - \frac{1}{2} \gamma_{\alpha \beta} \tensor{T}{^\rho_\rho}$, which is the traceless part of the stress tensor (not to be confused with the \emph{trace-reversed} stress tensor $\Th_{\alpha \beta}$), and 
\begin{align}
    \mathcal{R}^{(0)} &= \sqrt{ \frac{1}{2} T^{(0) \alpha \beta} T_{\alpha \beta}^{(0)} - \frac{1}{4} \left( \tensor{T}{^{(0)}^\alpha_\alpha} \right)^2 } \nonumber \\
    &= \sqrt{ - \det \left( \Tt_{\alpha \beta}^{(0)} \right) } \, , 
\end{align}
is the root-$\TT$ operator as before.

This means that -- from the viewpoint of holography -- the root-$\TT$ deformation plays a similar role as the $\TT$ deformation (or other $f(T)$ deformations), insofar as it imposes certain mixed boundary conditions where some function of the metric $\gamma_{\alpha \beta}$ and stress tensor $T_{\alpha \beta}$ is held fixed. However, the mixed boundary conditions which appear in the root-$\TT$ case are considerably more exotic because they involve the expression $\mathcal{R}^{(0)}$ which is non-analytic in the stress tensor $T^{(0)}_{\alpha \beta}$. Despite this unusual feature, we will show that the root-$\TT$ deformed boundary conditions have several surprisingly nice properties: for instance, various combinations of deformed quantities, like $T_{\alpha \beta}^{(\mu)} \, \delta \gamma^{(\mu) \alpha \beta}$ and $\det \left( \gamma_{\alpha \beta}^{(\mu)} \right)$, are equal to their undeformed values, and the root-$\TT$ deformed boundary conditions commute with the $\TT$-deformed boundary conditions, in a sense which we will make precise below. These unexpectedly simple relations, along with the pressing need to more deeply understand theories of root-$\TT$ type, motivates us to undertake a detailed study of the boundary conditions (\ref{root_TT_deformed_bcs}) in the remainder of the present work.

The layout of this paper is as follows. In Section \ref{sec:dictionary}, we review the holographic dictionary under general multi-trace deformations and apply these results to stress tensor deformations of $\mathrm{AdS}_3 / \mathrm{CFT}_2$. In Section \ref{sec:consistency}, we use consistency conditions, such as commutativity between $\TT$ and root-$\TT$, to identify the root-$\TT$ deformed boundary conditions and the flow equation for the finite-volume spectrum of the field theory under a root-$\TT$ deformation. In Section \ref{sec:AdS3RootTT}, we study $\mathrm{AdS}_3$ gravity with these root-$\TT$ deformed boundary conditions in both the metric and Chern-Simons formalisms and perform a holographic computation of the deformed spacetime mass which agrees with our flow equation for the root-$\TT$ deformed spectrum. In Section \ref{sec:conclusion}, we conclude and identify directions for future research.

\section{Holographic Dictionary for Stress Tensor Deformations}\label{sec:dictionary}

The connection between deformations of a field theory by local operators and modified boundary conditions for the gravity dual was pointed out in the early days of $\mathrm{AdS}$/$\mathrm{CFT}$. For double-trace deformations, the effect on the CFT partition function was discussed in \cite{Gubser:2002vv} and its relation to modified boundary conditions was explored in \cite{Klebanov:1999tb,Witten:2001ua,Berkooz:2002ug,Mueck:2002gm,Diaz:2007an,Hartman:2006dy,Gubser:2002zh}. A generalization to multi-trace deformations, which we will follow in section \ref{subsec:multi-trace}, was laid out in \cite{Papadimitriou:2007sj}. Although earlier work focused on relevant or marginal deformations, the analysis of irrelevant $\TT$ and $J \Tb$ deformations is described in \cite{Bzowski:2018pcy,Guica:2019nzm} and the lecture notes \cite{monica_notes}. See also \cite{Nguyen:2021pdz} for a recent discussion of the generating functional of connected stress tensor correlators in holography (without $\TT$-like deformations).

In this section we will review some of this well-known material with the goal of applying it to more general stress tensor deformations in $\mathrm{AdS}_3$/$\mathrm{CFT}_2$. An arbitrary scalar constructed from the stress tensor $T_{\alpha \beta}$ for a two dimensional field theory can be written as a function of two independent invariants,
\begin{align}\label{two_invariants}
    f \left( T_{\alpha \beta} \right) = f \left( \tensor{T}{^\alpha_\alpha} , T^{\alpha \beta} T_{\alpha \beta} \right) \, , 
\end{align}
since all higher traces of $T_{\alpha \beta}$ are related to these two by trace identities. At the classical level, any such function can be used to generate a deformation of a quantum field theory. The usual $\TT$ deformation corresponds to
\begin{align}
    f = T^{\alpha \beta} T_{\alpha \beta} - \left( \tensor{T}{^\alpha_\alpha}\right)^2 = \mathcal{O}_{\TT} \, , 
\end{align}
whereas the root-$\TT$ deformation is
\begin{align}
    f = \sqrt{ \frac{1}{2} T^{\alpha \beta} T_{\alpha \beta} - \frac{1}{4} \left( \tensor{T}{^\alpha_\alpha} \right)^2 } = \mathcal{R} \, .
\end{align}
As we will explain, for any operator $f$ which is chosen as a deformation of the two-dimensional field theory, one can find the modified generating functional in the large-$N$ limit by a path integral argument. For certain choices of $f$, it is then possible to explicitly solve for the modified boundary conditions in the $3d$ bulk gravity theory.

The surprising feature of a deformation by a \emph{marginal} combination of stress tensors, such as the root-$\TT$ operator, is that the additive shift in the generating functional of connected CFT correlators vanishes to leading order in $\frac{1}{N}$. Although such a deformation still has non-trivial effects on observables, this feature means that we will not be able to find the corresponding modified boundary conditions in the usual way. We will instead need to use a different argument which will be the subject of Section \ref{sec:consistency}.

\subsection{Multi-trace Deformations}\label{subsec:multi-trace}

We first review the reasoning which is used to find the change in the generating functional under a general multi-trace deformation of the $\mathrm{CFT}$. We follow \cite{Papadimitriou:2007sj} except for the mild generalization that we allow deformations by general scalar quantities constructed from operators carrying arbitrary indices, which allows us to include the case of stress tensor deformations. The analysis of this subsection applies in any dimension, so we will temporarily work in general spacetime dimension $d$ before specializing to $d = 3$ in later subsections. In this section we will also explicitly retain factors of $N$ in order to make the role of the large-$N$ limit more transparent. Although in later sections we will always implicitly work in a large-$N$ or large-$c$ limit in order to have a classical bulk gravity dual, we will typically not emphasize the central charge dependence of quantities appearing in path integrals.

Consider a $\mathrm{CFT}_d$ dual to a bulk $\mathrm{AdS}_{d+1}$ gravity theory. Let $\mathcal{O}_A$ be a collection of local operators in the conformal field theory which are single trace in the sense that each is dual to a fundamental field of the bulk gravity theory. For instance, one can imagine each $\mathcal{O}_A$ as being dual to a light scalar field in the $3d$ bulk, in which case $A$ is an internal index. When we specialize to stress tensor deformations in $\mathrm{AdS}_3$/$\mathrm{CFT}_2$, we will instead think of $\mathcal{O}_A$ as some component $T_{\alpha \beta}$ of the energy-momentum tensor, in which case $A$ is a multi-index of spacetime indices. For now, we will treat both cases uniformly by using an abstract index $A$ which may transform under the action of some unspecified Lie group $G$.

We will deform the action by adding $N^2 \mu \int d^d x \, \sqrt{\gamma} \, f ( \mathcal{O} )$. Here $f$ is a scalar function of $\mathcal{O}_A$, in the sense that it is invariant under the action of $G$, $\mu$ is a coupling constant with the appropriate dimension, and $\gamma_{\alpha \beta}$ is the boundary metric. We will assume that $f ( 0 ) = 0$ but make no further assumptions about the function $f$. For simplicity, in the remainder of this subsection we will assume $\gamma_{\alpha \beta} = \eta_{\alpha \beta}$ and thus omit factors of $\sqrt{\gamma}$. Quantities in the deformed theory will be decorated by a $\mu$ superscript or subscript, whereas quantities in the undeformed theory will carry a $(0)$ label. Our goal will be to find a relationship between the generating functionals of connected  $\mathcal{O}_A$  correlators in the deformed and undeformed theories, which we write as $W^{(\mu)} [ J^{(\mu)} ]$ and $W^{(0)} [ J^{(0)} ]$, respectively, and which are defined by the path integrals
\begin{align}\begin{split}\label{two_generating_functionals}
    e^{- W^{(0)} [ J^{(0)} ] } &= \int \mathcal{D} \psi \, \exp \left( - S_0 - N^2 \int d^d x \, J^{(0) A} ( x ) \mathcal{O}_A ( x ) \right) \, , \\
    e^{- W^{(\mu)} [ J^{(\mu)} ] } &= \int \mathcal{D} \psi \, \exp \left( - S_0 - N^2 \int d^d x \left( \mu f ( \mathcal{O} ) + J^{(\mu) A} (x) \mathcal{O}_A ( x ) \right) \right) \, .
\end{split}\end{align}
Here $J^{(0) A}$ and $J^{(\mu) A}$ are sources which are linearly coupled to the operators $\mathcal{O}_A$ in the undeformed and deformed theories, respectively. For simplicity, we suppress the $A$ indices on the sources $J^{(0) A}$, $J^{(\mu) A}$ when they appear as arguments in generating functionals. Correlators of the operators $\mathcal{O}_A$ are obtained from functional derivatives with respect to the source; for instance, the one-point function in the undeformed theory is given by
\begin{align}\label{OA_one_point}
    \langle \mathcal{O}_A \rangle_0 = \frac{1}{N^2} \frac{\delta W [ J^{(0)} ]}{\delta J^{(0) A}} \equiv \sigma_A ( x ) \, , 
\end{align}
where we introduce the shorthand $\sigma_A$ for convenience.

In the large-$N$ limit, all multi-point functions of operators $\mathcal{O}_A$ factorize into products of one-point functions of the form (\ref{OA_one_point}). This fact implies a simple relation between the two generating functionals in equation (\ref{two_generating_functionals}). To see this, we begin by changing variables in the path integral expression for $\exp \left( - W^{(\mu)} [ J^{(\mu)} ] \right)$, defining
\begin{align}\label{source_shift}
    \widetilde{J}_A = J^{(\mu)}_A + \mu \frac{\partial f ( \mathcal{O} ) }{\partial \mathcal{O}^A} \, .
\end{align}
This shift is performed because a general function $f ( \mathcal{O} )$ will have a term linear in $\mathcal{O}$ in its Taylor series expansion. Such a linear term in the effective action obstructs us from directly applying the results of large-$N$ factorization.\footnote{One way of understanding this, which is nicely explained in chapter 8 of \cite{coleman}, is to consider diagrammatics. For an effective action with a term linear in $\mathcal{O}$, there are infinitely many tree graphs that can be constructed with two external lines, since any lines may end on linear vertices. This complicates the large $N$ analysis, which usually proceeds by noting that the leading contribution at large $N$ comes from tree graphs with a minimal number of external lines (of which there should be finitely many). Performing the shift (\ref{source_shift}) removes the linear vertex and repairs this undesirable feature.} After implementing this shift to remove the linear term, the generating functional becomes
\begin{align}\label{generating_functional_multi_trace}
    \exp \left( - W^{(\mu)} \left( J^{(\mu)} \right) \right) = \int \mathcal{D} \psi \, \exp \left( - S_{0} - N^2 \int d^d x \, \left( \mu f ( \mathcal{O} ) + \widetilde{J}^{A} \mathcal{O}_A - \mu  \mathcal{O}_A \partial^A f \right) \right) \, ,  
\end{align}
where we introduce the shorthand $\partial^A f = \frac{\partial f ( \mathcal{O} ) }{\partial \mathcal{O}_A}$. We may now relate this expression to the undeformed generating functional evaluated on $\widetilde{J}$ as
\begin{align}\label{generating_functional_intermediate}
    &\exp \left( - W^{(\mu)} \left[ J^{(\mu)} \right] \right)  \nonumber \\
    &\quad = \int \mathcal{D} \psi \, \exp \left( - S_{0} - N^2 \int d^d x \, \left( \widetilde{J}^{A} \mathcal{O}_A \right) \right) \exp \left( - \mu N^2 \int d^2 x \, \left( f ( \mathcal{O} ) - \mathcal{O}_A \partial^A f \right) \right) \, \nonumber \\
    &\quad = e^{- W^{(0)} [ \widetilde{J} ] } \exp \left( - \mu N^2 \int d^d x \, \left( f ( \sigma ) - \sigma_A \partial^A f ( \sigma ) \right) \right) + O \left( \frac{1}{N} \right) \, .
\end{align}
The key observation is that the path integral on the second line of (\ref{generating_functional_intermediate}) defines a certain expectation value, namely of the second exponential factor, but in the large $N$ limit we may use factorization to evaluate this expectation value by replacing all instances of $\mathcal{O}_A$ with its one-point function $\sigma_A$. When $\mu = 0$, the argument of the second exponential factor vanishes and the two generating functions are equal, as expected.

The upshot of this manipulation is that, by taking logarithms of the first and last expressions of (\ref{generating_functional_intermediate}) and discarding subleading terms as $N \to \infty$, we conclude
\begin{align}
    - W^{(\mu)} \left[ J^{(\mu)} \right] = - W^{(0)} \big[ \,  \widetilde{J} \, \big] - \mu N^2 \int d^d x \, \left( f ( \sigma ) - \sigma_A \partial^a f ( \sigma ) \right) \, , 
\end{align}
or in terms of the rescaled generating functionals $w [ J ] = \frac{1}{N^2} W [ J ]$,
\begin{align}\label{final_generating_functional}
    w^{(\mu)} \left[ J^{(\mu)} \right] = w^{(0)} \big[ \,  \widetilde{J} \, \big] + \mu \int d^d x \, \left( f ( \sigma ) - \sigma_A \partial^A f ( \sigma ) \right) \, ,
\end{align}
where now $\sigma ( x ) = \frac{\delta w^{(0)} [ \tilde{J} ]}{\delta \tilde{J} (x)}$.

Equation (\ref{final_generating_functional}) is the main result which allows us to find the change in the generating functional under an arbitrary multi-trace deformation, including by non-analytic operators like root-$\TT$. The deformation by any such operator has two separate effects. First, the generating functional $w^{(\mu)}$ is shifted by a term involving an integral of $f(\sigma) - \sigma_A \partial^A f ( \sigma )$. Second, the effective source $J^{(\mu) A}$ which is used for computing one-point functions is shifted by a term proportional to the derivative of $f$, as in equation (\ref{source_shift}).

It will sometimes be convenient to use a varied form of equation (\ref{final_generating_functional}). The variations of the two generating functionals are defined by varying the sources and holding the corresponding one-point functions fixed:
\begin{align}\begin{split}\label{defn_variations}
    \delta W^{(0)} \left[ J^{(0)} \right] &= \int d^d x \, \langle O_A \rangle_0 \, \delta J^{(0) A} \, , \\
    \delta W^{(\mu)} \left[ J^{(\mu)} \right] &= \int d^d x \, \langle O_A \rangle_\mu \, \delta J^{(\mu) A } \, .
\end{split}\end{align}
Varying equation (\ref{final_generating_functional}) and substituting for $\delta w^{(0)}$ then gives
\begin{align}\label{generating_functional_varied}
    \delta w^{(\mu)} \left[ J^{(\mu)} \right] &= \delta w^{(0)} \big[ \,  \widetilde{J} \, \big] + \mu \int d^d x \, \delta \left( f ( \sigma ) - \sigma_A \partial^A f ( \sigma ) \right) \, \nonumber \\
    &= \int d^d x \, \left( \langle O_A \rangle_0 \, \delta \widetilde{J}^{A} + \mu \delta \left( f ( \sigma ) - \sigma_A \partial^A f ( \sigma ) \right) \right) \, .
\end{align}
Finally, equating this result with the expression for $\delta w^{(\mu)}$ in terms of $\delta J^{(\mu) A}$ gives
\begin{align}\label{varied_generating_funcational_matching}
    \int d^d x \, \langle O_A \rangle_\mu \, \delta J^{(\mu) A } = \int d^d x \, \left( \langle O_A \rangle_0 \, \delta \widetilde{J}^{A} + \mu \delta \left( f ( \sigma ) - \sigma_A \partial^A f ( \sigma ) \right) \right) \, .
\end{align}
Equation (\ref{varied_generating_funcational_matching}) will be useful for finding the modified boundary conditions for bulk fields after deforming the boundary field theory by some operator $f$. In particular, for a given deformation, one can match the coefficients of independent variations in (\ref{varied_generating_funcational_matching}) to obtain differential equations whose solution gives the deformed boundary conditions. This is especially helpful for studying more general deforming operators $f$ which depend both on the operators $\mathcal{O}_A$ and their sources $J^A$. Deformations by scalars constructed from the stress tensor $T_{\alpha \beta}$ are of this more complicated form, since they involve contractions with the boundary metric $\gamma^{\alpha \beta}$ which plays the role of the source for $T_{\alpha \beta}$. It is shown in \cite{Papadimitriou:2007sj} that an analysis of the varied equation (\ref{varied_generating_funcational_matching}) yields the correct modification to the stress tensor $T_{\alpha \beta}$ after a multi-trace deformation, which is convenient because this analysis is more straightforward than a direct computation from the deformed generating functional.

As a sanity check, it is useful to consider the case of a double-trace deformation,
\begin{align}
    \mu f ( \sigma ) = \frac{1}{2} \mu^{AB} \sigma_A \sigma_B \, , 
\end{align}
where $\mu^{AB}$ is a field-independent symmetric tensor. In this case,
\begin{align}
    \mu \partial^A f = \mu^{AB} \sigma_B \, ,
\end{align}
This means that the source $J_\mu^{(A)}$ satisfies
\begin{align}\label{double_trace_source_shift}
    J^{(\mu) A} = \widetilde{J}^A - \mu^{AB} \sigma_B \, , 
\end{align}
and thus the source has been shifted by a term which is linear in the corresponding expectation value. The deformed and undeformed generating functionals are related by
\begin{align}\label{double_trace_generating_shift}
    w^{(\mu)} \left[ J^{(\mu)} \right] = w^{(0)} \big[ \,  \widetilde{J} \, \big] - \frac{1}{2} \int d^d x \, \mu^{AB} \sigma_A \sigma_B \, .
\end{align}
Therefore, we see that a double-trace deformation is especially simple: although we deformed the action by \emph{adding} an integrated quantity proportional to $\int d^d x \, \mu^{AB} \sigma_A \sigma_B$, the generating functional has been deformed by \emph{subtracting} such a quantity.

\subsection{Compatibility with Hubbard-Stratonovich}\label{subsec:hub-strat}

In the case of a double-trace deformation, the general analysis of Section \ref{subsec:multi-trace} is equivalent to another common technique for deriving the modified holographic dictionary, namely the Hubbard-Stratonovich transformation. This method exploits the fact that a double-trace deformation is quadratic in fields and therefore can be decoupled by integrating in an appropriate auxiliary field. The Hubard-Stratonovich technique has a long history and was already used in \cite{Gubser:2002vv} to study the effect of a double-trace deformation on the dual CFT, which is nicely reviewed in \cite{Bzowski:2018pcy,Guica:2019nzm}. We note that a similar strategy was used in \cite{Cardy:2018sdv} in order to replace the $\TT$ operator with a coupling to a metric-like field $h_{\alpha \beta}$ and interpret the deformation as random geometry. However, this decoupling procedure does not straightforwardly apply to more general multi-trace deformations, such as the square-root-type deformation by $\mathcal{R}$. For completeness, we now briefly review this alternative derivation and confirm that the resulting modification to the generating functional is identical.

We again work in general spacetime dimension $d$ and focus on a deformation of the CFT action which takes the form
\begin{align}
    S_0 \longrightarrow S_0 + \frac{N^2}{2} \int d^d x \, \mu^{AB} \mathcal{O}_A \mathcal{O}_B \, , 
\end{align}
where the $\mathcal{O}_A$ are single-trace operators as before, and consider the deformed generating functional defined in equation (\ref{two_generating_functionals}),
\begin{align}\label{generating_functional_double_trace}
    e^{- W^{(\mu)} [ J^{(\mu)} ] } &= \int \mathcal{D} \psi \, \exp \left( - S_0 - N^2 \int d^d x \left( \frac{\mu^{AB}}{2} \mathcal{O}_A \mathcal{O}_B + J^{(\mu) A} (x) \mathcal{O}_A ( x ) \right) \right) \, .
\end{align}
We now integrate in an auxiliary field. To emphasize the similarity with the random geometry analysis of \cite{Cardy:2018sdv}, we will use the notation $h_A$ for this Hubbard-Stratonovich field. One has the path integral identity
\begin{align}\label{HS_identity}
    1 = \mathcal{N} \int \mathcal{D} h \, \exp \left( \frac{N^2}{2} \int d^2 x \, h^A \left( \mu^{-1} \right)_{AB} h^B \right) \, , 
\end{align}
Here the quantity $\mathcal{N}$ is a normalization factor which is defined by the property that it normalizes the path integral on the right side of (\ref{HS_identity}) to one. It can also be formally written as $\mathcal{N} = \frac{1}{\sqrt{ \det \left( \frac{\mu}{N^2} \right) } }$, although we will use the shorter expression $\mathcal{N}$ to avoid cluttering formulas. Inserting this identity into the expression (\ref{generating_functional_double_trace}) for the generating functional,
\begin{align}
    &\exp \left( - W^{(\mu)} \left( J^{(\mu)} \right) \right)  \nonumber \\
    &\quad = \mathcal{N} \int \mathcal{D} \psi \, \mathcal{D} h \, \exp \Bigg[ - S_{0} - N^2 \int d^d x \, \left( J^{(\mu) A} \mathcal{O}_A  + \frac{\mu^{AB}}{2} \mathcal{O}_A \, \mathcal{O}_B - \frac{1}{2} \, h^A \left( \mu^{-1} \right)_{AB} h^B \right) \Bigg] \, \nonumber \\
    &\quad = \mathcal{N} \int \mathcal{D} \psi \, \mathcal{D} h \, \exp \Bigg[ - S_{0} - N^2 \int d^d x \, \left( J^{(\mu) A} + \widehat{h}^A \right) \mathcal{O}_A - \frac{1}{2} \widehat{h}^A \mu^{-1}_{AB} \widehat{h}^B \Bigg]
 \, ,
\end{align}
where in the last step we have completed the square in the integrand by writing quantities in terms of a shifted auxiliary field
\begin{align}
    \widehat{h}^A = h^A + \mu^{AB} \mathcal{O}_B \, .
\end{align}
Seeing that the combination $\widehat{h}^A + J^{(\mu) A}$ now acts as the source for $\mathcal{O}_A$, we perform a second change of variables to
\begin{align}
    \widetilde{h}^A = \widehat{h}^A + J^{(\mu) A}
\end{align}
to find
\begin{align}\label{HS_intermediate}
    &\exp \left( - W^{(\mu)} \left( J^{(\mu)} \right) \right)  \nonumber \\
    &\quad =  \mathcal{N} \int \mathcal{D} \psi \, \mathcal{D} h \, \exp \Bigg[ - S_{0} - N^2 \int d^d x \, \left( \widetilde{h}^A \mathcal{O}_A + \frac{1}{2} \left( \widetilde{h}^A - J^{(\mu) A} \right) \mu^{-1}_{AB} \left( \widetilde{h}^B - J^{(\mu) B} \right) \right) \Bigg] \, \nonumber \\
    &\quad =  \mathcal{N} \int \mathcal{D} h \, e^{- W^{(0)} [ \widetilde{h} ] } \, \exp \Bigg[ - N^2 \int d^d x \, \left( \frac{1}{2} \left( \widetilde{h}^A - J^{(\mu) A} \right) \mu^{-1}_{AB} \left( \widetilde{h}^B - J^{(\mu) B} \right) \right) \Bigg] \, .
\end{align}
In the second step we have noted that performing the path integral including the first two terms in the exponential, $S_0$ and the coupling $\widetilde{h}^A \mathcal{O}_A$, defines the undeformed generating functional $\exp \left( - W^{(0)} \big[ \, \widetilde{h} \, \big] \right)$, since $\widetilde{h}^A$ acts as the source for $\mathcal{O}_A$. We have also implicitly used large-$N$ factorization, since the third term in the exponential also depends on the operators $\mathcal{O}_A$. In the last line of (\ref{HS_intermediate}), all implicit instances of such operators are understood to be replaced with the corresponding one-point functions.

In the large $N$ limit, the remaining path integral over $h$ can be performed using the saddle point approximation. The saddle occurs at the point $\widetilde{h}^A$ which satisfies
\begin{align}\label{htilde_saddle}
    - \frac{\delta W^{(0)} [ \widetilde{h} ] }{\delta \widetilde{h}^A} - \mu^{-1}_{AB} \left( \widetilde{h}^B - J^{(\mu) B} \right) = 0 \,.
\end{align}
On the other hand, the quantity $- \frac{\delta W^{(0)} [ \widetilde{h} ] }{\delta h^A}$ defines the one-point function $\langle \mathcal{O}_A \rangle_0 \equiv \sigma_A$, where we again introduce the shorthand $\sigma_A$ for the undeformed expectation value of $\mathcal{O}_A$.

Because we are modifying the field theory, the local operator $\mathcal{O}_A^{(0)}$ in the undeformed theory could correspond to some different operator $\mathcal{O}_A^{(\mu)}$ in the deformed theory. Therefore, in principle we should distinguish between deformed and undeformed operators, in addition to distinguishing between deformed and undeformed \emph{expectation values} which we have written as $\langle \; \cdot \; \rangle_\mu$ and $\langle \; \cdot \;  \rangle_0$, and which differ in that they are computed using path integrals weighted by different actions. However, we will see that for both the $\TT$ deformation and the root-$\TT$ deformation, the deformed and undeformed operators agree:
\begin{align}\label{same_operators}
    \mathcal{O}_A^{(0)} = \mathcal{O}_A^{(\mu)} \, .
\end{align}
More precisely, we will see that the derivatives of the operators $\mathcal{O}^{(\lambda)}_{\TT}$ and $\mathcal{R}^{(\mu)}$ with respect to the appropriate flow parameters $\lambda$ and $\mu$, respectively, both vanish. Thus we will simply assume that (\ref{same_operators}) holds in the present analysis; one can view this as an extra condition one might impose in order to single out a preferred class of deforming operators.

Solving equation (\ref{htilde_saddle}) then yields $\widetilde{h}^A = J^{(\mu) B} + \mu^{AB} \sigma_B$. We therefore find that
\begin{align}\label{HS_saddle}
    \exp \left( - W^{(\mu)} [ J^{(\mu) } ] \right) \sim \exp \left( - W^{(0)} \left[ J^{(\mu) B} + \mu^{AB} \sigma_B \right] - \frac{N^2}{2} \int d^d x \, \mu^{AB} \sigma_A \sigma_B \right) \, . 
\end{align}
Here we write $\sim$ to indicate both overall proportionality, since the saddle point integral introduces an additional prefactor which we will not track, and also the approximation to leading order in $\frac{1}{N}$. Taking logarithms and discarding the normalization, we conclude
\begin{align}
    W^{(\mu)} [ J^{(\mu)} ] &= W^{(0)} \big[ \, \widetilde{J} \, \big] - \frac{N^2}{2} \int d^d x \, \mu^{AB} \sigma_A \sigma_B \, , \nonumber \\
    \widetilde{J}^A &= J^{(\mu) A} + \mu^{AB} \sigma_B \, .
\end{align}
We see that this exactly reproduces equations (\ref{double_trace_source_shift}) and (\ref{double_trace_generating_shift}), after shifting to the re-scaled generating functionals $w$ by dividing through by $N^2$.

Therefore, the two approaches that we have described are equivalent for the case of double-trace deformations. Both computations use the assumption of large $N$ in a key way. In the first method we used large $N$ factorization in equation (\ref{generating_functional_intermediate}), and in the Hubbard-Stratonovich approach we used both large-$N$ factorization in equation (\ref{HS_intermediate}) and a saddle-point approximation in equation (\ref{HS_saddle}).

However, it is important to emphasize that the first method is more general since it applies to arbitrary multi-trace deformations. The Hubbard-Stratonovich technique crucially relies on the path integral identity (\ref{HS_identity}), which is a Gaussian integral and can therefore only introduce a quadratic dependence on $h^A$. Such a quadratic auxiliary field term is sufficient to decouple a double-trace deformation like the usual $\TT$, but for more general operators such as root-$\TT$, we will instead resort to the multi-trace analysis.

\subsection{Stress Tensor Deformations of $\mathrm{AdS}_3 / \mathrm{CFT}_2$}\label{sec:stress_tensor_AdS/CFT}

In the remainder of this work, we will focus on deformations which are constructed from the energy-momentum tensor rather than from general operators $\mathcal{O}_A$. It is worth pointing out that such deformations are qualitatively different in three bulk spacetime dimensions, which is our primary case of interest. In $\mathrm{AdS}_3$, the bulk metric has no local propagating degrees of freedom. As a result, we do not need to impose the usual restrictions that a deforming operator built from $\mathcal{O}_A$ be relevant or marginal in order to retain analytic control.

An irrelevant deformation built from an operator $\mathcal{O}_A$ which is dual to a dynamical field, such as a light scalar, would generically backreact on the metric and therefore become difficult to study. In contrast, an irrelevant deformation constructed from the $2d$ stress tensor $T_{\alpha \beta}$, such as the $\TT$ deformation, does not lead to any backreaction because the dual field is the (non-dynamical) bulk metric $g_{\alpha \beta}$. This means that we are free to consider deformations by \emph{any} scalar function $f ( T )$ of the stress tensor, even those with arbitrarily large dimension, and study the resulting mixed boundary conditions in the $\mathrm{AdS}_3$ bulk.

As we mentioned around equation (\ref{two_invariants}), the most general Lorentz scalar which can be constructed from a $2d$ stress tensor $T_{\alpha \beta}$ is
\begin{align}
    f \left( T_{\alpha \beta} \right) = f \left( x_1, x_2 \right) \, , \qquad  x_1 = \tensor{T}{^\alpha_\alpha} \, , \qquad x_2 = T^{\alpha \beta} T_{\alpha \beta}  \, ,
\end{align}
where we introduce $x_1 = \Tr ( T )$ and $x_2 = \Tr ( T^2 )$.

In the notation of Section \ref{subsec:multi-trace}, this corresponds to $\mathcal{O}_A = T_{\alpha \beta}$ and $\sigma_A = \langle T_{\alpha \beta} \rangle_0$, where $A$ is a multi-index of two boundary spacetime indices. We note that
\begin{align}
    \frac{\partial f}{\partial T_{\alpha \beta}} = \frac{\partial f}{\partial x_1} \gamma^{\alpha \beta} + 2 \frac{\partial f}{\partial x_2} T^{\alpha \beta} \, , 
\end{align}
where $\gamma_{\alpha \beta}$ is the $2d$ metric.

We may now import the general results for the shift in the generating functional under a multi-trace deformation defined by
\begin{align}
    \frac{\partial S^{(\mu)}}{\partial \mu} = \int d^2 x \, \sqrt{\gamma} \, f ( x_1, x_2 ) \, .
\end{align}
Because the boundary metric $\gamma_{\alpha \beta}$ now plays a more important role, we restore factors of $\sqrt{\gamma}$ in integrals, which were omitted in Sections \ref{subsec:multi-trace} and \ref{subsec:hub-strat}.

Using equation (\ref{final_generating_functional}), we find
\begin{align}\label{fT_generating_functional}
    w^{(\mu)} \left[ J^{(\mu)} \right] = w^{(0)} \big[ \,  \widetilde{J} \, \big] + \mu \int d^2 x \, \sqrt{\gamma} \, \left( f ( x_1, x_2 ) - \left( x_1 \frac{\partial f}{\partial x_1} + 2 x_2 \frac{\partial f}{\partial x_2} \right) \right) \, ,
\end{align}
where now we use $x_1, x_2$ interchangeably for the operators $\tensor{T}{^\alpha_\alpha}$, $T^{\alpha \beta} T_{\alpha \beta}$ and the expectation values $\langle \tensor{T}{^\alpha_\alpha} \rangle $, $\langle T^{\alpha \beta} \rangle \langle T_{\alpha \beta} \rangle$, as justified by large-$N$ factorization.

In these formulas, the source $J^{(\mu)}$ which couples to the deformed stress tensor $T_{\alpha \beta}^{(\mu)}$ is the deformed metric $\gamma_{\alpha \beta}^{(\mu)}$. This means that the deformation by $f$ involves both single-trace operators and their sources, which makes the behavior of this deformation more complicated. While a deformation by a function which depends only on operators $\mathcal{O}_A$ (but not their sources $J^A$) shifts the sources and leaves the expectation values $\langle \mathcal{O}_A \rangle$ unchanged, a deformation which depends on both $\mathcal{O}_A$ and $J^A$ will shift both the sources and the one-point functions. In this case, as we discussed above, it is more convenient to use the varied equation (\ref{varied_generating_funcational_matching}), which allows us (in principle, at least) to find expressions for both the deformed sources and the deformed expectation values. In this context, the appropriate varied equation for a stress tensor deformation is
\begin{align}
    \hspace{-10pt} \int d^d x \, \sqrt{\gamma^{(\mu)}} \, T_{\alpha \beta}^{(\mu)} \, \delta \gamma^{(\mu) \alpha \beta} = \int d^d x \, \sqrt{\gamma^{(\mu)}} \left[ T^{(0)}_{\alpha \beta} \, \delta \gamma^{(0) \alpha \beta} + \mu \, \delta \left( f ( x_1, x_2 ) - \left( x_1 \frac{\partial f}{\partial x_1} + 2 x_2 \frac{\partial f}{\partial x_2} \right) \right) \right] \, ,
\end{align}
or after taking the limit as $\mu \to 0$ to obtain a differential equation,
\begin{align}\label{general_variational_method}
    \frac{\partial}{\partial \mu} \int d^d x \, \sqrt{\gamma^{(\mu)}} \, T_{\alpha \beta}^{(\mu)} \, \delta \gamma^{(\mu) \alpha \beta} = \int d^d x \, \sqrt{\gamma^{(\mu)}} \, \delta \left( f ( x_1, x_2 ) - \left( x_1 \frac{\partial f}{\partial x_1} + 2 x_2 \frac{\partial f}{\partial x_2} \right) \right) \, .
\end{align}
The operator $\delta$ appearing on the right side acts on both $T_{\alpha \beta}$ and $\gamma_{\alpha \beta}$. Given a particular choice of deformation $f(x_1, x_2)$, one can then attempt to match the $\delta \gamma_{\alpha \beta}$ and $\delta T_{\alpha \beta}$ terms on both sides of (\ref{general_variational_method}) and solve the resulting coupled differential equations in $\mu$ to obtain solutions for the deformed quantities $T_{\alpha \beta}^{(\mu)}$ and $\gamma_{\alpha \beta}^{(\mu)}$.

The known results for the $\TT$ deformation can be recovered by setting
\begin{align}
    f ( x_1, x_2 ) = - \frac{1}{2} \left( x_2 - x_1^2 \right) = - \frac{1}{2} \mathcal{O}_{\TT} \, ,
\end{align}
where the factor of $- \frac{1}{2}$ is a choice of normalization for the operator. Substituting this deformation $f$ into (\ref{general_variational_method}) and stripping off the integrals gives the condition
\begin{align}\label{TT_varied_flow}
    \partial_\lambda \left( \sqrt{\gamma^{(\lambda)}} T_{\alpha \beta}^{(\lambda)} \, \delta \gamma^{(\lambda) \alpha \beta} \right) = \delta \left( \sqrt{\gamma^{(\lambda)}} \left( T^{(\lambda) \alpha \beta} T_{\alpha \beta}^{(\lambda)} - \left( \tensor{T}{^{(\lambda)}^\alpha_\alpha} \right)^2 \right) \right) \, ,
\end{align}
where we have changed the label for the deformation parameter from $\mu$ to $\lambda$ to emphasize that this flow is associated to the $\TT$ deformation (see footnote \ref{notation_footnote}). The indices in equation (\ref{TT_varied_flow}) are raised and lowered with the deformed metric $\gamma^{(\lambda)}_{\alpha \beta}$. One can solve this equation with the initial conditions $\gamma^{(\lambda)}_{\alpha \beta} \to \gamma^{(0)}_{\alpha \beta}$, $T_{\alpha \beta}^{(\lambda)} \to T_{\alpha \beta}^{(0)}$ as $\lambda \to 0$, as described in \cite{Guica:2019nzm} and reviewed in appendix \ref{app:TT_bc_pde_soln}. The solution to this differential equation can be expressed in terms of the trace-reversed stress tensor, $\widehat{T}_{\alpha \beta} = T_{\alpha \beta} - \gamma_{\alpha \beta} \tensor{T}{^\rho_\rho}$, in terms of which one finds
\begin{align}\begin{split}\label{TT_def_bc_later}
    \gamma^{(\lambda)}_{\alpha \beta} &= \gamma^{(0)}_{\alpha \beta} - 2 \lambda \widehat{T}_{\alpha \beta}^{(0)} + \lambda^2 \widehat{T}^{(0)}_{\alpha \rho} \,  \widehat{T}^{(0)}_{\sigma \beta} \, \gamma^{(0) \rho \sigma} \, , \\
    \widehat{T}^{(\lambda)}_{\alpha \beta} &= \widehat{T}^{(0)}_{\alpha \beta} - \lambda \widehat{T}^{(0)}_{\alpha \rho} \, \widehat{T}^{(0)}_{\sigma \beta} \gamma^{(0) \rho \sigma} \, ,
\end{split}\end{align}
which reproduces equation (\ref{TT_deformed_gamma_T}) for the $\TT$-deformed boundary conditions which we quoted in the introduction.

One might ask whether there are other choices for the deforming operator $f$ which are distinguished in some sense. For instance, it is natural to ask whether there is any choice of $f$ for which the shift in the generating functional appearing in equation (\ref{fT_generating_functional}) vanishes. Such a function $f$ satisfies the differential equation
\begin{align}
    f ( x_1, x_2 ) = x_1 \frac{\partial f}{\partial x_1} + 2 x_2 \frac{\partial f}{\partial x_2} \, , 
\end{align}
which has the general solution
\begin{align}
    f ( x_1, x_2 ) = x_1 g \left( \frac{x_2}{x_1^2} \right)
\end{align}
where $g$ is an arbitrary function. We demand that this deformation is well-defined if the seed theory is a CFT, for which $x_1 = \tensor{T}{^\alpha_\alpha} = 0$. The only way for the argument of the function $g$ to remain finite when $x_1 = 0$ is if $g ( y ) = \sqrt{c_1 y}$, in which case
\begin{align}
    f ( x_1, x_2 ) = x_1 \sqrt{ c_1 \frac{x_2}{x_1^2} } = \sqrt{ c_1 x_2 } \, .
\end{align}
Choosing the normalization factor $c_1 = \frac{1}{2}$, we find
\begin{align}
    f ( x_1, x_2 ) = \sqrt{ \frac{1}{2} x_2 } = \sqrt{ \frac{1}{2} T^{\alpha \beta} T_{\alpha \beta} } = \mathcal{R} \, \big\vert_{\tensor{T}{^\alpha_\alpha} = 0 } \, .
\end{align}
Therefore, the only physical sensible stress tensor deformation of a CFT with a vanishing shift in (\ref{fT_generating_functional}) is, up to proportionality, the root-$\TT$ operator $\mathcal{R}$ defined in (\ref{root_TT_def}). Note that this argument fixes the dependence of $f$ on $x_2$ but not on $x_1$, since we have restricted to the case of a conformal theory for which $x_1 = 0$. We will determine the dependence on $x_1$ by demanding that this deformation commute with the $\TT$ deformation in Section \ref{sec:consistency}.

Suppose that we wish to identify the deformed metric $\gamma_{\alpha \beta}^{(\mu)}$ and stress tensor $T_{\alpha \beta}^{(\mu)}$ associated with a deformation by this operator $\mathcal{R}$. One immediately encounters the subtlety that the differential equation (\ref{general_variational_method}) reduces to
\begin{align}\label{rtt_variational_pde}
    \frac{\partial}{\partial \mu} \left( \sqrt{\gamma^{(\mu)}} T_{\alpha \beta}^{(\mu)} \, \delta \gamma^{(\mu) \alpha \beta } \right) = 0 \, .
\end{align} 
This means that the operator $\mathcal{R}$ is in the kernel of the map which sends deformations to sources on the right side of the differential equation (\ref{general_variational_method}). There are multiple solutions to equation (\ref{rtt_variational_pde}). The most obvious one is the trivial solution $\gamma^{(\mu)}_{\alpha \beta} = \gamma^{(0)}_{\alpha \beta}$ and $T^{(\mu)}_{\alpha \beta} = T^{(0)}_{\alpha \beta}$. Another, less obvious, solution can be conveniently written in terms of the traceless part of the stress tensor, $\widetilde{T}_{\alpha \beta} = T_{\alpha \beta} - \frac{1}{2} \gamma_{\alpha \beta} \tensor{T}{^\rho_\rho}$. That solution is
\begin{align}\begin{split}\label{root_TT_deformed_bcs_later}
    \gamma_{\alpha \beta}^{(\mu)} &= \cosh ( \mu ) \gamma_{\alpha \beta}^{(0)} + \frac{\sinh ( \mu )}{\mathcal{R}^{(0)}} \widetilde{T}_{\alpha \beta}^{(0)} \, , \\
    \Tt_{\alpha \beta}^{(\mu)} &= \cosh ( \mu ) \Tt_{\alpha \beta}^{(0)} + \sinh ( \mu )\mathcal{R}^{(0)} \gamma^{(0)}_{\alpha \beta} \, ,
\end{split}\end{align}
where $\mathcal{R}^{(0)}$ is the root-$\TT$ operator constructed from the undeformed metric and stress tensor. One can verify that the expressions (\ref{root_TT_deformed_bcs_later}) solve the differential equation (\ref{rtt_variational_pde}). In fact, several quantities of interest remain individually undeformed along this flow:
\begin{align}
    \det \left( \gamma_{\alpha \beta}^{(\mu)} \right)  = \det \left( \gamma_{\alpha \beta}^{(0)} \right) \, , \qquad T_{\alpha \beta}^{(\mu)} \, \delta \gamma^{(\mu) \alpha \beta } = T_{\alpha \beta}^{(0)} \, \delta \gamma^{(0) \alpha \beta } \, , \qquad \mathcal{R}^{(\mu)} = \mathcal{R}^{(0)} \, .
\end{align}
If the only condition we impose is that our deformed metric and stress tensor satisfy (\ref{rtt_variational_pde}), then there is no way to distinguish between the trivial solution and the $\mu$-dependent solution (\ref{root_TT_deformed_bcs_later}). Furthermore, it is not immediately obvious whether there are other solutions $\gamma^{(\mu)}_{\alpha \beta}$, $T^{(\mu)}_{\alpha \beta}$ which also satisfy this flow. For this reason, from the perspective of the deformed generating functional, we cannot uniquely identify a single solution for the deformed metric and stress tensor which corresponds to the root-$\TT$ deformation.

In order to circumvent this ambiguity, we will pursue a complementary analysis which does not rely on the deformed generating functional. Instead, we will stipulate a set of consistency conditions which we expect the root-$\TT$ deformed metric and stress tensor to satisfy, and demonstrate that (\ref{root_TT_deformed_bcs_later}) is the only solution with these properties. This gives an independent piece of evidence that these deformed boundary conditions are the correct ones which correspond to a root-$\TT$ deformation of the boundary theory. We turn to this argument in the next section.

\section{Root-\texorpdfstring{$\TT$}{TT} from Consistency Conditions}\label{sec:consistency}

We have seen that the root-$\TT$ deformation is subtle to treat holographically because it belongs to a class of deformations for which the combination
\begin{align}\label{root_TT_class_deformations}
    \int d^2 x \, \sqrt{\gamma^{(\mu)}} T_{\alpha \beta}^{(\mu)} \, \delta \gamma^{(\mu) \alpha \beta }
\end{align}
is independent of $\mu$. This class also includes trivial deformations, such as boundary diffeomorphisms or scale transformations, which leave the theory unchanged.

However, we expect that the root-$\TT$ deformation is \emph{not} such a trivial deformation, and should modify the behavior of the theory in some way. One piece of evidence for this is that the $2d$ root-$\TT$ deformation of a collection of bosons is the dimensional reduction of the $4d$ root-$\TT$ deformation of the free Maxwell theory \cite{Conti:2022egv}, which gives rise to the ModMax theory. This ModMax theory represents a genuine modification of the Maxwell theory, in that physical observables are modified; one example is that the ModMax theory exhibits birefringence whereas the Maxwell theory does not.

We would therefore like to distinguish the root-$\TT$ deformed theory from other deformations in the same class which obey (\ref{root_TT_class_deformations}). To do this, we will enumerate a list of properties which we expect the root-$\TT$ deformed theory to obey and search for the most general deformation which satisfies these properties. This will allow us to identify both a candidate set of deformed boundary conditions $\gamma^{(\mu)}_{\alpha \beta}$, $T^{(\mu)}_{\alpha \beta}$ and a proposal for the deformed finite-volume spectrum of a root-$\TT$ deformed CFT on a cylinder.

An important ingredient in this analysis is the assumption that the root-$\TT$ deformation commutes with the ordinary $\TT$ deformation, in a sense which we will make precise. This expectation is motivated by the observation that classical $\TT$ and root-$\TT$ flows for the Lagrangian exhibit this property in many examples \cite{Ferko:2022cix,Borsato:2022tmu}. The property of commuting with $\TT$ is \emph{not} shared by generic marginal stress tensor deformations. A simple example is the marginal deformation generated by the trace of the stress tensor,
\begin{align}\label{trace_deformation}
    \frac{\partial S}{\partial \mu} = \int d^2 x \, \sqrt{\gamma} \, \tensor{T}{^a_a} \, .
\end{align}
The flow (\ref{trace_deformation}) simply generates scale transformations, so a conformal field theory is invariant under such a deformation. However, a $\TT$-deformed field theory is not scale-invariant because the theory has a dimensionful scale set by $\lambda$. Thus, scale transformations do not commute with $\TT$. Deforming a CFT first by (\ref{trace_deformation}) and then $\TT$-deforming with parameter $\lambda$ is the same as only performing the $\TT$ step, whereas first deforming the CFT by $\TT$ and then performing the scale transformation (\ref{trace_deformation}) is \emph{not} the same as $\TT$-deforming by $\lambda$.

\subsection{Derivation of Deformed Boundary Conditions}\label{sec:boundary_condition_derivation}

We aim to find a one-parameter family of modified boundary conditions $\gamma^{(\mu)}_{\alpha \beta}$, $T_{\alpha \beta}^{(\mu)}$ with the following properties:

\begin{enumerate}[label=(\roman*)]

    \item\label{trace_assumption} The deformed boundary conditions should correspond to a classically marginal deformation of the dual field theory. This means that the parameter $\mu$ is dimensionless and that, if the undeformed stress tensor is traceless so that
    \begin{align}
        \gamma^{(0) \alpha \beta} T^{(0)}_{\alpha \beta} = 0 \, ,
    \end{align}
    then the deformed stress tensor is also traceless with respect to the deformed metric,
    \begin{align}
        \gamma^{(\mu) \alpha \beta} T^{(\mu)}_{\alpha \beta} = 0 \, .
    \end{align}

    \item\label{group_assumption} The deformations of the metric and stress tensor form a group. In particular, deformations compose. If we deform an initial configuration by $\mu_1$,
    \begin{align}
        \gamma^{(0)}_{\alpha \beta} \, , T^{(0)}_{\alpha \beta} \overset{\mu_1}{\longrightarrow} \gamma^{(\mu_1)}_{\alpha \beta} \, , T^{(\mu_1)}_{\alpha \beta} \, ,
    \end{align}
    and then use these quantities as the initial condition for a second deformation by $\mu_2$,
    \begin{align}
        \gamma^{(\mu_1)}_{\alpha \beta} \, , T^{(\mu_1)}_{\alpha \beta} \overset{\mu_2}{\longrightarrow} \gamma^{(\mu_1 + \mu_2)}_{\alpha \beta} \, , T^{(\mu_1 + \mu_2)}_{\alpha \beta} \, ,
    \end{align}
    then the doubly-deformed quantities are identical to those obtained from doing a single deformation by the total parameter $\mu_1 + \mu_2$,
    \begin{align}
        \gamma^{(0)}_{\alpha \beta} \, , T^{(0)}_{\alpha \beta} \xrightarrow{\mu_1 + \mu_2} \gamma^{(\mu_1 + \mu_2)}_{\alpha \beta} \, , T^{(\mu_1 + \mu_2)}_{\alpha \beta} \, .
    \end{align}
    Here we assume that $\mu = 0$ is the identity element, so that the deformed boundary conditions reduce to the undeformed boundary conditions as the deformation parameter is taken to zero. We further assume the group to be non-trivial, so a deformation by $\mu \neq 0$ must be different from the identity.
    
    \item\label{bc_commute_assumption} The root-$\TT$ deformation commutes with the ordinary $\TT$ deformation, in the following sense. If we first deform the metric and stress tensor using the $\TT$ deformed boundary conditions and flow by parameter $\lambda$, and then use these deformed quantities as the initial condition for a root-$\TT$ flow by parameter $\mu$, then the result is identical to first deforming by root-$\TT$ with parameter $\mu$ and then by $\TT$ with parameter $\lambda$.
    
    \begin{center}
    \begin{tikzcd}
    	{\gamma^{(0)}_{\alpha \beta} , T^{(0)}_{\alpha \beta} } && {\gamma^{(\lambda)}_{\alpha \beta} , T^{(\lambda)}_{\alpha \beta} } \\
    	\\
    	{\gamma^{(\mu)}_{\alpha \beta} , T^{(\mu)}_{\alpha \beta} } & {} & { \gamma^{(\lambda , \mu)}_{\alpha \beta} , T^{(\lambda , \mu)}_{\alpha \beta}  }
    	\arrow["{\mathcal{O}_{\TT}}", from=1-1, to=1-3]
    	\arrow["{\mathcal{R}}"', from=1-1, to=3-1]
    	\arrow["{{\mathcal{O}_{\TT}}}"', from=3-1, to=3-3]
    	\arrow["{\mathcal{R}}", from=1-3, to=3-3]
    \end{tikzcd}

\end{center}
    
\end{enumerate}

We will first use assumptions \ref{trace_assumption} and \ref{group_assumption} to determine the nature of the modified boundary conditions when the seed theory is conformal, and then use the third assumption to extend this procedure to the case when the undeformed theory is non-conformal.

\subsubsection*{\ul{\it Conformal Seed Theory}}

For a conformal seed theory satisfying $\gamma^{(0) \alpha \beta} T^{(0)}_{\alpha \beta} = 0$, the only independent dimensionful Lorentz scalar quantity in the problem is $T^{(0) \alpha \beta} T^{(0)}_{\alpha \beta}$. Ordinarily, there are two independent scalars that can be constructed from a general $2 \times 2$ matrix $M$ -- for instance, $\tr ( M )$ and $\tr ( M^2 )$ -- but we have assumed that the trace of the stress tensor vanishes. We can equivalently say that any Lorentz scalar built from a traceless stress tensor is a function of
\begin{align}
    \mathcal{R}^{(0)} = \sqrt{ \frac{1}{2} T^{(0) \alpha \beta} T_{\alpha \beta}^{(0)} } \, .
\end{align}
On the other hand, there are also only two functionally independent symmetric $2$-tensors available in this problem, namely $\gamma_{\alpha \beta}^{(0)}$ and $T_{\alpha \beta}^{(0)}$. Again, one could attempt to form a new independent $2$-tensor by taking products of the form
\begin{align}
    \left( T^2 \right)_{\alpha \beta} = T^{(0)}_{\alpha \gamma} \tensor{T}{^{(0)}^{\gamma}_\beta} \, , 
\end{align}
but because of the tracelessness condition, one has the identity
\begin{align}\label{traceless_simp}
    \left( T^2 \right)_{\alpha \beta} = \left( \mathcal{R}^{(0)} \right)^2 \gamma_{\alpha \beta} \, , 
\end{align}
so this combination does not actually give an independent tensor structure. Obviously, all higher powers of the stress tensor will also be proportional to either $\gamma_{\alpha \beta}^{(0)}$ or $T_{\alpha \beta}^{(0)}$ with coefficients that are functions of $\mathcal{R}^{(0)}$.

We therefore find that the most general ansatz for deformed symmetric tensors $\gamma_{\alpha \beta}^{(\mu)}$ and $T_{\alpha \beta}^{(\mu)}$ with the correct scaling dimensions is
\begin{align}\label{root_TT_cft_seed_ansatz}
    \gamma^{(\mu)}_{\alpha \beta} &= f_1 ( \mu ) \gamma_{\alpha \beta}^{(0)} + \frac{f_2 ( \mu )}{\mathcal{R}^{(0)}} T_{\alpha \beta}^{(0)} \, , \\
    T^{(\mu)}_{\alpha \beta} &= f_3 ( \mu ) T_{\alpha \beta}^{(0)} + f_4 ( \mu ) \mathcal{R}^{(0)} \gamma^{(0)}_{\alpha \beta} \, .
\end{align}
All that remains is to fix the four functions $f_i ( \mu )$. First we will use the assumption that the deformed stress tensor remains traceless with respect to the deformed metric, so that
\begin{align}
    \gamma^{(\mu) \alpha \beta} T^{(\mu)}_{\alpha \beta} = 0 \, .
\end{align}
This condition is satisfied if and only if
\begin{align}\label{f4_constraint}
    f_4 ( \mu ) = \frac{f_2 ( \mu ) f_3 ( \mu )}{f_1 ( \mu )} \, ,
\end{align}
which fixes one of the functions. 

Next we impose that subsequent deformations form a group, which is listed as assumption \ref{group_assumption} above. One the one hand, we can first deform the metric and stress tensor by parameter $\mu_1$ to obtain
\begin{align}\label{root_TT_cft_sequential}
    \gamma^{(\mu_1)}_{\alpha \beta} &= f_1 ( \mu_1 ) \gamma_{\alpha \beta}^{(0)} + \frac{f_2 ( \mu_1 )}{\mathcal{R}^{(0)}} T_{\alpha \beta}^{(0)} \, , \\
    T^{(\mu_1)}_{\alpha \beta} &= f_3 ( \mu_1 ) T_{\alpha \beta}^{(0)} + \frac{f_2 ( \mu_1 ) f_3 ( \mu_1 )}{f_1 ( \mu_1 )} \mathcal{R}^{(0)} \gamma^{(0)}_{\alpha \beta} \, ,
\end{align}
where we have used (\ref{f4_constraint}), and then use (\ref{root_TT_cft_sequential}) as the initial condition for a second deformation by parameter $\mu_2$. This gives one set of deformed quantities, $\gamma^{(\mu_1 + \mu_2)}_{\alpha \beta}$ and $T_{\alpha \beta}^{(\mu_1 + \mu_2)}$. On the other hand, we can deform all at once by a combined parameter $\mu_1 + \mu_2$, to yield
\begin{align}\begin{split}\label{root_TT_two_step}
    \gamma^{\prime (\mu_1 + \mu_2)}_{\alpha \beta} &= f_1 ( \mu_1 + \mu_2 ) \gamma_{\alpha \beta}^{(0)} + \frac{f_2 ( \mu_1 + \mu_2 )}{\mathcal{R}^{(0)}} T_{\alpha \beta}^{(0)} \, , \\
    T^{\prime (\mu_1 + \mu_2)}_{\alpha \beta} &= f_3 ( \mu_1 + \mu_2 ) T_{\alpha \beta}^{(0)} + \frac{f_2 ( \mu_1 + \mu_2 ) f_3 ( \mu_1 + \mu_2 )}{f_1 ( \mu_1 + \mu_2 )} \mathcal{R}^{(0)} \gamma^{(0)}_{\alpha \beta} \, .
\end{split}\end{align}
We have decorated the quantities in (\ref{root_TT_two_step}) with primes to emphasize that they may differ from the results $\gamma^{(\mu_1 + \mu_2)}_{\alpha \beta}$ and $T_{\alpha \beta}^{(\mu_1 + \mu_2)}$ of performing the two deformations sequentially. We then impose the constraints
\begin{align}\label{group_constraints}
    \gamma^{(\mu_1 + \mu_2)}_{\alpha \beta} = \gamma^{\prime (\mu_1 + \mu_2)}_{\alpha \beta} \, , \qquad T^{ (\mu_1 + \mu_2)}_{\alpha \beta} = T_{\alpha \beta}^{\prime (\mu_1 + \mu_2)} \, .
\end{align}
In performing the algebra to find the implications of equations (\ref{group_constraints}), we will assume that $f_1 > 0$ and $f_3 > 0$, which is convenient for simplifying expressions like $\sqrt{ f_1^2 }$ which appear in intermediate steps. This sign choice is reasonable because we are interested in deformations for which $f_1 ( 0 ) = f_3 ( 0 ) = 1$, so these functions should remain positive at least for sufficiently small deformation parameter.

After making this assumption, one finds that these equations hold if and only if
\begin{align}\begin{split}\label{function_groups_constraints}
    f_1 ( \mu_1 + \mu_2 ) &= f_1 ( \mu_1 ) f_1 ( \mu_2 ) + f_2 ( \mu_1 ) f_2 ( \mu_2 ) \, , \\
    f_2 ( \mu_1 + \mu_2 ) &= f_1 ( \mu_2 ) f_2 ( \mu_1 ) + f_1 ( \mu_1 ) f_2 ( \mu_2 ) \, , \\
    f_3 \left( \mu_1 + \mu_2 \right) &= f_3 ( \mu_1 ) f_3 ( \mu_2 ) \left( 1 + \frac{f_2 ( \mu_1 ) f_2 ( \mu_2 )}{f_1 ( \mu_1 ) f_1 ( \mu_2 ) } \right) \, .
\end{split}\end{align}
We can turn the first two conditions in (\ref{function_groups_constraints}) into differential equations for $f_1$ and $f_2$ with the initial condition that $f_1(0) = 1$ and $f_2 ( 0 ) = 0$, which is required so that the deformation reproduces the undeformed theory as $\mu \to 0$. For instance, taking a derivative of the first line of (\ref{function_groups_constraints}) with respect to $\mu_2$ and then taking $\mu_2 = 0$ yields
\begin{align}
    f_1' ( \mu_1 ) = f_1' ( 0 ) f_1 ( \mu_1 ) + f_2' ( 0 ) f_2 ( \mu_1 ) \, . 
\end{align}
To ease notation, let $f_1' ( 0 ) = a$ and $f_2' ( 0 ) = b$. Differentiating the second line of (\ref{function_groups_constraints}) with respect to $\mu_1$ and then taking $\mu_1$ to zero gives $f_2' ( \mu_2 ) = b f_1 ( \mu_2 ) + a f_2 ( \mu_2 )$. Thus we have a system of differential equations
\begin{align}
    f_1' ( x ) = a f_1 ( x ) + b f_2 ( x ) \, , \qquad f_2 ' ( x ) = b f_1 ( x ) + a f_2 ( x ) \, , 
\end{align}
whose general solution with the initial conditions $f_1 ( 0 ) = 1$, $f_2 ( 0 ) = 0$ is
\begin{align}
    f_1 ( x ) = e^{a x} \cosh ( b x ) \, , \qquad f_2 ( x ) = e^{a x} \sinh ( b x ) \, .
\end{align}
For this class of solutions, the constraint in the third line of equation (\ref{function_groups_constraints}) then imposes
\begin{align}
    f_3 \left( \mu_1 + \mu_2 \right) &= f_3 ( \mu_1 ) f_3 ( \mu_2 ) \left( 1 + \tanh ( b \mu_1 ) \tanh ( b \mu_2 ) \right) \, ,
\end{align}
which can be turned into a differential equation with the initial condition $f_3 ( 0 ) = 1$ as above. The result of this procedure is $f_3 ( x ) = e^{c x} \cosh ( b x )$, where $c$ is another constant.

Therefore, the most general $\mu$-dependent deformation of the metric and stress tensor consistent with our assumptions is
\begin{align}\begin{split}\label{root_TT_cft_a_b}
    \gamma^{(\mu)}_{\alpha \beta} &= e^{a \mu} \left( \cosh ( b \mu ) \gamma_{\alpha \beta}^{(0)} + \frac{\sinh ( b \mu ) }{\mathcal{R}^{(0)}} T_{\alpha \beta}^{(0)} \right) \, , \\
    T^{(\mu)}_{\alpha \beta} &= e^{c \mu} \left( \cosh ( b \mu ) T_{\alpha \beta}^{(0)} + \sinh ( b \mu ) \mathcal{R}^{(0)} \gamma^{(0)}_{\alpha \beta} \right) \, ,
\end{split}\end{align}
where $a$, $b$, $c$ are arbitrary constants.

Some comments are in order. First, the deformations associated with the parameters $a$ and $c$ are simply the freedom to re-scale the metric or stress tensor by a constant $\mu$-dependent factor, which is expected since such a scaling respects conformal symmetry and forms a group. However, any such change in coordinates can be un-done by a diffeomorphism along with a redefinition of the stress tensor by a multiplicative factor (which does not affect conservation). Therefore we will set $a = c = 0$ in what follows.

Second, the choice of the parameter $b$ corresponds to the scaling of the dimensionless flow parameter $\mu$, or equivalently to our choice of normalization for the operator $\mathcal{R}$. If $b = 0$, then there is no change in the metric or stress tensor (up to diffeomorphisms) for any value of $\mu$, and the group structure of our deformation is the trivial group. This violates our assumption \ref{group_assumption}, where we demand that the deformations form a non-trivial group, so $b = 0$ is forbidden. For simplicity, we will choose $b = 1$. With these choices, our modified boundary conditions for the case of a conformal seed theory are
\begin{align}\begin{split}\label{root_TT_cft_final}
    \gamma^{(\mu)}_{\alpha \beta} &= \cosh ( \mu ) \gamma_{\alpha \beta}^{(0)} + \frac{\sinh ( \mu ) }{\mathcal{R}^{(0)}} T_{\alpha \beta}^{(0)} \, , \\
    T^{(\mu)}_{\alpha \beta} &= \cosh ( \mu ) T_{\alpha \beta}^{(0)} + \sinh ( \mu ) \mathcal{R}^{(0)} \gamma^{(0)}_{\alpha \beta}  \, .
\end{split}\end{align}
We conclude that, up to diffeomorphisms and normalization, the unique choice of modified $\mathrm{AdS}_3$ boundary conditions which implement a marginal deformation of a $\mathrm{CFT}_2$ satisfying our assumptions are (\ref{root_TT_cft_final}). This is perhaps not too surprising, since there is only a single Lorentz invariant that be constructed from a traceless stress tensor, so we had only one choice of scalar $\mathcal{R}^{(0)}$ which could appear in the modified boundary conditions. However, when both $T^{\alpha \beta} T_{\alpha \beta}$ and $\tensor{T}{^\alpha_\alpha}$ are non-zero, there is more freedom in the deformation and we will require additional input to uniquely identify the appropriate analogue of (\ref{root_TT_cft_final}).

\subsubsection*{\ul{\it Non-Conformal Seed Theory}}

Typically one would not be interested in seed theories for which $\tensor{T}{^{(0)}^\alpha_\alpha} \neq 0$, since a generic non-conformal $2d$ QFT will not have any $\mathrm{AdS}_3$ dual. An important exception is if the seed theory \emph{itself} was obtained through applying an irrelevant stress tensor deformation, such as the $\TT$ deformation, to a conformal seed theory. Such a $\TT$-deformed CFT has a stress tensor with a non-vanishing trace,\footnote{\label{trace_flow_footnote} To wit, the trace satisfies the trace flow equation $\tensor{T}{^\alpha_\alpha} = - 2 \lambda \mathcal{O}_{\TT}$.} and yet it is dual to an $\mathrm{AdS}_3$ bulk with modified boundary conditions as we have described. One might therefore ask what happens if we use such a $\TT$-deformed theory as the input for a second deformation by root-$\TT$.

First we consider the most general expression for the modified boundary conditions $\gamma^{(\mu)}_{\alpha \beta}$ and $T^{(\mu)}_{\alpha \beta}$, which depend both on $\mu$ and on the undeformed quantities $\gamma^{(0)}_{\alpha \beta}$ and $T^{(0)}_{\alpha \beta}$, with the property that these expressions reduce to (\ref{root_TT_cft_final}) in the special case where $\tensor{T}{^{(0)}^\alpha_\alpha} = 0$. Because the stress tensor is no longer traceless, its square will not be proportional to the metric. As a result, one might believe that there are now three independent tensor structures in the problem, namely
\begin{align}
    \gamma_{\alpha \beta}^{(0)} \, , \quad T_{\alpha \beta}^{(0)} \, , \, \text{and} \quad \left( T^{(0)} \right)^2_{\alpha \beta} = T_{\alpha \rho}^{(0)} \gamma^{(0) \rho \sigma} T_{\sigma \beta}^{(0)} \, , 
\end{align}
and that the most general deformed metric $\gamma^{(\mu)}_{\alpha \beta}$ and stress tensor $T^{(\mu)}_{\alpha \beta}$ will each be a linear combination of three different tensor structures, with appropriate coefficients.

However, this is not the case and there are in fact still only two tensor structures. One can see this by writing quantities in terms of the traceless part of the stress tensor,
\begin{align}
    \widetilde{T}_{\alpha \beta} = T_{\alpha \beta} - \frac{1}{2} \tensor{T}{^\rho_\rho} \gamma_{\alpha \beta} \, .
\end{align}
Then $T_{\alpha \beta} = \widetilde{T}_{\alpha \beta} + \frac{1}{2} \gamma_{\alpha \beta} T$, where we write $T = \tensor{T}{^\alpha_\alpha}$ for the trace of the stress tensor to lighten notation. We also suppress the $(0)$ superscripts for the moment. Then the putative new tensor structure arising from the square of the stress tensor is
\begin{align}
    T^2_{\alpha \beta} &= T_{\alpha \sigma} \gamma^{\sigma \rho} T_{\rho \beta} \nonumber \\
    &= \left( \widetilde{T}_{\alpha \sigma} + \frac{1}{2} \gamma_{\alpha \sigma} T \right) \gamma^{\sigma \rho} \left( \widetilde{T}_{\rho \beta} + \frac{1}{2} \gamma_{\rho \beta} T \right) \nonumber \\
    &= \widetilde{T}^2_{\alpha \beta} + \widetilde{T}_{\alpha \beta} T + \frac{1}{4} T^2 \gamma_{\alpha \beta} \, .
\end{align}
We have already seen in equation (\ref{traceless_simp}) that $\widetilde{T}^2_{\alpha \beta} = \mathcal{R}^2 \gamma_{\alpha \beta}$. We conclude that there are still only two independent tensor structures $\gamma_{\alpha \beta}$ and $\widetilde{T}_{\alpha \beta}$, and that a generic candidate expression for a deformed symmetric tensor like $\gamma^{(\mu)}_{\alpha \beta}$ or $T^{(\mu)}_{\alpha \beta}$ must still be a linear combination of these two structures with appropriate scalar coefficients.

However, what \emph{has} changed is that there are now two Lorentz scalars that can be constructed from $T^{(0)}_{\alpha \beta}$ rather than just one. One way of parameterizing the two functionally independent scalars is $x_1 = \tensor{T}{^{(0)}^\alpha_\alpha}$, $x_2 = T^{(0) \alpha \beta} T^{(0)}_{\alpha \beta}$, as we have done above. It will be more useful to instead use $x_1$ and $\mathcal{R}^{(0)} = \sqrt{ \frac{1}{2} x_2 - \frac{1}{4} x_1^2 }$. Clearly any function of $x_1$ and $x_2$ can also be expressed as a function of $x_1$ and $\mathcal{R}^{(0)}$.

A convenient way to write most general deformed boundary conditions is
\begin{align}\begin{split}\label{root_TT_noncft_general}
    \gamma^{(\mu)}_{\alpha \beta} &= f_1 ( \mu , y ) \cosh ( \mu ) \gamma_{\alpha \beta}^{(0)} + f_2 ( \mu , y ) \frac{\sinh ( \mu ) }{\mathcal{R}^{(0)}}  \widetilde{T}_{\alpha \beta}^{(0)} \, , \\
    \widetilde{T}^{(\mu)}_{\alpha \beta} &= f_3 ( \mu , y )  \cosh ( \mu ) \widetilde{T}_{\alpha \beta}^{(0)} + f_4 ( \mu , y ) \sinh ( \mu )  \mathcal{R}^{(0)} \gamma^{(0)}_{\alpha \beta} \, .
\end{split}\end{align}
The functions $f_i$ may depend on $\mu$ and on the dimensionless combination 
\begin{align}
    y \equiv \frac{x_1}{\mathcal{R}^{(0)}} \, .
\end{align}
The expressions (\ref{root_TT_noncft_general}) give the most general deformed boundary conditions that can be constructed from a non-conformal seed theory. In order to reduce to the earlier results (\ref{root_TT_cft_final}) in the case of a CFT seed, we impose
\begin{align}
    f_i ( \mu, 0 ) = 1 \, .
\end{align}
We now expect that there should be many solutions for the functions $f_i$ which correspond to boundary deformations by different marginal operators. For instance, one could deform by some operator of the form
\begin{align}
    \mathcal{O} = \sqrt{ c_1 \tensor{T}{^{(0) \alpha \beta}} T^{(0)}_{\alpha \beta} + c_2 \left( \tensor{T}{^{(0)}^\alpha_\alpha} \right)^2 } + c_3 \tensor{T}{^{(0)}^\alpha_\alpha} \, , 
\end{align}
for any choice of $c_i$. There should exist some choice of modified boundary conditions corresponding to any such operator, and we expect that any such deformation will satisfy the group property described in assumption \ref{group_assumption} above.

Rather than perform a systematic investigation of all such allowed deformed boundary conditions, we will attempt to single out a unique deformation within this family by imposing our assumption \ref{bc_commute_assumption}, namely that this deformation commute with $\TT$. More precisely, we can first substitute a metric $\gamma_{\alpha \beta}^{(0)}$ and traceless stress tensor $T_{\alpha \beta}^{(0)}$ into the expressions (\ref{TT_def_bc_later}) to obtain the $\TT$-deformed boundary conditions $\gamma_{\alpha \beta}^{(\lambda)}$ and $T_{\alpha \beta}^{(\lambda)}$, and then substitute these two expressions into (\ref{root_TT_noncft_general}) to obtain
\begin{align}
    \gamma_{\alpha \beta}^{(\lambda, \mu)} \, , \quad T_{\alpha \beta}^{(\lambda, \mu)} \, .
\end{align}
On the other hand, we could instead first substitute $\gamma_{\alpha \beta}^{(0)}$, $T_{\alpha \beta}^{(0)}$ into (\ref{TT_def_bc_later}) to obtain $\gamma_{\alpha \beta}^{(\mu)}$ and $T_{\alpha \beta}^{(\mu)}$, and then plug these into (\ref{TT_def_bc_later}) to find
\begin{align}
    \gamma_{\alpha \beta}^{(\mu, \lambda)} \, , \quad T_{\alpha \beta}^{(\mu, \lambda)} \, .
\end{align}
We then impose the two constraints
\begin{align}\label{commute_constraint_algebraic}
    \gamma_{\alpha \beta}^{(\mu, \lambda)} = \gamma_{\alpha \beta}^{(\lambda, \mu)} \, , \qquad T_{\alpha \beta}^{(\mu, \lambda)} = T_{\alpha \beta}^{(\lambda, \mu)} \, .
\end{align}
This equation can be analyzed explicitly in components, by beginning with a general metric with entries $\gamma_{zz}$, $\gamma_{\overbar{z} \overbar{z}}$, $\gamma_{z \overbar{z}} = \gamma_{\overbar{z} z}$ and a general stress tensor compatible with the tracelessness constraints, and then evaluating both sides of (\ref{commute_constraint_algebraic}). We will omit the general expressions resulting from this procedure, which are not especially enlightening, and proceed to the implications of (\ref{commute_constraint_algebraic}). The constraint arising from demanding that $\gamma^{(\mu, \lambda)}_{z \overbar{z}} = \gamma^{(\lambda, \mu)}_{z \overbar{z}}$ is
\begin{align}
    f_2 ( \mu , y ) = 1 + \frac{4 + y^2}{4 y} \coth ( \mu ) \left( 1 - f_1 \right) \, .
\end{align}
Substituting this result and demanding that $\gamma^{(\mu, \lambda)}_{z z} = \gamma^{(\lambda, \mu)}_{z z}$ then yields
\begin{align}
    f_1 = 1 \, .
\end{align}
Therefore $f_1 = f_2 = 1$. Using these constraints and requiring that $T^{(\mu,\lambda)}_{z \overbar{z}} = T^{(\lambda, \mu)}_{z \overbar{z}}$ gives
\begin{align}
    f_4 = 1 + \frac{4 y \coth ( \mu ) }{4 + y^2} \left( 1 - f_3 \right) \, ,
\end{align}
and substituting this back into the equation $T^{(\mu,\lambda)}_{z z} = T^{(\lambda, \mu)}_{z z}$ yields
\begin{align}
    f_3 = 1 \, .
\end{align}
Therefore all four of the undetermined functions must satisfy $f_i = 1$ in order to commute with $\TT$. We conclude that the only expressions for modified boundary conditions which are consistent with our assumptions are
\begin{align}\begin{split}\label{root_TT_deformed_metric_bcs_later}
    \gamma_{\alpha \beta}^{(\mu)} &= \cosh ( \mu ) \gamma_{\alpha \beta}^{(0)} + \frac{\sinh ( \mu )}{\mathcal{R}^{(0)}} \widetilde{T}_{\alpha \beta}^{(0)} \, , \\
    \Tt_{\alpha \beta}^{(\mu)} &= \cosh ( \mu ) \Tt_{\alpha \beta}^{(0)} + \sinh ( \mu ) \mathcal{R}^{(0)} \gamma^{(0)}_{\alpha \beta} \, ,
\end{split}\end{align}
which are exactly the ones which we claim correspond to the root-$\TT$ deformation.

\subsection{Derivation of Deformed Energy Levels}\label{sec:Energy_Levels}

The $\TT$ deformation of a QFT on a cylinder of radius $R$ has a well-known effect on the spectrum of the theory \cite{Zamolodchikov:2004ce,Smirnov:2016lqw,Cavaglia:2016oda}. For an energy eigenstate $|n\rangle$ with energy $E_n ( R, \lambda )$ and momentum $P_n$, the deformed energy satisfies the inviscid Burgers' equation,
\begin{align}\label{inviscid_burgers}
    \frac{\partial E_n}{\partial \lambda} = E_n \frac{\partial E_n}{\partial R} + \frac{P_n^2}{R} \, , 
\end{align}
and the momentum $P_n$ remains unchanged. If the undeformed theory is a CFT, so that all of the undeformed energy levels are of the form $E_n^{(0)} \equiv E_n ( R, 0 ) = \frac{a_n}{R}$ for constants $a_n$, then (\ref{inviscid_burgers}) can be solved in closed form to obtain
\begin{align}
    E_n ( R , \lambda ) = \frac{R}{2 \lambda} \left( \sqrt{ 1 + \frac{4 \lambda E_n^{(0)} }{R} + \frac{4 \lambda^2 P_n^2}{R^2} } - 1  \right) \, .
\end{align}
The flow equation (\ref{inviscid_burgers}) can be derived by using the point-splitting definition of the local $\TT$ operator in any translation-invariant QFT and then expressing the components of the energy-momentum tensor in terms of $E_n$, $R$, $\frac{\partial E_n}{\partial R}$, and $P_n$. 

In the case of the root-$\TT$ deformation, it is not known whether one can define a local operator $\mathcal{R}$ by point-splitting. Therefore we cannot give a rigorous derivation of a flow equation like (\ref{inviscid_burgers}) for a quantum field theory deformed by root-$\TT$. However, in the spirit of the preceding subsection, we can attempt to list the properties that such a flow equation would necessarily possess and then see whether there exists a unique differential equation satisfying these properties. We stress that this type of argument does not constitute a proof that a root-$\TT$ deformed QFT exists and has a particular spectrum. It would merely show that, \emph{assuming} that the root-$\TT$ deformation is well-defined quantum-mechanically and behaves in the expected way, then there is only one possible flow equation that the spectrum could satisfy.\footnote{Note that such a differential equation for the cylinder spectrum is distinct from a flow equation for the classical Hamiltonian density, which has been studied in \cite{Tempo:2022ndz}.}

Before enumerating the desired properties of such a flow equation, it is useful to obtain a rough expectation for what the result might look like. Suppose, for the sake of argument, that there exists a local operator $\mathcal{R} ( x )$ in the spectrum of a QFT with the property that
\begin{align}\label{root_TT_optimistic}
    \langle \mathcal{R} \rangle = \sqrt{ \frac{1}{2} \langle T^{\alpha \beta} \rangle \langle T_{\alpha \beta} \rangle - \frac{1}{4} \langle \tensor{T}{^\alpha_\alpha} \rangle^2 } \, ,
\end{align}
and consider a deformation of the action given by $\frac{\partial S}{\partial \mu} = \int d^2 x \, \mathcal{R}$. By expressing the components of the stress tensor for the theory on a cylinder of radius $R$ in terms of energies and momenta, exactly as in the derivation of the inviscid Burgers' equation for $\TT$, one would arrive at a putative root-$\TT$ flow equation
\begin{align}
    \frac{\partial E_n}{\partial \mu} = \sqrt{ \frac{1}{4} \left( E_n - R \frac{\partial E_n}{\partial R} \right)^2 - P_n^2 } \, , 
\end{align}
or equivalently
\begin{align}\label{root_TT_energy_flow}
    \left( \frac{\partial E_n}{\partial \mu} \right)^2 - \frac{1}{4} \left( E_n - R \frac{\partial E_n}{\partial R} \right)^2 + P_n^2 = 0  \, . 
\end{align}
If the initial condition for this flow is a CFT, so $E_n \sim \frac{1}{R}$ and $P_n \sim \frac{1}{R}$, then the solution is
\begin{align}\label{root_TT_deformed_CFT_energies}
    E_n ( R, \mu ) = \cosh ( \mu ) E_n ( R, 0 ) + \sinh ( \mu ) \sqrt{ \left( E_n^{(0)} \right)^2 - P_n^2 } \, , 
\end{align}
where $E_n^{(0)} = E_n ( R , 0 )$, and again the momenta $P_n$ are unaffected.

Much like the root-$\TT$ flow equation for the Lagrangian, this candidate deformation of the energy levels forms a two-parameter family of commuting deformations along with the $\TT$ flow. Beginning with a CFT, the solution for the doubly-deformed spectrum is
\begin{align}\label{doubly_deformed}
    E_n ( R, \mu , \lambda ) = \frac{R}{2 \lambda} \left( \sqrt{ 1 + 4 \lambda \left( \cosh ( \mu ) E_n^{(0)} + \sinh ( \mu ) \sqrt{ \left( E_n^{(0)} \right)^{2} - P_n^2 } \right) + \frac{4 \lambda^2}{R^2} P_n^2 } -  1 \right) \, , 
\end{align}
where $E_n^{(0)} = E_n ( R, 0, 0)$. The spectrum (\ref{doubly_deformed}) satisfies the two commuting flow equations
\begin{align}
    \left( \frac{\partial E_n}{\partial \mu} \right)^2 - \frac{1}{4} \left( E_n - R \frac{\partial E_n}{\partial R} \right)^2 + P_n^2 = 0 \, , \qquad \frac{\partial E_n}{\partial \lambda} - E_n \frac{\partial E_n}{\partial R} - \frac{P_n^2}{R} = 0 \, , 
\end{align}
corresponding to the root-$\TT$ and $\TT$ deformations, respectively.

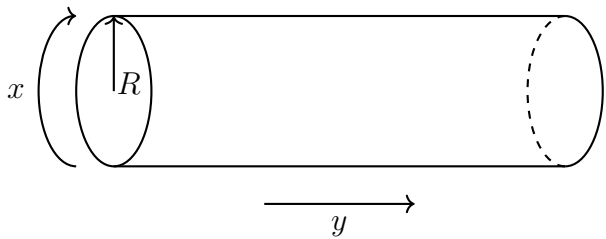
\begin{figure}[H]
\begin{tikzpicture}
\draw[rounded corners=35pt, thick](5.5,0)--(11.5,0);
\draw[rounded corners=35pt, thick](5.5,-2)--(11.5,-2);
\draw[thick] (11.5,-2) arc (-90:90:0.5cm and 1cm);
\draw[dashed, thick] (11.5,-2) arc (-90:-270:0.5cm and 1cm);
\draw[thick] (6,-1) arc (0:360:0.5cm and 1cm);
\node (a) at (0,0.5) {\phantom{$N$}};
\node (b) at (4,-2.5) {\phantom{$\mathbb{R}^+$}};
  \draw[->, thick] (7.5,-2.5) -- (9.5,-2.5);
    \draw (8.5,-2.8) node {$y$};
\draw[->, thick] (5,-2) arc (-90:90:-0.5 and 1);
\draw[->,thick] (5.5,-1)--(5.5,0);
  \draw (4.2,-1) node {$x$};
    \draw (5.7,-.9) node {$R$};
\end{tikzpicture}
\caption{We denote the compact direction by $x \sim x + R$, where $R$ is the radius of the spatial $S^1$, and write $y$ for the non-compact Euclidean time direction.}
\label{fig:cylinder}
\end{figure}

In Section \ref{sec:AdS3RootTT} we will give an argument that (\ref{root_TT_energy_flow}) is the correct flow equation using holography. For now, we would like to argue that this partial differential equation is the only reasonable possibility. To that end, we would like to look for the most general flow equation for a spectrum with the following properties:

\begin{enumerate}[label=(\alph*)]

    \item\label{marginal_flow_assumption} The flow is generated by a marginal stress tensor deformation. This means that it is a partial differential equation for $\frac{\partial E_n}{\partial \mu}$, where $\mu$ is a dimensionless parameter, which arises from a deformation of the Euclidean action by a Lorentz scalar constructed from the stress-energy tensor.

    \item The momentum $P_n$ is undeformed along the flow, so $P_n ( \mu ) = P_n ( 0 )$.

    \item\label{commuting_energy_assumption} The flow equation forms a two-parameter family of commuting flows with the inviscid Burgers' equation associated with the $\TT$ deformation.
\end{enumerate}
We will show that the only flow equation consistent with \ref{marginal_flow_assumption} - \ref{commuting_energy_assumption} is (\ref{root_TT_energy_flow}). First, it will be useful to express the possible Lorentz scalars constructed from $T_{\alpha \beta}$ in terms of energies and momenta. We work in Euclidean signature with coordinates $(x, y)$, where $x \sim x + R$ is the compact direction of the cylinder and $y$ is the Euclidean time direction. See figure \ref{fig:cylinder}.
In an energy eigenstate $| n \rangle$, the components of the stress tensor are related to the energy $E_n$ and momentum $P_n$ of the state as follows:
\begin{align}
    T_{yy} = - \frac{1}{R} E_n ( R ) \, , \qquad T_{xx} = - \frac{\partial E_n ( R )}{\partial R} \, , \qquad T_{xy} = \frac{i}{R} P_n ( R ) \, .
\end{align}
Furthermore, as we have described above, any Lorentz scalar constructed from $T_{\alpha \beta}$ is a function of the two independent invariants
\begin{align}
    x_1 = \tensor{T}{^\alpha_\alpha} = - \frac{2 E_n}{R} - 2 \frac{\partial E_n}{\partial R} \, , \qquad x_2 = T^{\alpha \beta} T_{\alpha \beta} = \left( \frac{\partial E_n}{\partial R} \right)^2 + \frac{E_n^2}{R^2} - \frac{2 P_n^2}{R^2} \, .
\end{align}
We are interested in a deformation of the form $\partial_\mu S_E = \int d^2 x \, f ( x_1, x_2 )$. The Euclidean Lagrangian density is the Hamiltonian density, whose integral over a spatial slice gives the energy $E_n$ of a state. Therefore this deformation of the Euclidean action can be written as
\begin{align}
    \partial_\mu E_n = \int_{0}^{R} \, dx \, f ( x_1, x_2 ) = R f ( x_1, x_2 ) \, .
\end{align}
We have assumed that this flow is generated by a marginal deformation, which means that $f ( x_1, x_2 )$ has mass dimension $2$. The function $f$ must therefore be homogeneous of degree $\frac{1}{2}$ in $x_2$ and degree $1$ in $x_1$. This allows us to scale the factor of $R$ into the arguments of $f$:
\begin{align}
    \frac{\partial E_n}{\partial \mu} = f \left( R x_1, R^2 x_2 \right) \, .
\end{align}
Next we use the second assumption, that the momenta $P_n$ are undeformed along the flow. This means that the theory is connected to some conformal field theory by a flow along which the momenta are constant, and therefore the dependence of momenta on the radius is fixed to be $P_n \sim \frac{1}{R}$ as in a CFT. It is convenient to define dimensionless momenta $p_n = R P_n$. We will also re-scale $x_1$ by a factor of $-\frac{1}{2}$ for convenience and write
\begin{align}\begin{split}
    \frac{\partial E_n}{\partial \mu} &= f \left( \tilde{x}_1, \tilde{x}_2 \right) \, , \\
    \tilde{x}_1 = E_n + R \frac{\partial E_n}{\partial R} \, , &\qquad \tilde{x}_2 = R^2 \left( \frac{\partial E_n}{\partial R^2} \right)^2 + E_n^2 - \frac{2 p_n^2}{R^2} \, .
\end{split}\end{align}
This is the most general ansatz for a flow equation consistent with our first two assumptions. We will now fix the dependence of $f$ on $\tilde{x}_1, \tilde{x}_2$ using the third assumption.

Consider a two-parameter family of theories with energies $E_n ( R, \lambda, \mu )$ which satisfy the simultaneous partial differential equations
\begin{align}
    \frac{\partial E_n}{\partial \mu} = f \left( \tilde{x}_1, \tilde{x}_2 \right) \, , \qquad \frac{\partial E_n}{\partial \lambda} = E_n \frac{\partial E_n}{\partial R} + \frac{p_n^2}{R^3} \, .
\end{align}
Differentiating the $\mu$ flow equation with respect to $R$ gives
\begin{align}\label{d2_mu_R}
    \frac{\partial^2 E_n}{\partial \mu \, \partial R} = \frac{\partial f}{\partial \tilde{x}_1} \left( 2 \frac{\partial E_n}{\partial R} + R \frac{\partial^2 E_n}{\partial R^2} \right) + \frac{2}{R^3} \frac{\partial f}{\partial \tilde{x}_2} \left( 2 p_n^2 + R^3 \frac{\partial E_n}{\partial R} \left( E_n + R \frac{\partial E_n}{\partial R} + R^2 \frac{\partial^2 E_n}{\partial R^2 } \right) \right) \, ,
\end{align}
while the derivative of the $\lambda$ flow equation with respect to $R$ is
\begin{align}
    \frac{\partial^2 E_n}{\partial \lambda \, \partial R} = E_n \frac{\partial^2 E_n}{\partial R^2} + \left( \frac{\partial E_n}{\partial R} \right)^2 - \frac{3 p_n^2}{R^4} \, .
\end{align}
We may compute the mixed second partial derivative $\frac{\partial^2 E_n}{\partial \mu \, \partial \lambda}$ in two ways. Taking a $\mu$ derivative of the expression for $\frac{\partial E_n}{\partial \lambda}$ and simplifying using (\ref{d2_mu_R}) yields
\begin{align}\label{lambda_then_mu}
    \frac{\partial^2 E_n}{\partial \mu \, \partial \lambda} &= f ( \tilde{x}_1 , \tilde{x}_2 ) \frac{\partial E_n}{\partial R} + E_n \frac{\partial f}{\partial \tilde{x}_1} \left( 2 \frac{\partial E_n}{\partial R} + R \frac{\partial^2 E_n}{\partial R^2 } \right) \nonumber \\
    &\quad + \frac{2 E_n}{R^3} \frac{\partial f}{\partial \tilde{x}_2} \left( 2 p_n^2 + R^3 \frac{\partial E_n}{\partial R} \left( E_n + R \frac{\partial E_n}{\partial R} + R^2 \frac{\partial^2 E_n}{\partial R^2 } \right) \right) \, .
\end{align}
On the other hand, the $\lambda$ derivative of $\frac{\partial E_n}{\partial \mu}$ is
\begin{align}\label{mu_then_lambda}
    \frac{\partial^2 E_n}{\partial \lambda \, \partial \mu} &= \frac{1}{R^3} \frac{\partial f}{\partial \tilde{x}_1} \left( - 2 p_n^2 + R^4 \left( \frac{\partial E_n}{\partial R} \right)^2 + R^3 E_n \frac{\partial E_n}{\partial R} + R^4 \frac{\partial^2 E_n}{\partial R^2} \right) \nonumber \\
    &\qquad + \frac{\partial f}{\partial \tilde{x}_2} \left( 2 E_n \left( \frac{p_n^2}{R^3} + E_n \frac{\partial E_n}{\partial R} \right) + \frac{2}{R^2} \frac{\partial E_n}{\partial R} \left( - 3 p_n^2 + R^4 \left( \frac{\partial E_n}{\partial R} \right)^2 + R^4 E_n \frac{\partial^2 E_n}{\partial R^2 } \right) \right) \, .
\end{align}
By our third assumption, the $\lambda$-flow must commute with the $\mu$-flow and hence the two mixed second partial derivatives must be equal. We set (\ref{lambda_then_mu}) equal to (\ref{mu_then_lambda}) and eliminate the variables $p_n$ and $\frac{\partial E_n}{\partial R}$ in favor of $y_1, y_2$. This leads to the differential equation
\begin{align}\label{mixed_second_pde}
    0 = \tilde{x}_1 f + \tilde{x}_1^3 \frac{\partial f}{\partial \tilde{x}_2} - 3 \tilde{x}_1 \tilde{x}_2 \frac{\partial f}{\partial \tilde{x}_2} - \tilde{x}_2 \frac{\partial f}{\partial \tilde{x}_1} + E_n \left( \tilde{x}_1 \frac{\partial f}{\partial \tilde{x}_1} + 2 \tilde{x}_2 \frac{\partial f}{\partial \tilde{x}_2} - f \right) \, .
\end{align}
The function $f$ can depend on the variables $\tilde{x}_1$ and $\tilde{x}_2$ but not on the function $E_n ( R, \lambda, \mu )$ directly. Thus in order for the equation (\ref{mixed_second_pde}) to be consistent, the $E_n$-dependent and $E_n$-independent terms must vanish separately:
\begin{align}\begin{split}\label{two_pieces_pde}
    0 &= \tilde{x}_1 \frac{\partial f}{\partial \tilde{x}_1} + 2 \tilde{x}_2 \frac{\partial f}{\partial \tilde{x}_2} - f \, , \\
    0 &= \tilde{x}_1 f + \tilde{x}_1^3 \frac{\partial f}{\partial \tilde{x}_2} - 3 \tilde{x}_1 \tilde{x}_2 \frac{\partial f}{\partial \tilde{x}_2} - \tilde{x}_2 \frac{\partial f}{\partial \tilde{x}_1} \, .
\end{split}\end{align}
The solution to the first line of (\ref{two_pieces_pde}) is
\begin{align}
    f = \tilde{x}_1 g \left( \frac{\tilde{x}_2}{\tilde{x}_1^2} \right) \, , 
\end{align}
where $g$ is an arbitrary function. Letting $X = \frac{\tilde{x}_2}{\tilde{x}_1^2}$ and substituting this ansatz into the second line of (\ref{two_pieces_pde}) gives
\begin{align}
    \left( X - 1 \right) \left( g ( X ) + \left( 1 - 2 X \right) g' ( X ) \right) = 0 \, .
\end{align}
There are two possibilities. The first factor vanishes if $X = 1$, which is the case when
\begin{align}
    T^{\alpha \beta} T_{\alpha \beta} = \frac{1}{4} \left( \tensor{T}{^\alpha_\alpha} \right)^2 \, .
\end{align}
This is a trivial case in which the stress tensor for the theory is degenerate and the two trace structures become dependent. We will discard this solution and require that $X$ is not identically equal to $1$. This leaves us with the second possibility, $g ( X ) + \left( 1 - 2 X \right) g' ( X ) = 0$, which has the solution
\begin{align}
    g ( X ) = c_1 \sqrt{ 2 X - 1 } \, , 
\end{align}
where $c_1$ is an arbitrary constant. Tracing back through the changes of variables, this corresponds to a deformation of the form
\begin{align}
    f ( x_1, x_2 ) = c_1 \sqrt{ 2 x_2 - x_1^2 } \, .
\end{align}
Choosing the normalization to be $c_1 = \frac{1}{2}$, we conclude that the function $f$ is
\begin{align}
    f \left( x_1, x_2 \right) = \sqrt{ \frac{1}{2} x_2 - \frac{1}{4} x_1^2 } = \mathcal{R} \, , 
\end{align}
which is precisely the root-$\TT$ operator. The flow equation for the energies is
\begin{align}
    \frac{\partial E_n}{\partial \mu} = \sqrt{ \frac{1}{4} \left( E_n - R \frac{\partial E_n}{\partial R} \right)^2 - P_n^2 } \, , 
\end{align}
and taking the square of this equation recovers (\ref{root_TT_energy_flow}).

Our conclusion is that there is only a single marginal deformation of the cylinder spectrum for a $2d$ quantum field theory which is constructed from the energy-momentum tensor and which commutes with the irrelevant $\TT$ flow. This unique deformation is the one which corresponds to the combination of stress tensors which appears in the classical root-$\TT$ deformation. We reiterate that this does not represent a proof that the root-$\TT$ operator is necessarily well-defined at the quantum level. However, if there exists \emph{any} deformation of the quantum theory with the properties that we listed, it must lead to exactly the flow equation which one would have na\"ively guessed would correspond to the root-$\TT$ deformation, as we did around equation (\ref{root_TT_optimistic}).

\section{AdS$_3$ Gravity with Root-$\TT$ Deformed Boundary Conditions}
\label{sec:AdS3RootTT}

In this section, we aim to show that the root-$\TT$ deformed boundary conditions derived in section \ref{sec:boundary_condition_derivation} are compatible with our proposed flow equation for the spectrum in section \ref{sec:Energy_Levels}. To do this, we will compute the mass (or energy) of a spacetime with root-$\TT$ deformed boundary conditions and compare this deformed mass to its undeformed value.

It is well-known that the notion of mass is subtle in a generally covariant theory, and there are many definitions of the total mass of a spacetime which are applicable in different contexts. In our case, since we are interested in asymptotically $\mathrm{AdS}_3$ spacetimes, it will be most convenient to define the spacetime mass as the integral of the quasi-local Brown-York stress tensor \cite{PhysRevD.47.1407}. In a $d$-dimensional spacetime  $\mathcal{M}$, this mass integral is given by
\begin{align}\label{spacetime_mass}
    M = \int_{\Sigma} d^{d-1} x \, \sqrt{\sigma} \, u^\mu T_{\mu \nu} \xi^\nu \, , 
\end{align}
where $\Sigma$ is a spacelike surface in the boundary $\partial \mathcal{M}$ with metric $\sigma_{\alpha \beta}$, $u^\mu$ is a timelike unit normal, and $\xi^\nu$ is the Killing vector associated with time translations. Our strategy will be to compute the mass (\ref{spacetime_mass}) by choosing a convenient coordinate system generated by a field-dependent change of variables which implements our root-$\TT$ deformed boundary conditions. Such field-dependent diffeomorphisms have also appeared in various works in the context of the ordinary $\TT$ deformation \cite{Conti:2018tca,Guica:2019nzm,Conti:2022egv}. 

It would be very interesting to study the mass of $\mathrm{AdS}_3$ spacetimes subject to modified boundary conditions using a more general prescription such as the covariant phase space formalism \cite{ Iyer:1994ys,Wald:1999wa}. The result of a mass calculation in this formalism is guaranteed to agree with (\ref{spacetime_mass}), but because this machinery maintains covariance, it may be possible to obtain mass flow equations associated with $\TT$ and root-$\TT$ deformations (or even more general stress tensor deformations) without resorting to a field-dependent diffeomorphism. 

We will also obtain the corresponding root-$\TT$ deformed boundary conditions in the Chern-Simons description of AdS$_3$ gravity. In this formalism, the definition of the deformed spacetime mass is not immediately obvious. As we will review around equation (\ref{eq:boundaryundeformedCS}), in the undeformed theory with conventional boundary conditions, it is straightforward to show that the mass (\ref{spacetime_mass}) is equal to the value of the Chern-Simons boundary term which imposes the appropriate boundary conditions. We will see by explicit computation that this remains true when this boundary term is modified to the one which imposes the root-$\TT$ deformed boundary conditions. This provides evidence that the Chern-Simons boundary term continues to yield the spacetime mass even with modified boundary conditions, which one might attempt to prove more generally by a computation in the canonical formulation.

\subsection{Metric Formalism}

First, we will briefly review the salient details in AdS$_3$ gravity to set the stage for the deformed energy spectrum computation.\footnote{For useful reviews on the AdS$_3$/CFT$_2$ correspondence, see \cite{Kraus:2006wn,Donnay:2016iyk,Compere:2018aar}.} Pure three-dimensional general relativity contains no local degrees of freedom, but is nontrivial enough to have black hole solutions and is a useful arena to study interesting phenomena in a controllable manner. A general solution of AdS$_3$ gravity can be written in the Fefferman-Graham expansion \cite{fefferman1985conformal}
\begin{equation}
\label{eq:AdS3Background}
ds^2 = \frac{\ell^2 d\rho^2}{4 \rho^2} + \left( \frac{ g^{(0)}_{\alpha \beta} (x^\alpha)}{\rho} + g^{(2)}_{\alpha \beta} (x^\alpha) + \rho g^{(4)}_{\alpha \beta} (x^\alpha)  \right) dx^\alpha \, dx^\beta\,,
\end{equation}
which terminates at second order \cite{Skenderis:1999nb} and where $\rho = 0$ is the AdS$_3$ boundary. The AdS$_3$ radius is $\ell$ and the three-dimensional Einstein equations determine $g^{(4)}_{\alpha \beta}$ in terms of the other two Fefferman-Graham expansion coefficients as
\begin{equation}
\label{eq:g4}
    g^{(4)}_{\alpha \beta} = \frac{1}{4} g^{(2)}_{\alpha \gamma} g^{(0) \gamma \delta} g^{(2)}_{\delta \beta}\,.
\end{equation}
Asymptotically AdS$_3$ solutions realize two copies of the Virasoro algebra, which are generated by Brown-Henneaux diffeomorphisms \cite{Brown:1986nw} that preserve the leading asymptotics of the metric (\ref{eq:AdS3Background}). Such diffeomorphisms correspond to conformal transformations in the $2d$ boundary theory. From the AdS/CFT dictionary \cite{Balasubramanian:1999re,deHaro:2000vlm}, the Fefferman-Graham quantity $g^{(2)}_{\alpha \beta}$ is proportional to the expectation value of the boundary CFT stress tensor
\begin{equation}
g^{(2)}_{\alpha \beta} = 8 \pi G \ell \left( T_{\alpha \beta} - g^{(0)}_{\alpha \beta} T^{ \alpha}_\alpha \right) \equiv 8 \pi G \ell \widehat{T}_{\alpha \beta} \, , 
\end{equation}
and $g^{(0)}_{\alpha \beta}$ is the metric on the boundary where the dual CFT lives. To derive the energy spectrum of this background \eqref{eq:AdS3Background} with the root-$TT$ deformed boundary conditions, we borrow some of the key methods developed to study holographic aspects of the double-trace $\TT$ deformation in the metric formalism \cite{Guica:2019nzm} and Chern-Simons formalism \cite{He:2020hhm} at large $N$. See appendix \ref{sec:AdS3TTApp} for a review of these methods. As a consequence of our analysis, we will also find that the bulk spacetime exhibits superluminal propagation for one sign of 
the root-$\TT$ deformation parameter $\mu$, which is also the case for the bad-sign $\TT$ deformation.

\subsubsection*{\ul{\it Root-$\TT$ Deformed Theory}}

In section \ref{sec:boundary_condition_derivation}, we argued that the root-$\TT$ deformed boundary metric and stress tensor are
\begin{align}\begin{split}\label{root_TT_deformed_bcs_repeat}
    \gamma_{\alpha \beta}^{(\mu)} = \cosh ( \mu ) \gamma_{\alpha \beta}^{(0)} + \frac{\sinh ( \mu )}{\mathcal{R}^{(0)}} \widetilde{T}_{\alpha \beta}^{(0)} \,, \quad
    \Tt_{\alpha \beta}^{(\mu)} = \cosh ( \mu ) \Tt_{\alpha \beta}^{(0)} + \sinh ( \mu ) \mathcal{R}^{(0)} \gamma^{(0)}_{\alpha \beta} \, .
\end{split}\end{align}
The strategy we follow here is motivated by the analysis of $\TT$-deformed AdS$_3$/CFT$_2$ in \cite{Guica:2019nzm}. As in the holographic analysis of the $\TT$ deformation, we will make two assumptions about the root-$\TT$ flow. The first is that this deformation is smooth and therefore preserves the boundary theory's degeneracy of states; for a black hole solution, this corresponds to the statement that the black hole horizon area is unchanged. The second assumption is that the deformation does not affect the momentum quantum number $P_n$ in the boundary field theory, which is quantized in units of $\frac{1}{R}$, where $R$ is the cylinder radius. 
Thus we will equate the deformed and undeformed areas and angular momenta. These assumptions imply the root-deformed energy spectrum which was derived by consistency conditions in section \ref{sec:Energy_Levels}. Unlike the $\TT$ deformation, the trace of the root-$\TT$ theory does not flow, as expected for a classically marginal deformation. 

We will now focus on the case of a Ba\~nados geometry  \cite{Banados:1998gg} which is parameterized by two quantities $\mathcal{L} ( u )$ and $\overbar{\mathcal{L}} ( v )$. A Ba\~nados geometry's Fefferman-Graham quantities are defined
\begin{equation}
    \begin{aligned}
    \label{eq:FG1}
    g^{(0)}_{\alpha \beta} dx^\alpha \, dx^\beta = du \, dv\,, \quad
        g^{(2)}_{\alpha \beta} dx^\alpha dx^\beta = \mathcal{L}(u) du^2 + \overbar{\mathcal{L}}(v) dv^2\,, \quad
        g^{(4)}_{\alpha \beta} dx^\alpha dx^\beta = \mathcal{L}(u) \overbar{\mathcal{L}}(v) dudv
    \end{aligned}
\end{equation}
implying that the metric \eqref{eq:AdS3Background} becomes
\begin{align}
\label{eq:BanadosMetric}
    ds^2 &= \frac{\ell^2 d\rho^2}{4\rho^2} + \frac{du \, dv}{\rho} + \mathcal{L}(u) du^2 + \overbar{\mathcal{L}}(v) dv^2 + \rho \mathcal{L}(u) \overbar{\mathcal{L}}(v) \, du \, dv \, . 
\end{align}
The root-$\TT$ deformed boundary metric and stress tensor given in \eqref{root_TT_deformed_bcs_repeat} are therefore
\begin{equation}
\begin{aligned}
\label{eq:root-TTstress}
\gamma^{(\mu)}_{\alpha \beta} &= (\cosh  \mu ) g_{\alpha \beta}^{(0)} + \frac{\sinh \mu }{2\sqrt{\mathcal{L}(u) \overbar{\mathcal{L}}(v) }}  g^{(2)}_{\alpha \beta} =   \frac{1}{2} \left( \begin{array}{cc}
  \sqrt{\frac{\mathcal{L}(u)}{\overbar{\mathcal{L}}(v)}} \sinh \mu  & \quad  \cosh \mu  \\
 \cosh \mu  & \quad   \sqrt{\frac{\overbar{\mathcal{L}}(v)}{\mathcal{L}(u)}} \sinh \mu
    \end{array}\right)\,, 
\\ \widetilde{T}^{(\mu)}_{\alpha \beta} &=
 \frac{\cosh \mu}{2} g^{(2)}_{\alpha \beta} +\sqrt{\mathcal{L}(u) \overbar{\mathcal{L}}(v)  } (\sinh \mu) g^{(0)}_{\alpha \beta} =  \frac{1}{2} \left( \begin{array}{cc}
      \mathcal{L}(u) \cosh \mu   & \quad \sqrt{\mathcal{L}(u) \overbar{\mathcal{L}}(v)}  \sinh \mu \\
    \sqrt{\mathcal{L}(u) \overbar{\mathcal{L}}(v)} \sinh \mu     & \quad \overbar{\mathcal{L}}(v) \cosh \mu
    \end{array} \right)\,,
 \end{aligned}
\end{equation}
where we work in conventions such that $4\pi G \ell = 1$ and substituted \eqref{eq:FG1} into \eqref{root_TT_deformed_bcs_repeat}. We also used $g^{(2)}_{\alpha \beta} =2 T_{\alpha \beta}$ and computed the operator $\displaystyle{\mathcal{R} = \sqrt{\mathcal{L}(u) \overbar{\mathcal{L}}(v)}}$. We now identify a field-dependent diffeomorphism to new coordinates $U, V$ defined by
\begin{equation}
\label{eq:rootTTcoordtransformation}
dU = \left( \cosh \frac{\mu}{2} \right) du + \sqrt{\frac{\overbar{\mathcal{L}}(v)}{\mathcal{L}(u)}} \left( \sinh \frac{\mu}{2} \right) dv\,, \quad dV = \left( \cosh \frac{\mu}{2} \right) dv + \sqrt{\frac{\mathcal{L}(u)}{\overbar{\mathcal{L}}(v)}} \left( \sinh \frac{\mu}{2} \right) du\,.  
\end{equation}
which has the property that the metric, when written in these new variables, returns to the standard form:
\begin{equation}
\gamma^{(\mu)}_{\alpha \beta} dx^\alpha \, dx^\beta = dU \, dV \, .
\end{equation}
Expressing \eqref{eq:rootTTcoordtransformation} in matrix notation, we may write the field dependent coordinate transformation and its inverse as
\begin{equation}
\begin{aligned}
\label{eq:inverserootTT}
\left( \begin{array} {c}
dU\\dV
\end{array}
\right) &= \left( \begin{array}{cc}
    \cosh \frac{\mu}{2}   & \quad \sqrt{\frac{\overbar{\mathcal{L}} (v) }{\mathcal{L}(u)  }} \sinh \frac{\mu}{2}  \\
   \sqrt{\frac{\mathcal{L}(u) }{\overbar{\mathcal{L}}(v)  }} \sinh \frac{\mu}{2}   & \quad \cosh \frac{\mu}{2} 
    \end{array} \right) \left( \begin{array} {c}
du\\dv
\end{array}
\right)\,, \\ 
\left( \begin{array} {c}
du\\dv
\end{array}
\right) &= \left( \begin{array}{cc}
    \cosh \frac{\mu}{2}   & \quad -\sqrt{\frac{\overbar{\mathcal{L}}(v) }{\mathcal{L}(u) }} \sinh \frac{\mu}{2}  \\
  - \sqrt{\frac{\mathcal{L}(u)}{\overbar{\mathcal{L}}(v)}} \sinh \frac{\mu}{2}   & \quad \cosh \frac{\mu}{2} 
    \end{array} \right) \left( \begin{array} {c}
dU\\dV
\end{array}
\right).
\end{aligned}
\end{equation}
For black hole solutions with constant $(\mathcal{L}(u), \overbar{\mathcal{L}}(v)) \equiv \left( \mathcal{L}_\mu, \overbar{\mathcal{L}}_\mu\right)$, the Fefferman-Graham quantities in the $(U, V)$ coordinates are
\begin{equation}
\begin{aligned}
\label{eq:FGRTT}
g^{(0)}_{\alpha \beta} dx^\alpha \, dx^\beta&= du \, dv
= -\frac{1}{2} \sinh \mu \left( \sqrt{\frac{\mathcal{L}_\mu}{\overbar{\mathcal{L}}_\mu}} dU^2 + \sqrt{\frac{\overbar{\mathcal{L}}_\mu}{\mathcal{L}_\mu}} dV^2 \right) + \cosh \mu dU \, dV\,,
\\ 
g^{(2)}_{\alpha \beta} dx^\alpha \, dx^\beta&= \mathcal{L}_\mu du^2 + \overbar{\mathcal{L}}_\mu dv^2 = \cosh \mu \left( \mathcal{L}_\mu dU^2 + \overbar{\mathcal{L}}_\mu dV^2 \right) - 2 \sqrt{\mathcal{L}_\mu \overbar{\mathcal{L}}_\mu} \sinh \mu dU \, dV\,,
\\
g^{(4)}_{\alpha \beta} dx^\alpha \, dx^\beta &= \mathcal{L}_\mu \overbar{\mathcal{L}}_\mu  du \, dv = \mathcal{L}_\mu \overbar{\mathcal{L}}_\mu \left( -\frac{1}{2} \sinh \mu \left( \sqrt{\frac{\mathcal{L}_\mu}{\overbar{\mathcal{L}}_\mu}} dU^2 + \sqrt{\frac{\overbar{\mathcal{L}}_\mu}{\mathcal{L}_\mu}} dV^2 \right) + \cosh \mu dU \, dV \right)\,.
\end{aligned}
\end{equation}
%
%
%
%
%
The metric \eqref{eq:AdS3Background} in terms of these Fefferman-Graham quantities \eqref{eq:FGRTT} at the event horizon $\rho_h = \left(\mathcal{L}_\mu \overbar{\mathcal{L}}_\mu \right)^{-\frac{1}{2}}$ is
\begin{equation}\hspace{-10pt}
\label{eq:metricdeformedrootTT}
ds^2 \big\vert_{\rho = \rho_h} = \frac{\ell^2 \mathcal{L}_\mu \overbar{\mathcal{L}}_\mu}{4} d\rho^2 + e^{-\mu} \left( \left( \sqrt{\mathcal{L}_\mu} + \sqrt{\overbar{\mathcal{L}}_\mu } \right)^2 d\phi^2 + \left( \sqrt{\mathcal{L}_\mu} - \sqrt{\overbar{\mathcal{L}}_\mu } \right)^2 dT^2 + 2 \left( \mathcal{L}_\mu - \overbar{\mathcal{L}}_\mu \right) dT \, d\phi \right)\,,
\end{equation}
where $(U, V) = (\phi + T, \phi - T)$. The undeformed and deformed event horizon areas are read off from \eqref{eq:metricdeformedrootTT}
\begin{equation}
A^{(0)} = \int^R_0 d\phi\sqrt{g_{\phi \phi}} = R\left( \sqrt{\mathcal{L}_0} + \sqrt{\overbar{\mathcal{L}}_0} \right), \quad A^{(\mu)} = \int^R_0 d\phi\sqrt{g_{\phi \phi}}|_{\rho_h} = R e^{-\frac{\mu}{2}} \left( \sqrt{\mathcal{L}_\mu} + \sqrt{\overbar{\mathcal{L}}_\mu} \right)\,. 
\end{equation}
Now to extract the deformed energy and angular momentum. Using \eqref{eq:root-TTstress} and \eqref{eq:FGRTT}, we write the components of the stress tensor
\begin{equation}
\label{eq:ttrootcomponentsenergymomentum}
T^{(\mu)}_{\alpha \beta} \, dx^\alpha \, dx^\beta = \frac{1}{2} (\mathcal{L}_\mu dU^2 + \overbar{\mathcal{L}}_\mu dV^2) 
= \frac{1}{2} ( \mathcal{L}_\mu +\overbar{\mathcal{L}}_\mu) (dT^2 + d\phi^2) + (\mathcal{L}_\mu - \overbar{\mathcal{L}}_\mu) \, d\phi \, dT \, .
\end{equation}
Restoring factors of $4 \pi G \ell$, the deformed energy and angular momentum from \eqref{eq:ttrootcomponentsenergymomentum} are
\begin{equation}
\label{eq:E&J}
E_\mu = \int^R_0 d\phi \, T^{(\mu)}_{TT} = \frac{R}{8\pi G \ell} (\mathcal{L}_\mu + \overbar{\mathcal{L}}_\mu), \quad J_\mu = \int^R_0 d\phi T^{(\mu)}_{T\phi} = \frac{R}{8 \pi G \ell}(\mathcal{L}_\mu - \overbar{\mathcal{L}}_\mu)\,.
\end{equation}
The root-$\TT$ deformed energy \eqref{eq:E&J} is a simple sum $\mathcal{L}_\mu +\overbar{\mathcal{L}}_\mu$, reminiscent of a CFT's energy, which is a sign that the root-$\TT$ deformed theory remains a CFT. This simplicity of energy ceases to exist for the $\TT$ deformation due to the deformed theory being non-conformal. In the $\TT$ deformation of AdS$_3$/CFT$_2$, the energy is
\begin{equation}
\label{eq:EnergyTT}
E_\lambda = \frac{R}{8\pi G \ell} \frac{\mathcal{L}_\lambda + \overbar{\mathcal{L}}_\lambda - 2 \rho_c \mathcal{L}_\lambda \overbar{\mathcal{L}}_\lambda }{1-\rho_c^2 \mathcal{L}_\lambda \overbar{\mathcal{L}}_\lambda  } \, , 
\end{equation}
with $(\mathcal{L}_\lambda, \overbar{\mathcal{L}}_\lambda)$ defined in \cite{Guica:2019nzm} and \eqref{eq:solsTT1}. The next ingredients are the areas and angular momenta, which obey
\begin{equation}
\label{eq:areaangularmomentumTTroot}
\sqrt{\mathcal{L}_0} + \sqrt{\overbar{\mathcal{L}}_0} = e^{- \frac{\mu}{2}} \left( \sqrt{\mathcal{L}_\mu} + \sqrt{\overbar{\mathcal{L}}_\mu} \right), \quad \mathcal{L}_0 - \overbar{\mathcal{L}}_0 = \mathcal{L}_\mu - \overbar{\mathcal{L}}_\mu \, , 
\end{equation}
and the solutions of \eqref{eq:areaangularmomentumTTroot} are
\begin{equation}
\label{eq:rootTTLLb}
\mathcal{L}_\mu = \left( \sqrt{\mathcal{L}_0} \cosh \frac{\mu}{2} + \sqrt{\overbar{\mathcal{L}}_0} \sinh \frac{\mu}{2} \right)^2\,, \quad \overbar{\mathcal{L}}_\mu = \left( \sqrt{\overbar{\mathcal{L}}_0} \cosh \frac{\mu}{2} + \sqrt{\mathcal{L}_0} \sinh \frac{\mu}{2} \right)^2\,.
\end{equation}
%
%
%
Using the relations (\ref{eq:rootTTLLb}) to express the deformed energy in terms of $\mathcal{L}_0$, $\overbar{\mathcal{L}}_0$ then yields
\begin{equation}
\begin{aligned}
\label{eq:energyrootTT}
E_\mu = \frac{R}{8 \pi G \ell} \left( (\mathcal{L}_0 + \overbar{\mathcal{L}}_0 ) \cosh \mu+ 2 \sqrt{\mathcal{L}_0 \overbar{\mathcal{L}}_0 }\sinh \mu \right)\,.
\end{aligned}
\end{equation}
We can rewrite \eqref{eq:energyrootTT} in terms of the undeformed energy and angular momentum by recalling that 
\begin{equation}
\begin{aligned}
E_0 &= \frac{R}{8 \pi G \ell} \left( \mathcal{L}_0 + \overbar{\mathcal{L}}_0 \right)\,, \quad J_0 = \frac{R}{8 \pi G \ell} \left( \mathcal{L}_0 - \overbar{\mathcal{L}}_0 \right) \, , \\
\mathcal{L}_0&= \frac{4\pi G \ell}{R} \left( E_0 +J_0 \right)\,, \quad \overbar{\mathcal{L}}_0= \frac{4\pi G \ell}{R} \left( E_0 -J_0 \right) \, .
\end{aligned}
\end{equation}
After identifying the bulk angular momentum $J_0$ with the CFT momentum $P_0$, this gives the same energy spectrum (\ref{root_TT_deformed_CFT_energies}) we found from consistency conditions, namely
\begin{equation}
\label{eq:root-TT-energyspectrum}
E_\mu = E_0 \cosh \mu+  \sqrt{E_0^2 - P_0^2 }\sinh \mu\,.
\end{equation}

\subsubsection*{\ul{\it Propagation Speed}}

In the case of the usual $\TT$ deformation, there is a sharp distinction between the two signs of the deformation parameter $\lambda$. In our conventions, $\lambda > 0$ corresponds to the ``good sign'' of the flow. With this choice of sign, so long as $\lambda$ is not too large, all of the energy eigenvalues in a $\TT$-deformed CFT remain real. For the ``bad sign'' $\lambda < 0$, however, all but finitely many of the energies in the deformed theory become complex.\footnote{In some cases, these complex energies can be removed by performing multiple $\TT$ deformations in a row. For instance, one can deform a pair of CFTs by the bad sign of $\lambda$ and then subsequently deform the tensor product of these theories by a good-sign flow to cure the spectrum \cite{Ferko:2022dpg}.} This signals a pathology in the bad-sign-deformed theory which appears to be quite robust to the type of $\TT$-deformation one uses. For instance, a single-trace $\TT$ deformation with the bad sign corresponds to a bulk dual with closed timelike curves \cite{Chakraborty:2019mdf}. The conventional double-trace $\TT$ deformation, which is the version considered in this work, with the bad choice of sign is dual to a bulk spacetime which exhibits superluminal propagation \cite{McGough:2016lol,Hartman:2018tkw}.

It is natural to ask whether the root-$\TT$ deformation has a similar pathology for one choice of the sign. Such a pathology would not be visible at the level of the formula (\ref{eq:root-TT-energyspectrum}) for the root-$\TT$ deformed spectrum, which appears to yield real energies for either sign of $\mu$. However, we will now show that the sign choice $\mu < 0$ leads to a bulk spacetime which allows superluminal propagation. This suggests that, as with $\TT$, only the positive sign of the root-$\TT$ flow parameter may lead to a sensible deformed theory.

To demonstrate this superluminal propagation, we begin with a diagonal stress tensor $\widetilde{T}_{\alpha \beta} (0) = \operatorname{diag}(\widetilde{T}_{tt}(0), \widetilde{T}_{xx}(0))$ on a two-dimensional space equipped with $(t, x)$ coordinates and flat metric $\eta_{\alpha \beta} = \operatorname{diag} (-1, 1)$. The boundary deformed metric \eqref{root_TT_deformed_bcs_repeat} in this setting is 
\begin{equation}
\begin{aligned}
ds^2 &= \left( - dt^2 + dx^2 \right) \cosh \mu + \frac{\sinh \mu}{\mathcal{R}^{(0)}} \left( \widetilde{T}_{tt}(0) dt^2 + \widetilde{T}_{xx}(0)dx^2 \right)
\\&=-  e^{-\mu } dt^2 + e^\mu dx^2\,,
\end{aligned}
\end{equation}
where we have used $ \widetilde{T}_{tt} = \widetilde{T}_{xx} = \mathcal{R}^{(0)}$.

Null geodesics obey $ds^2 = 0$ which have the following propagation speed 
\begin{equation}
v=  e^{-\mu}\,, 
\end{equation}
%
%
%
and in particular we see that $v > 1$ if $\mu <0$. This confirms that the bulk supports superluminal propagation for the negative sign of the root-$\TT$ deformation parameter.

This result might have been anticipated because the root-$\TT$ deformation is closely connected to the Modified Maxwell or ModMax theory of electrodynamics in four spacetime dimensions. In particular, the $4d$ root-$\TT$ deformation of the free Maxwell theory yields the ModMax theory \cite{Babaei-Aghbolagh:2022uij, Ferko:2023ruw}, and the dimensional reduction of this theory to two spacetime dimensions is the Modified Scalar theory which is obtained from a root-$\TT$ flow of free scalars \cite{Conti:2022egv}. It was already pointed out in \cite{Bandos:2020jsw} that the $4d$ ModMax theory also allows for superluminal propagation when $\gamma < 0$, which corresponds to $\mu < 0$ in our notation. This gives another reason to suspect that the root-$\TT$ deformation may be ill-behaved for $\mu < 0$.

\subsection{Chern-Simons Formalism}
The three-dimensional Einstein-Hilbert action is expressible semi-classically as the difference of two Chern-Simons actions \cite{Achucarro:1986uwr,Witten:1988hc}
\begin{equation}
    S_{\text{EH}} [g_{\alpha \beta}]= S_{\text{CS}}[A] - S_{\text{CS}}[\overbar{A}] \, ,
\end{equation}
where 
\begin{equation}
    S_{\text{CS}}[A] = \frac{\ell}{16 \pi G} \int \operatorname{Tr} \left( A\wedge dA + \frac{2}{3} A \wedge A \wedge A \right) \, , 
\end{equation}
and the bulk one-form Chern-Simons connections $(A, \overbar{A})=( A_\alpha dx^\alpha, \overbar{A}_\alpha dx^\alpha)$ are expressed in terms of the vielbein $E^a = E^a{}_\alpha dx^\alpha$ and spin connection $\Omega^a = \frac{1}{2} \epsilon^{abc} \Omega_{\alpha b c} dx^\alpha$:
\begin{equation}
A^a = \Omega^a + \frac{1}{\ell} E^a, \quad \overbar{A}^a = \Omega^a - \frac{1}{\ell} E^a \, , 
\end{equation}
where $a=-1, 0, 1$ are SL$(2, \mathbb{R})$ group indices. The equations of motion imply flatness
\begin{equation}
F= dA+ A \wedge A = 0, \quad \overbar{F} = d \overbar{A} + \overbar{A} \wedge \overbar{A} = 0 \, , 
\end{equation}
and the bulk metric is related to the Chern-Simons gauge fields as
\begin{equation}
g_{\alpha \beta} = \frac{\ell^2}{2} \operatorname{Tr} \left( (A_\alpha - \overbar{A}_\alpha)(A_\beta - \overbar{A}_\beta) \right)\,,
\end{equation}
where the trace is over SL$(2, \mathbb{R})$ indices. The connections associated to the  Ba\~{n}ados geometry \eqref{eq:BanadosMetric} are
\begin{equation}
\begin{aligned}
\label{eq:connectionsundeformed}
A& =-\frac{1}{2\rho} L_0 d\rho + \frac{1}{\ell} \left(- \sqrt{\rho} \mathcal{L}_0 L_{-1} + \frac{1}{\sqrt{\rho}} L_{1}  \right)du =    \left(\begin{array}{ll}
-\frac{d\rho}{4\rho} & \quad -\frac{\sqrt{\rho} \mathcal{L}_0 du }{\ell }\\
-\frac{du}{\ell \sqrt{\rho}  } & \quad \frac{d\rho}{4\rho}
\end{array}\right)   \,,\\
\overbar{A} &= \frac{1}{2\rho} L_0 d\rho + \frac{1}{\ell} \left( \frac{1}{\sqrt{\rho}} L_{-1}-\sqrt{\rho} \overbar{\mathcal{L}}_0 L_1 \right)dv =    \left(\begin{array}{ll}
\frac{d\rho}{4\rho} & \quad \frac{dv}{\ell \sqrt{\rho} } \\
 \frac{\sqrt{\rho} \overbar{\mathcal{L}}_0  dv}{\ell } & \quad -\frac{d\rho}{4\rho}
\end{array}\right) \, , 
\end{aligned}
\end{equation}
where the SL$(2, \mathbb{R})$ generators are 
\begin{equation}
    L_{-1}=\left(\begin{array}{ll}
0 & 1 \\
0 & 0
\end{array}\right), \quad L_0=\frac{1}{2}\left(\begin{array}{cc}
1 & 0 \\
0 & -1
\end{array}\right), \quad L_1=\left(\begin{array}{ll}
0 & 0 \\
-1 & 0
\end{array}\right) \, .
\end{equation}
These generators satisfy the standard commutation relations
\begin{equation}
\begin{aligned}
[L_m, L_n] = (m-n) L_{m+n}\,.
    \end{aligned}
\end{equation}
%
%
%
In radial gauge, we may extract the radial dependence from the bulk connections as
\begin{equation}
\begin{aligned}
A(\rho, u) &= b^{-1}(\rho) (d + a(u)) b(\rho), \quad \overbar{A}(\rho, v) = b (d + \overbar{a}(v)) b^{-1}(\rho)\,, \\ b(\rho) &= e^{- \frac{1}{2} L_0 \ln \rho} =  \left( \begin{array}{cc}
 \rho^{-\frac{1}{4}}& \quad 0 \\
0 & \quad \rho^{\frac{1}{4}}
\end{array}
\right) \, , 
\end{aligned}
\end{equation}
where the boundary connections are
\begin{equation}\label{CS_boundary_banados}
    a(u) =\frac{1}{\ell} \left(-  \mathcal{L}_0(u) L_{-1} + L_{1}  \right)du, \quad \overbar{a}(v) =  \frac{1}{\ell} \left( L_{-1} -  \overbar{\mathcal{L}}_0(v) L_1 \right)dv \,.
\end{equation}
To compare more easily with our metric formalism analysis, we work in the same temporal and periodic coordinates\footnote{We distinguish between the undeformed coordinates $(t, \varphi)$ and the deformed coordinates $(T, \phi)$.}
\begin{equation}
\label{eq:tphivars}
t= \frac{1}{2} \left( u+v \right)\,, \quad \varphi = \frac{1}{2} \left( u-v \right)\,, \quad \varphi \sim \varphi + R\,.
\end{equation}
In these variables \eqref{eq:tphivars}, the chiral boundary conditions are $A_t = A_\varphi\,, \overbar{A}_t = - \overbar{A}_\varphi$. To have a variational principle which realizes these chiral boundary conditions, we add the following boundary term to the total Chern-Simons action:
\begin{equation}
\label{eq:boundaryundeformedCS}
S =  S_{\text{CS}}[A] - S_{\text{CS}}[\overbar{A}] + \frac{\ell}{16 \pi G} \int_{\partial M} dt \, d\varphi \, \operatorname{Tr} \left(A_\varphi^2 + \overbar{A}_\varphi^2\right)\,.
\end{equation}
In the undeformed theory with conventional boundary conditions, it can be shown that the boundary term in \eqref{eq:boundaryundeformedCS} that imposes the chiral boundary conditions is related to the mass of the bulk spacetime, which can be defined via other means in the metric formalism \cite{Regge:1974zd, Coussaert:1995zp}. In our case, this mass is simply the black hole's total energy. We can see this explicitly by substituting \eqref{eq:connectionsundeformed} into the boundary action, which yields
\begin{equation}
S_{\text{bdry}} = \frac{\ell}{16 \pi G} \int_{\partial M} dt \,  d\varphi \, \operatorname{Tr} \left(A_\varphi^2 + \overbar{A}_\varphi^2\right) =  \frac{1}{8\pi G \ell} \int_{\partial M} dt \, d\varphi \left( \mathcal{L}_0 + \overbar{\mathcal{L}}_0 \right) \, , 
\end{equation}
since the undeformed energy is
\begin{equation}
E_0 = \frac{R}{8 \pi G \ell} \left( \mathcal{L}_0 + \overbar{\mathcal{L}}_0 \right)\,.
\end{equation}
It is not obvious, without performing a computation in the canonical formulation, that the Chern-Simons boundary action will continue to yield the mass of the bulk spacetime in the presence of a boundary deformation. However, we will find that this is indeed the case when the boundary is deformed by the root-$\TT$ operator.

Next we will understand the mixed boundary conditions imposed by the root-$\TT$ deformation in Chern-Simons variables. There are two equivalent approaches that one might use in order to find the deformed boundary conditions. One strategy to find the deformed Chern-Simons connections is to use the field dependent coordinate transformation \eqref{eq:inverserootTT}. In describing this approach, we will work with an explicit choice of coordinate system.

The second method is to work with a covariant expansion of the boundary connections in terms of vielbeins $e_i^a$ and their dual expectation values $f_i^a$, which are related to the stress tensor. One can then work out the mixing of sources and expectation values in Chern-Simons variables, either by imposing consistency conditions of the kind discussed in section \ref{sec:boundary_condition_derivation}, or by taking the results for $\gamma_{\alpha \beta}^{(\mu)}$ and $\widetilde{T}_{\alpha \beta}^{(\mu)}$ in equation (\ref{root_TT_deformed_metric_bcs_later}) as given and then finding the modification in Chern-Simons variables which reproduce these results. 

\subsubsection*{\ul{\it Root-$\TT$ Deformed Chern-Simons: Coordinate Approach}}

The first way to find the deformed Chern-Simons connections is to use the field dependent coordinate transformation \eqref{eq:inverserootTT}. In describing this approach, we will work directly with boundary coordinates $U, V$ for the deformed theory and $u, v$ for the undeformed theory. Transforming the connections $A$ and $\overbar{A}$ using this change of coordinates yields
\begin{equation}
\begin{aligned}
A(\mu)&= -\frac{1}{2\rho} L_0 d\rho + \frac{1}{\ell} \left( -\sqrt{\rho} \mathcal{L}_\mu  L_{-1} + \frac{1}{\sqrt{\rho}} L_1 \right) \left( \cosh \frac{\mu}{2} dU - \sqrt{\frac{  \overbar{\mathcal{L}}_\mu  }{\mathcal{L}_\mu}} \sinh \frac{\mu}{2} dV \right)\,,\\
\overbar{A}(\mu) &= \frac{1}{2\rho} L_0 d\rho + \frac{1}{\ell} \left( \frac{1}{\sqrt{\rho}}  L_{-1} - \sqrt{\rho} \overbar{\mathcal{L}}_\mu L_1 \right) \left( - \sqrt{\frac{\mathcal{L}_\mu}{\overbar{\mathcal{L}}_\mu}} \sinh \frac{\mu}{2}  dU + \cosh \frac{\mu}{2} dV \right)\,.
\end{aligned}
\end{equation}
It is straightforward to see the mixed boundary conditions in this root-$\TT$ deformed setting:
\begin{equation}
\sqrt{\frac{\overbar{\mathcal{L}}_\mu}{\mathcal{L}_\mu}} \left(\sinh \frac{\mu}{2} \right) A_U(\mu) + \left(\cosh \frac{\mu}{2} \right) A_V(\mu) = 0\,, \quad \sqrt{\frac{\overbar{\mathcal{L}}_\mu}{\mathcal{L}_\mu}} \left( \cosh \frac{\mu}{2} \right) \overbar{A}_U(\mu) + \left( \sinh \frac{\mu}{2} \right)   \overbar{A}_V (\mu) = 0\,.
\end{equation}
%
%
%
%
%
Moreover, we can extract the deformed boundary Chern-Simons connections
\begin{equation}\label{deformed_cs_coordinate_approach}
\begin{aligned}
a(\mu)&= \frac{1}{\ell} \left(-\mathcal{L}_\mu  L_{-1} + L_1 \right) \left( \left( \cosh \frac{\mu}{2} - \sqrt{\frac{\overbar{\mathcal{L}}_\mu}{\mathcal{L}_\mu} }  \sinh \frac{\mu}{2} \right) d\phi  + \left( \cosh \frac{\mu}{2} + \sqrt{\frac{\overbar{\mathcal{L}}_\mu}{\mathcal{L}_\mu}} \sinh \frac{\mu}{2} \right) dT
  \right)\,,\\
  \overbar{a}(\mu) &= \frac{1}{\ell} \left(   L_{-1} -  \overbar{\mathcal{L}}_\mu L_1 \right) \left( \left( \cosh \frac{\mu}{2} -\sqrt{\frac{\mathcal{L}_\mu}{\overbar{\mathcal{L}}_\mu}} \sinh \frac{\mu}{2}\right) d\phi - \left(\cosh \frac{\mu}{2}  +\sqrt{\frac{\mathcal{L}_\mu}{\overbar{\mathcal{L}}_\mu}} \sinh \frac{\mu}{2}    \right) dT \right)  \, , 
  \end{aligned}
\end{equation}
which obey 
\begin{equation}
\label{eq:BCsroot-TT}
a_T(\mu) = \frac{\cosh \frac{\mu}{2} +   \sqrt{\frac{\overbar{\mathcal{L}}_\mu}{\mathcal{L}_\mu}} \sinh \frac{\mu}{2} }{ \cosh \frac{\mu}{2} - \sqrt{\frac{\overbar{\mathcal{L}}_\mu}{\mathcal{L}_\mu}} \sinh \frac{\mu}{2}   } a_\phi (\mu)\,, \quad \overbar{a}_{T}(\mu) = - \frac{ \cosh \frac{\mu}{2} + \sqrt{\frac{\mathcal{L}_\mu}{\overbar{\mathcal{L}}_\mu}} \sinh \frac{\mu}{2}   }{\cosh \frac{\mu}{2}    - \sqrt{ \frac{\mathcal{L}_\mu}{\overbar{\mathcal{L}}_\mu}  } \sinh \frac{\mu}{2} } \overbar{a}_\phi(\mu)\,.
\end{equation}
To make contact with our discussion of the horizon area in the metric formalism, we note that one may compute the BTZ black hole's Bekenstein-Hawking entropy (and thus its horizon area) directly in the Chern-Simons formalism. Following \cite{deBoer:2013gz}, the black hole entropy is given in terms of Chern-Simons quantities as
\begin{equation}
\label{eq:Entropy}
    S = C \operatorname{Tr} \left( (\lambda_\phi - \overbar{\lambda}_\phi)L_0 \right) \,,
\end{equation}
where $C$ is a constant which depends on the central charge $c$, but whose precise value is not important for this discussion, $\lambda_\phi$ and $\overbar{\lambda}_\phi$ are diagonal traceless matrices containing the eigenvalues of $a_\phi$ and $\overbar{a}_\phi$.

Equation (\ref{eq:Entropy}) was derived in \cite{deBoer:2013gz} using a particular boundary term which is appropriate for the Drinfeld-Sokolov form of the connections, which in our case corresponds to a Ba\~nados type solution. We note that the root-$\TT$ deformed connections are \emph{not} of this form when written in terms of the original coordinates, which will become clear when we obtain covariant expressions for the deformed in equation (\ref{TT_CS_mixed_bcs}). However, when we write the deformed connections in new coordinates $(T, \phi)$ using the field-dependent diffeomorphism (as we have done above), the connections \emph{are} of Ba\~nados type, albeit characterized by deformed parameters $\mathcal{L}_\mu$ and $\overbar{\mathcal{L}}_\mu$. Therefore it is justified to use the expression (\ref{eq:Entropy}) so long as we work in the transformed coordinates.

Diagonalizing the connections given in \eqref{deformed_cs_coordinate_approach}, one finds
\begin{equation}\label{root_TT_lambda_matrices}
\begin{aligned}
    \lambda_\phi &= \frac{1}{\ell} \left( \begin{array}{cc}
       \sqrt{\mathcal{L}_\mu} \cosh \frac{\mu}{2} - \sqrt{\overbar{\mathcal{L}}_\mu} \sinh \frac{\mu}{2}  & \quad   0\\
   0      & \quad  - \sqrt{\mathcal{L}_\mu} \cosh \frac{\mu}{2} + \sqrt{\overbar{\mathcal{L}}_\mu} \sinh \frac{\mu}{2} 
    \end{array} \right)\,, \\ 
    \overbar{\lambda}_\phi  &= \frac{1}{\ell}  \left( \begin{array}{cc}
     -  \sqrt{\overbar{\mathcal{L}}_\mu} \cosh \frac{\mu}{2} + \sqrt{\mathcal{L}_\mu} \sinh \frac{\mu}{2}  & \quad   0\\
   0      & \quad    \sqrt{\overbar{\mathcal{L}}_\mu} \cosh \frac{\mu}{2} - \sqrt{\mathcal{L}_\mu} \sinh \frac{\mu}{2}
    \end{array} \right)\,.
    \end{aligned}
\end{equation}
Therefore
\begin{equation}\label{CS_two_entropies}
    S^{(0)} = \frac{C}{\ell} \left(\sqrt{\mathcal{L}_0} + \sqrt{\overbar{\mathcal{L}}_0}\right)\,, \quad S^{(\mu)} = \frac{C}{\ell} e^{- \frac{\mu}{2}}  \left( \sqrt{\mathcal{L}_\mu  }  + \sqrt{\overbar{\mathcal{L}}_\mu} \right) \, .
\end{equation}
Equating the two entropies in equation (\ref{CS_two_entropies}) then gives the same area equation which we found in \eqref{eq:areaangularmomentumTTroot} using a metric-formalism analysis.

The corresponding boundary term which we must add in order to have a well-defined variational principle with respect to these root-$\TT$ deformed boundary conditions is
\begin{align}
\delta S_{\text{bdry}} =\frac{\ell}{8\pi G} \int_{\partial M} dT \, d\phi \, &\Bigg( \operatorname{Tr} \left[  \frac{\cosh \frac{\mu}{2} +   \sqrt{\frac{\overbar{\mathcal{L}}_\mu}{\mathcal{L}_\mu}} \sinh \frac{\mu}{2} }{ \cosh \frac{\mu}{2} - \sqrt{\frac{\overbar{\mathcal{L}}_\mu}{\mathcal{L}_\mu}} \sinh \frac{\mu}{2}   } a_\phi(\mu)\ \delta a_\phi (\mu) 
 \right] \nonumber \\
&\qquad + \Tr \left[ \frac{ \cosh \frac{\mu}{2} + \sqrt{\frac{\mathcal{L}_\mu}{\overbar{\mathcal{L}}_\mu}} \sinh \frac{\mu}{2}   }{\cosh \frac{\mu}{2}    - \sqrt{ \frac{\mathcal{L}_\mu}{\overbar{\mathcal{L}}_\mu}  } \sinh \frac{\mu}{2} } \overbar{a}_\phi(\mu)\ \delta  \overbar{a}_\phi (\mu) \right] \Bigg) \, .
\end{align}
%
%
%
We substitute the boundary connections \eqref{eq:BCsroot-TT} and their variations to find
%
\begin{equation}
\delta S_{\text{bdry}} = \frac{1}{8\pi G \ell} \int_{\partial M} dT \, d\phi \left( \delta \mathcal{L}_\mu + \delta \overbar{\mathcal{L}}_\mu \right) \, , 
\end{equation}
from which $S_{\text{bdry}}$ is easily read off
\begin{equation}
\label{eq:boundaryactionroot-tt}
S_{\text{bdry}}= \frac{1}{8 \pi G \ell} \int_{\partial M} dT \, d\phi \, \left(  \mathcal{L}_\mu + \overbar{\mathcal{L}}_\mu \right)\,.
\end{equation}
In summary, we have shown that the final expression (\ref{eq:boundaryactionroot-tt}) for the deformed boundary action in Chern-Simons variables is identical to that of the root-$\TT$ deformed energy, given in (\ref{eq:E&J}), of the spacetime computed in the metric formalism.\footnote{In \cite{He:2020hhm} it was shown that the corresponding Chern-Simons boundary action for $\TT$-deformed boundary conditions also matches the $\TT$-deformed spacetime energy \eqref{eq:EnergyTT}.}

Note that we have not given any \emph{a priori} justification that the deformed Chern-Simons boundary action yields the spacetime mass when the boundary theory is deformed by a general multi-trace operator. Although it is easy to show that this is true in the undeformed theory, a general proof that the Chern-Simons boundary action computes the spacetime mass in the presence of modified boundary conditions would require a computation of the Hamiltonian using an analysis of the canonical structure. We will not pursue such an analysis here. However, the fact that the deformed boundary action (\ref{eq:boundaryactionroot-tt}) does agree with the energy computed in the metric formulation may be viewed as an \emph{a posteriori} argument that such an analysis in the canonical formulation would conclude that the boundary term equals the spacetime energy in the case of root-$\TT$ deformed boundary conditions.

\subsubsection*{\ul{\it Root-$\TT$ Deformed Chern-Simons: Covariant Approach}}

We now describe the second approach. In order to make the sources and expectation values in Chern-Simons variables explicit, it is convenient to expand the boundary gauge fields as
\begin{align}\label{llabres_expansion}
    a_i &= 2 e_i^+ L_{1} - f_i^- L_{-1} + \omega_i L_0 \, , \nonumber \\
    \overbar{a}_i &= f_i^+ L_{1} - 2 e_i^- L_{-1} + \omega_i L_0 \, .
\end{align}
In the case of a Ba\~nados-type geometry, this expansion reduces to the one given in (\ref{CS_boundary_banados}). In these expansions, $e_i^a$ plays the role of the boundary vielbein, where we use middle Latin letters $i, j, k$ for curved boundary indices and early Latin letters $a, b, c$ for flat boundary indices. We have chosen the numerical factors appearing in the expansions (\ref{llabres_expansion}) to simplify our final results, but they will lead to some unfamiliar factors of $2$ in certain expressions. For instance, the boundary metric in these conventions is
\begin{align}\label{metric_factors_2}
    \gamma_{ij} = 2 e_i^a \eta_{ab} e_j^b \, , 
\end{align}
which has an additional factor of $2$ compared to the standard definition. We also define
\begin{align}
    e = \det ( e_j^b ) \, , 
\end{align}
so that $\det ( \gamma_{ij} ) = - 4 e^2$, and the Levi-Civita symbols with flat and curved indices are
\begin{align}
    \epsilon_{a b} = \begin{bmatrix} 0 & 1 \\ -1 & 0 \end{bmatrix}_{ab} \, , \qquad \epsilon^{ij} = \frac{1}{2 e} \begin{bmatrix} 0 & 1 \\ -1 & 0 \end{bmatrix}^{ij} \, , 
\end{align}
These satisfy various identities such as $g^{ij} = - \epsilon^{i k} \epsilon^{j l} g_{kl}$, $\epsilon^{ab} = 2 \epsilon^{ij} e_i^a e_j^b$, and so on, with factors of $2$ that can be traced back to the definition (\ref{metric_factors_2}). Flat indices are raised and lowered with $\eta_{ab}$, where we take $\eta_{+-} = \eta_{-+} = - 1$ in this subsection. We refer the reader to section 2 of \cite{Ebert:2022ehb}, or to \cite{Llabres:2019jtx}, for more details on these notational conventions.

In the holographic dictionary, this vielbein $e_i^a$ is the source while the other expansion coefficients $f_i^a$ are the dual expectation values, which are related to the boundary stress tensor with one flat and one curved index according to the relation
\begin{align}
    T^i_a = \frac{1}{4 \pi G} \epsilon_{ab} \epsilon^{ij} f_j^b \, .
\end{align}
We will assume that the boundary spin connection $\omega$ vanishes in the undeformed theory, which is appropriate for a flat boundary.

We expect, based on the general analysis of section \ref{sec:dictionary}, that the addition of a multi-trace boundary term in Chern-Simons variables will impose modified boundary conditions in which some deformed source $e_i^a ( \mu )$ is now held fixed in the variational principle, rather than the undeformed source $e_i^a ( 0 )$. In the case of a boundary $\TT$ deformation, we recall from \cite{Ebert:2022ehb,Llabres:2019jtx} that the resulting modification of the sources and expectation values is simply
\begin{align}\label{TT_CS_mixed_bcs}
    e_i^a ( \lambda ) = e_i^a ( 0 ) + \frac{\lambda}{4 \pi G} f_i^a \, , \qquad f_i^a ( \lambda ) = f_i^a ( 0 ) \, .
\end{align}
One can determine the analogue of (\ref{TT_CS_mixed_bcs}) which corresponds to a boundary root-$\TT$ deformation by following the procedure of section \ref{sec:boundary_condition_derivation}. That is, we first write down the most general expression for deformed quantities $e_i^a ( \mu )$ and $f_i^a ( \mu )$ which depend on a dimensionless parameter $\mu$, preserve tracelessness for a conformal seed theory, and commute with the $\TT$-deformed boundary conditions (\ref{TT_CS_mixed_bcs}). We will not carry out these steps explicitly, since they are identical to those of section \ref{sec:boundary_condition_derivation} after changing from metric variables to Chern-Simons variables. Instead we simply quote the result, which for a CFT seed is
\begin{align}\label{CS_rTT_deformed_bcs}
    e_i^a ( \mu ) = \cosh \left( \frac{\mu}{2} \right) e_i^a ( 0 ) + \frac{\sinh \left( \frac{\mu}{2} \right)}{ \mathcal{R}^{(0)}} f_i^a ( 0 ) \, , \quad f_i^a ( \mu ) = \cosh \left( \frac{\mu}{2} \right) f_i^a ( 0 ) + \sinh \left( \frac{\mu}{2} \right) \mathcal{R}^{(0)} e_i^a ( 0 ) \, . 
\end{align}
Here $\mathcal{R}^{(0)}$ is the usual root-$\TT$ operator, which can be expressed purely in Chern-Simons variables. Again, these expressions will have some unusual numerical factors introduced by (\ref{metric_factors_2}). For instance, we can reproduce a general stress tensor on a standard flat metric via
\begin{align}
    \tensor{e}{_i^a} = \frac{1}{\sqrt{2}} \tensor{\begin{bmatrix} 0 & 1 \\ 1 & 0 \end{bmatrix}}{_i^a} \, , \qquad \tensor{f}{_i^a} = \frac{4 \pi G}{\sqrt{2}} \tensor{\begin{bmatrix} T_{zz} & - T_{z \overbar{z}} \\ - T_{z \overbar{z}} & T_{\overbar{z} \overbar{z}} \end{bmatrix}}{_i^a} \, , 
\end{align}
and then the stress tensor with two curved indices is
\begin{align}
    T_{ij} = \tensor{T}{^k_a} e^a_j g_{k i} = \begin{bmatrix} T_{zz} & T_{z \overbar{z}} \\ T_{z \overbar{z}} & T_{\overbar{z} \overbar{z}} \end{bmatrix}_{ij} \, , 
\end{align}
and its trace is $g^{ij} T_{ij} = - 2 T_{z \overbar{z}}$, while the invariant $T^{ij} T_{ij}$ is
\begin{align}
    T^{ij} T_{ij} = 2 \left( T_{zz} T_{\overbar{z} \overbar{z}} + T_{z \overbar{z}}^2 \right) \, .
\end{align}
It is straightforward to covariantize these statements and obtain expressions for $\mathcal{R}^{(0)}$. For instance, in the case of a conformal seed theory with a traceless stress tensor, we find
\begin{align}
    \mathcal{R}^{(0)} = \frac{1}{4 \pi G} \sqrt{ - f_i^a f_j^b \epsilon_{ab} \epsilon^{ij} } \, .
\end{align}
In the general case where the undeformed stress tensor is not traceless, we can define a traceless part of $f_i^a$, which is the analogue of $\widetilde{T}_{\alpha \beta}$, as
\begin{align}
    \widetilde{f}_i^a = f_i^a - 4 \pi G \, e_i^a \left( e_j^b T^j_b \right) \, ,  
\end{align}
and then express the root-$\TT$ operator as
\begin{align}
    \mathcal{R}^{(0)} = \frac{1}{4 \pi G} \sqrt{ - \widetilde{f}_i^a \widetilde{f}_j^b \epsilon_{ab} \epsilon^{ij} } \, .
\end{align}
One can then check that, after transforming from Chern-Simons variables to metric variables, the deformed quantities (\ref{CS_rTT_deformed_bcs}) reproduce the metric and stress tensor (\ref{TT_CS_mixed_bcs}) in the case of a conformal seed theory (or for a general seed, if we replace $f_i^a$ with $\widetilde{f}_i^a$).

In particular, the deformed connections computed with the $e_i^a ( \mu )$ in (\ref{CS_rTT_deformed_bcs}) agree with those in (\ref{deformed_cs_coordinate_approach}). One can see this by expressing the $f_i^a$ in terms of $\mathcal{L}$ and $\overbar{\mathcal{L}}$ and choosing coordinates $(\phi, T)$. Indeed it must have been the case that these agree, since the coordinate transformation which was used to obtain (\ref{deformed_cs_coordinate_approach}) is precisely the one that generates the root-$\TT$ deformed metric and stress tensor in the metric formalism, and the deformed vielbein (\ref{CS_rTT_deformed_bcs}) reproduces these quantities. Therefore the two methods are equivalent.

\section{Conclusion and Outlook}\label{sec:conclusion}

In this paper, we have investigated several properties of the root-$\TT$ operator in holography. Among our main results is the proposal (\ref{root_TT_energy_flow}) for the flow of the finite-volume spectrum of a root-$\TT$ deformed CFT. We have explicitly verified that this flow equation matches the deformed spacetime mass for a class of Ba\~nados-type $\mathrm{AdS}_3$ solutions subject to root-$\TT$ deformed boundary conditions. This represents the first calculation which may shed light on quantum aspects of the root-$\TT$ deformation. Although a quantum definition of the root-$\TT$ operator itself is still not known in the field theory, we have sidestepped this issue by working in a large $N$ limit and performing a holographic calculation in the bulk dual.

Besides the question of a quantum definition of the root-$\TT$ operator, there remain many other avenues for future research, two of which we outline below. We believe that a better understanding of these issues will offer new insights in non-analytic root-$\TT$-like (or ModMax-like) theories, and we hope to return to some of these questions in future work.

\subsubsection*{\ul{\it Correlation functions}}

An immediate, and important, next step would be to study correlation functions in a root-$\TT$ deformed $\mathrm{CFT}_2$. Due to the awkwardness of the square-root of an operator, calculating a root-$\TT$ deformed correlation function in perturbation theory seems difficult and ambiguous. However, because the root-$\TT$ operator exhibits some of the special properties of the $\TT$ operator, there might be hope that a perturbative calculation is feasible. In particular, we have seen that demanding commutativity of the $\TT$ and root-$\TT$ flows is a powerful constraint which allowed us to uniquely fix the deformed boundary conditions and flow equation for the spectrum. One might conjecture that perturbative corrections to correlation functions may also be fixed by imposing commutativity of the following diagram:
    \begin{center}
    \begin{tikzcd}
    	{\left\langle \prod_i \mathcal{O}_i (x_i) \right\rangle^{(0)} } && {\left\langle \prod_i \mathcal{O}_i (x_i) \right\rangle^{(\lambda)} } \\
    	\\
    	{\left\langle \prod_i \mathcal{O}_i (x_i) \right\rangle^{(\mu)} } & {} & { \left\langle \prod_i \mathcal{O}_i (x_i) \right\rangle^{(\lambda, \mu)}  }
    	\arrow["{\mathcal{O}_{\TT}}", from=1-1, to=1-3]
    	\arrow["{\mathcal{R}}"', from=1-1, to=3-1]
    	\arrow["{{\mathcal{O}_{\TT}}}"', from=3-1, to=3-3]
    	\arrow["{\mathcal{R}}", from=1-3, to=3-3]
    \end{tikzcd}
\end{center}
To be more concrete, the $\TT$-deformed two-point planar stress tensor correlators take the following form from dimensional analysis, translational and rotational symmetry \cite{Kraus:2018xrn,Ebert:2022cle}
\begin{equation}
    \begin{aligned}
    \label{eq:uhreh}
        \langle T_{zz} (x) T_{zz} (0) \rangle^{(\lambda)} &= \frac{1}{z^4} f_1 (y), \quad
            \langle T_{zz} (x) T_{z\overbar{z}} (0) \rangle^{(\lambda)} = \frac{1}{z^3 \overbar{z}} f_2 (y)\,,\\
                    \langle T_{zz} (x) T_{\overbar{z} \overbar{z}} (0) \rangle^{(\lambda)} &= \frac{1}{z^2 \overbar{z}^2} f_3 (y)\,,\quad 
                                     \langle T_{z\overbar{z}} (x) T_{z \overbar{z}} (0) \rangle^{(\lambda)}= \frac{1}{z^2 \overbar{z}^2} f_4 (y)\,,
    \end{aligned}
\end{equation}
where $y = z \overbar{z}/\lambda$ and the functions $f_i(y)$ are fixed by stress tensor conservation $\partial^\alpha T_{\alpha \beta} = 0$ and the trace flow equation $T_{z\overbar{z}} = -\pi \lambda \TT$ giving $T_{z\overbar{z}} = - \pi \lambda T_{zz} T_{\overbar{z} \overbar{z}} + O(\lambda^2)$. Using the trace flow equation, we can easily determine $f_4(y)$ at $O(\lambda^2)$:
\begin{equation}
\begin{aligned}
\langle T_{z\overbar{z}}(x) T_{z\overbar{z}}(0)  \rangle^{(\lambda)} &= \langle (-\pi \lambda T_{zz}(x) T_{\overbar{z} \overbar{z} } (x))(-\pi \lambda T_{zz}(0) T_{\overbar{z} \overbar{z} } (0)) \rangle^{(0)} + \cdots
\\&= \pi^2 \lambda^2 \langle T_{zz} (x) T_{zz}(0) \rangle^{(0)} \langle T_{\overbar{z}\overbar{z}} (x) T_{\overbar{z}\overbar{z}}(0) \rangle^{(0)}  + \cdots
\\&= \frac{\pi^2 \lambda^2 c^2}{4 z^4 \overbar{z}^4} + \cdots \nonumber \\
\implies f_4(y) &= \frac{\pi^2 c^2}{4y^2}  + \cdots\,.
\end{aligned}
\end{equation}
The correlators also obey $\partial^\alpha \langle T_{\alpha \beta} (x) T_{\rho \sigma}(0) \rangle^{(\lambda)} = 0$ which give three conservation equations:
\begin{equation}
\begin{aligned}
\label{conservation equations}
 &\beta = \rho = \sigma = z:   \partial_{\overbar{z}} \langle T_{zz} (x) T_{zz} (0) \rangle^{(\lambda)} + \partial_{z} \langle T_{\overbar{z} z} (x) T_{zz}(0) \rangle^{(\lambda)} =  \partial_{\overbar{z}}\left( \frac{f_1(y)}{z^4} \right) +   \partial_{z} \left( \frac{f_2(y)}{z^3 \overbar{z}} \right) = 0\,, \\& 
 \beta =\overbar{z}, \rho = \sigma = z: \partial_{\overbar{z}} \langle T_{z \overbar{z}} (x) T_{zz} (0) \rangle^{(\lambda)} + \partial_{z} \langle T_{\overbar{z} \overbar{z}} (x) T_{zz} (0) \rangle^{(\lambda)} =  \partial_{\overbar{z}} \left( \frac{f_2(y)}{z^3 \overbar{z}}\right) +  \partial_{z}\left(  \frac{f_3(y)}{z^2 \overbar{z}^2}\right) = 0\,, \\& \beta =z, \rho = z, \sigma = \overbar{z}: \partial_{\overbar{z}} \langle T_{z \overbar{z}} (x) T_{\overbar{z} z}(0) \rangle^{(\lambda)} + \partial_{z} \langle T_{\overbar{z} z} (x) T_{z \overbar{z}}(0) \rangle^{(\lambda)} =  \partial_{\overbar{z}}\left( \frac{f_2(y)}{z^3 \overbar{z}} \right) +  \partial_{z} \left( \frac{f_4(y)}{z^2 \overbar{z}^2} \right) = 0\,.
\end{aligned}
\end{equation}
Since $f_4(y)$ at $O(\lambda^2)$ is known, we can determine the other $f_i(y)$ from solving \eqref{conservation equations} with initial conditions that the seed theory's correlators are recovered when $\lambda = 0$:
\begin{equation}\hspace{-25pt}
    \begin{aligned}
        f_1 (y) &= \frac{c}{2} + \frac{5 \pi^2 \lambda^2 c^2}{6 z^2 \overbar{z}^2} + \cdots \, , \; f_2 (y) = - \frac{\pi^2 \lambda^2 c^2}{3 z^2 \overbar{z}^2} + \cdots \, , \; f_3(y) &= \frac{\pi^2 \lambda^2 c^2}{4 z^2 \overbar{z}^2} + \cdots \, , \;  f_4(y) =\frac{\pi^2 \lambda^2 c^2}{4 z^2 \overbar{z}^2} + \cdots \, .
    \end{aligned}
\end{equation}
For the root-$\TT$ case, one should follow similar logic as in the above $\TT$ example:
\begin{equation}
    \begin{aligned}
    \label{eq:uhreh1}
        \langle T_{zz} (x) T_{zz} (0) \rangle^{(\mu)} &= \frac{1}{z^4} g_1 (u), \quad \quad \; 
            \langle T_{zz} (x) T_{z\overbar{z}} (0) \rangle^{(\mu)} = \frac{1}{z^3 \overbar{z}} g_2 (u)\,,\\
                    \langle T_{zz} (x) T_{\overbar{z} \overbar{z}} (0) \rangle^{(\mu)} &= \frac{1}{z^2 \overbar{z}^2} g_3 (u)\,,\quad 
                                     \langle T_{z\overbar{z}} (x) T_{z \overbar{z}} (0) \rangle^{(\mu)} = \frac{1}{z^2 \overbar{z}^2} g_4 (u)\,,
    \end{aligned}
\end{equation}
    where the $g_i(u)$ obey the same stress tensor conservation equations \eqref{conservation equations}. It would be interesting to see whether one or more of the $g_i(u)$ can be fixed from demanding commutativity of the $\TT$ and root-$\TT$ flows. For example, perhaps commutativity may fix one of the $g_i ( u )$ and then conservation of the stress tensor may fix the others. It would also be interesting to understand this commutativity and correlators in the context of quantum corrections, such as the two-loop corrected $\TT$-deformed planar stress tensor correlators found in \cite{Ebert:2022cle}.

\subsubsection*{\ul{\it The Fate of Conformal Symmetry}}

The root-$\TT$ operator is classically marginal and thus preserves conformal invariance at the classical level. It is an important open question to determine the fate of conformal symmetry in the quantum theory, assuming that a quantum definition of the root-$\TT$ operator exists. Quantum corrections might make this operator marginally relevant or marginally irrelevant, which would mean that conformal invariance is broken at the quantum level. 

One way to probe this question is to investigate the high-energy density of states. In any two-dimensional CFT, the degeneracy of states for large energy and high temperature is described by the Cardy formula \cite{Cardy:1986ie}, which fixes the asymptotic scaling to be
\begin{equation}
\label{eq:Cardy}
    \rho (E_0) \sim \exp\left(2\pi \sqrt{\frac{cE_0}{3}}\right) \, , \quad S(E_0) \sim 2\pi \sqrt{\frac{cE_0}{3}} \, .
\end{equation}
Therefore, to investigate whether a root-$\TT$ deformed CFT remains a CFT, one might ask whether its high-energy behavior agrees with (\ref{eq:Cardy}). A sketch of an argument in support of this claim might proceed as follows. First, since the root-$\TT$ deformed energy spectrum
depends on both the energy $E_n^{(0)}$ and momentum $P_n$ of the corresponding state in the undeformed theory, we cannot immediately use the na\"ive Cardy formula (\ref{eq:Cardy}), which has already coarse-grained over all states with an energy near $E_n$ but with any momentum $P_n$. However, we may use a generalization of the Cardy formula which accounts for the spin of a CFT state \cite{Hartman:2014oaa,Pal:2019zzr}. In terms of left-moving and right-moving energies, this formula reads
\begin{equation}
    \rho(E_L,E_R) \sim \exp \left( 2\pi \sqrt{\frac{cE_L}{6}} + 2\pi \sqrt{\frac{cE_R}{6}} \right) \, .
\end{equation}
One can express our deformed spectrum (\ref{root_TT_deformed_CFT_energies}) in terms of the left-moving and right-moving energies, such that $E_\mu = \left( E_L \right)_\mu + \left( E_R \right)_\mu$ and $P_\mu = \left( E_L \right)_\mu - \left( E_R \right)_\mu = P_0$, which satisfy
\begin{equation}
\begin{aligned}
    \sqrt{(E_L)_0} &= \sqrt{(E_L)_\mu} \cosh(\frac{\mu}{2}) - \sqrt{(E_R)_\mu} \sinh(\frac{\mu}{2}) \, , \; \\\sqrt{(E_R)_0} &= \sqrt{(E_R)_\mu} \cosh(\frac{\mu}{2}) - \sqrt{(E_L)_\mu} \sinh(\frac{\mu}{2}) \, .
\end{aligned}
\end{equation}
The deformed density of states $\rho_\mu ( E_L, E_R)$ is then obtained by expressing the density of states of the undeformed theory in terms of the deformed left-moving and right-moving energies. Up to a factor which is unimportant for the leading exponential behavior, we find
\begin{equation}
    \rho_\mu(E_L,E_R) \sim \exp \left( 2\pi\sqrt{\frac{c E_L}{6}} e^{-\mu/2} +2\pi \sqrt{\frac{c E_R}{6}} e^{-\mu/2} \right) \, .
\end{equation}
It therefore appears that the high-energy density of states for the deformed theory still has the (generalized) Cardy behavior appropriate for a conformal field theory, although with a new effective central charge $c_{\text{eff}} = c e^{-\mu}$. In particular, this gives one hint that the root-$\TT$ deformation (if it indeed is well-defined quantum mechanically) may actually be marginally \emph{relevant}, since the central charge appears to decrease along the flow for positive $\mu$.\footnote{One can also see this by considering the behavior of the spectrum (\ref{root_TT_deformed_CFT_energies}) as $\mu \to \infty$. In this limit, it appears that all negative-energy states in the undeformed theory approach zero deformed energy, while all undeformed positive-energy states have deformed energies which grow without bound. This suggests that the large-$\mu$ root-$\TT$ deformed theory becomes a gapped system with only a finite number of states.}

Although suggestive, there are some subtleties which prevent this argument from being fully rigorous. One is that we have not, strictly speaking, demonstrated that the root-$\TT$ flow equation holds for an arbitrary state in the deformed theory. Our gravitational calculation only demonstrates that this flow equation holds for holographic states which are dual to Ba\~nados-type geometries, and only in the large $N$ regime. A robust quantum definition of the root-$\TT$ operator might allow one to more carefully analyze the high-energy behavior of the deformed theory and determine whether it still exhibits Cardy behavior.

Another way of probing the fate of conformal invariance is to investigate modular properties of the root-$\TT$ deformed torus partition function. If one could show that the deformed partition function remains modular invariant, this would offer further evidence that the theory is conformal. One strategy for doing this would be to derive a differential equation that the deformed partition function satisfies. In the case of the $\TT$ deformation, it is known \cite{Cardy:2018sdv,Aharony:2018bad} that the torus partition function obeys the flow equation
\begin{align}
    \partial_\lambda Z_\lambda ( \tau, \overbar{\tau} ) = \left( \tau_2 \partial_{\tau} \partial_{\overbar{\tau}} + \frac{1}{2} \left( \partial_{\tau_2} - \frac{1}{\tau_2} \right) \lambda \partial_\lambda \right) Z_\lambda ( \tau, \overbar{\tau} ) \, , 
\end{align}
and that $Z_\lambda$ is invariant under a modular transformation if the $\TT$ parameter $\lambda$ also transforms. It appears that a root-$\TT$ deformed theory obeys an analogous flow equation,
\begin{align}\label{Z_flow_root_TT}
    \partial_\gamma^2 Z_\gamma ( \tau , \overbar{\tau } )  =  \left( \tau_2^2 \left( \partial_\tau \partial_{\overbar{\tau}} \right) + \tau_2 \partial_{\tau_2} \right) Z_\gamma ( \tau , \overbar{\tau } ) \, ,
\end{align}
which suggests that the root-$\TT$ deformed theory may be modular invariant. The properties of the flow equation (\ref{Z_flow_root_TT}) will be investigated in more detail elsewhere \cite{wip}.

\section*{Acknowledgements}

We thank Per Kraus, Ruben Monten and Savdeep Sethi for helpful discussions. We also thank Hao-Yu Sun for conversations and for early collaboration on this project. Some of the results in this paper were presented on August 11, 2022 at the ``\href{https://sites.google.com/view/fundamental-aspects-of-gravity/home?authuser=0}{Fundamental aspects of gravity}'' workshop held at Imperial College, London, and at the APCTP workshop ``\href{https://apctpstring.wixsite.com/integrability2022}{Integrability, Duality, and Related Topics}'' on November 1, 2022 in Pohang, South Korea.
C.F. is grateful to the participants of both workshops for useful comments, and acknowledges APCTP for hospitality during a visit where part of this work was done. S.E. is supported by the Bhaumik Institute. C. F. is supported by U.S. Department of Energy grant DE-SC0009999 and by funds from the University of California. Z.S. is supported from the U.S. Department of Energy (DOE) under cooperative research agreement DE-SC0009919,  Simons Foundation award No. 568420 (K.I.) and the Simons Collaboration on Global Categorical Symmetries. 

\appendix
\section{AdS$_3$ Gravity with $\TT$-Deformed Boundary Conditions}
\label{sec:AdS3TTApp}

In section \ref{sec:AdS3RootTT}, we used several methods that have been developed for studying $\mathrm{AdS}_3$ gravity with $\TT$-deformed boundary conditions, both in the metric formalism \cite{Guica:2019nzm} and in the Chern-Simons formalism \cite{He:2020hhm}. To make the present work self-contained, we review some aspects of these methods in this appendix, which are also useful for our analysis of root-$\TT$ deformed boundary conditions. We refer the reader to the original works for further details, and to the related work \cite{Llabres:2019jtx} for additional results in the Chern-Simons formalism.

\subsection{Metric Formalism}\label{app:TT_bc_pde_soln}

We recall that the modified metric $\gamma_{\alpha \beta}^{(\lambda)}$ and stress tensor $T_{\alpha \beta}^{(\lambda)}$ corresponding to a boundary $\TT$ deformation satisfy the equation (\ref{TT_varied_flow}) which was re-derived in the main text. By equating the coefficients of the independent terms on both sides of this equation, one arrives at a set of partial differential equations for the deformed quantities. These differential equations were first analyzed in \cite{Guica:2008mu}, where it was shown that they can be written as
\begin{equation}
\label{eq:PDEsTT}
\frac{\partial \gamma_{\alpha \beta}}{\partial \lambda} = -2 \hat{T}_{\alpha \beta}, \quad \frac{\partial \hat{T}_{\alpha \beta}}{\partial \lambda} = - \hat{T}_{\alpha \gamma} \hat{T}_\beta{}^\gamma, \quad \frac{\partial ( \hat{T}_{\alpha \gamma} \hat{T}_\beta{}^\gamma  )}{\partial \lambda} = 0\,.
\end{equation}
Here we have omitted the $(\lambda)$ superscripts on $\gamma_{\alpha \beta}^{(\lambda)}$ and $\widehat{T}_{\alpha \beta}^{(\lambda)} = T_{\alpha \beta}^{(\lambda)} - \gamma_{\alpha \beta}^{(\lambda)} \tensor{T}{^{(\lambda)}^\rho_\rho}$.

The solutions of \eqref{eq:PDEsTT} are \eqref{TT_deformed_gamma_T}. In terms of the Fefferman-Graham quantities, the deformed boundary metric and stress tensor are
\begin{equation}\label{appendix_deformed_metric}
\begin{aligned}
\gamma^{(\lambda)}_{\alpha \beta} &= g^{(0)}_{\alpha \beta} - \frac{2\lambda}{8 \pi G \ell} g^{(2)}_{\alpha \beta} +  \frac{\lambda^2 }{\left(8 \pi G \ell \right)^2}  g^{(2)}_{\alpha \rho} g^{(2)}_{\sigma \beta} \gamma^{(0)\rho\sigma}
\\&= g^{(0)}_{\alpha \beta} - \lambda g^{(2)}_{\alpha \beta} + \lambda^2 g^{(4)}_{\alpha \beta}\,,
\end{aligned}
\end{equation}
and
\begin{equation}
\begin{aligned}\label{appendix_deformed_T}
\widehat{T}^{(\lambda)}_{\alpha \beta} &= \widehat{T}^{(0)}_{\alpha \beta} - \lambda \widehat{T}^{(0)}_{\alpha \rho} \, \widehat{T}^{(0)}_{\sigma \beta} \gamma^{(0) \rho \sigma}
\\&=\frac{1}{8 \pi G \ell} g^{(2)}_{\alpha \beta} - \frac{\lambda}{(8 \pi G \ell)^2} g^{(2)}_{\alpha \rho} g^{(2)}_{\sigma \beta} g^{(0)\rho \sigma}
\\&= \frac{1}{2} \left( g^{(2)}_{\alpha \beta} - 2\lambda g^{(4)}_{\alpha \beta} \right)\,,
\end{aligned}
\end{equation}
where we used \eqref{eq:g4} and work in conventions such that $4\pi G \ell=  1$. For the bad sign of the deformation parameter, these modified asymptotic boundary conditions can be interpreted as Dirichlet boundary conditions at a finite radial coordinate $\rho_c = - \frac{\lambda}{4 \pi G \ell}$.\footnote{One can see by straightforward algebra that the asymptotic conditions (\ref{appendix_deformed_metric}) are equivalent to fixing the induced metric to be $g_{\alpha \beta}^{(0)}$ at this value of $\rho_c$ if $\lambda < 0$. Another way to determine the relation between the bulk cutoff $\rho_c$ and the $\TT$ coupling $\lambda$ is using the trace flow equation $T^\alpha{}_{\alpha} \propto \lambda \det T_{\alpha \beta}$ \cite{McGough:2016lol,Kraus:2018xrn,Hartman:2018tkw}.}
Although we are primarily interested in the good sign of the deformation, it is convenient to express various quantities in terms of $\rho_c$, although we note that for $\lambda > 0$ we have $\rho_c < 0$ and in this context $\rho_c$ cannot be interpreted as a physical value of the coordinate $\rho$. Thus
\begin{equation}\label{app_T_rhoc}
\gamma^{(\lambda)}_{\alpha \beta} = g^{(0)}_{\alpha \beta} + \rho_c g^{(2)}_{\alpha \beta} + \rho_c^2 g^{(4)}_{\alpha \beta}\,, \quad \widehat{T}^{(\lambda)}_{\alpha \beta} =\frac{1}{2} \left( g^{(2)}_{\alpha \beta} + 2\rho_c g^{(4)}_{\alpha \beta} \right)\,.
\end{equation}
Specializing to a Ba\~{n}ados geometry \eqref{eq:BanadosMetric}, the boundary metric in Fefferman-Graham quantities is
\begin{equation}
\label{eq:kre2}
\gamma^{(\lambda)}_{\alpha \beta} dx^\alpha \, dx^\beta = du \, dv + \rho_c \left( \mathcal{L} (u) du^2 + \overbar{\mathcal{L}}(v) dv^2 \right) +\rho_c^2 \mathcal{L} (u) \overbar{\mathcal{L}}(v) du \, dv \, .
\end{equation}
We express \eqref{eq:kre2} as 
\begin{equation}
\gamma^{(\lambda)}_{\alpha \beta} dx^\alpha \, dx^\beta = dU \, dV \,,
\end{equation}
where $(U, V)$ are the undeformed coordinates
\begin{equation}
\label{eq:TTcoords}
dU = du + \rho_c \overbar{\mathcal{L}}(v) dv, \quad dV = dv + \rho_c \mathcal{L}(u) du\,.
\end{equation}
In matrix form, we can define the state dependent coordinate transformation in \eqref{eq:TTcoords} and its inverse as 
\begin{equation}
\begin{aligned}
\label{eq:statedepcoordTT}
\left( \begin{array} {c}
dU\\dV
\end{array}
\right) &= \left( \begin{array}{cc}
    1    & \quad \rho_c \overbar{\mathcal{L}}(v)  \\
   \rho_c \mathcal{L}(u)    & \quad 1 
    \end{array} \right) \left( \begin{array} {c}
du\\dv
\end{array}
\right), \\ \left( \begin{array} {c}
du\\dv
\end{array}
\right) &= \frac{1}{1-\rho_c^2 \mathcal{L}(u) \overbar{\mathcal{L}}(v)}  \left( \begin{array}{cc}
    1    & \quad -\rho_c \overbar{\mathcal{L}}(v)  \\
   -\rho_c \mathcal{L}(u)    & \quad 1 
    \end{array} \right)\left( \begin{array} {c}
dU\\dV
\end{array}
\right)\,.
\end{aligned}
\end{equation}
Using \eqref{eq:statedepcoordTT}, we can write the boundary metric $g^{(0)}_{\alpha \beta}$ in the $(U, V)$ coordinates
\begin{equation}
\begin{aligned}
\label{eq:g0TT}
g^{(0)}_{\alpha \beta} dx^\alpha \, dx^\beta &= du \, dv\\& = \frac{(dU - \rho_c \overbar{\mathcal{L}}(v) dV  ) (dV - \rho_c \mathcal{L}(u) dU  )}{(1-\rho_c^2 \mathcal{L}(u) \overbar{\mathcal{L}}(v) ) ^2} \, ,
\end{aligned}
\end{equation}
as well as the other Fefferman-Graham quantities:
\begin{equation}
\begin{aligned}
\label{eq:g2TT}
g^{(2)}_{\alpha \beta} dx^\alpha \, dx^\beta &= \mathcal{L}(u) du^2 + \overbar{\mathcal{L}}(v) dv^2 \\&=  \mathcal{L}(u) \left( \frac{dU - \rho_c \overbar{\mathcal{L}}(v) }{1-\rho_c^2 \mathcal{L}(u) \overbar{\mathcal{L}}(v)  } \right)^2 + \overbar{\mathcal{L}}(v) \left( \frac{dV - \rho_c \mathcal{L}(v)}{1-\rho_c^2 \mathcal{L}(u) \overbar{\mathcal{L}}(v)  } \right)^2
\\&= \frac{(1+\rho_c^2 \mathcal{L}(u) \overbar{\mathcal{L}}(v) ) (\mathcal{L} (u) dU^2 + \overbar{\mathcal{L}} dV^2  ) - 4\rho_c \mathcal{L}(u) \overbar{\mathcal{L}}(v) dU \, dV    }{(1-\rho_c^2 \mathcal{L} \overbar{\mathcal{L}}(v)  )^2} \, , 
\end{aligned}
\end{equation}
and
\begin{equation}
\begin{aligned}
\label{eq:g4TT}
g^{(4)}_{\alpha \beta} dx^\alpha \, dx^\beta &= \mathcal{L}(u) \overbar{\mathcal{L}}(v) du \, dv\\& =\mathcal{L}(u) \overbar{\mathcal{L}}(v) \frac{(dU - \rho_c \overbar{\mathcal{L}}(v) dV  ) (dV - \rho_c \mathcal{L}(u) dU  )}{(1-\rho_c^2 \mathcal{L}(u) \overbar{\mathcal{L}}(v) ) ^2}\,.
\end{aligned}
\end{equation}
Proving \eqref{eq:g4TT} is straightforward:
\begin{equation}
 g^{(4)}_{\alpha \beta} = \frac{1}{4} g^{(2)}_{\alpha \rho} g^{(2)}_{\sigma \beta} g^{(0)\rho \sigma} =\frac{1}{4} \left( \begin{array}{cc}
    \mathcal{L}(u)  & \quad  0 \\
    0  & \quad \overbar{\mathcal{L}}(v)
 \end{array} \right)   \left( \begin{array}{cc}
   0  & \quad  2 \\
   2  & \quad 0
 \end{array} \right)   \left( \begin{array}{cc}
    \mathcal{L}(u)  & \quad  0 \\
    0  & \quad \overbar{\mathcal{L}}(v)
 \end{array} \right)   =\mathcal{L}(u) \overbar{\mathcal{L}}(v) g^{(0)}_{ \alpha \beta}\,.
\end{equation}
Substituting the expressions for $g^{(2)}_{\alpha \beta}$ and $g^{(4)}_{\alpha \beta}$ in \eqref{eq:g2TT} and \eqref{eq:g4TT} into the result (\ref{app_T_rhoc}) for the trace-reversed deformed stress tensor $\widehat{T}^{(\lambda)}_{\alpha \beta}$, we find that
\begin{equation}
\begin{aligned}
\widehat{T}^{(\lambda)}_{\alpha \beta} dx^\alpha \, dx^\beta &= \frac{1}{2} \left( g^{(2)}_{\alpha \beta} +2 \rho_c g^{(4)}_{\alpha \beta} \right) dx^\alpha \, dx^\beta 
\\&= \frac{\mathcal{L}(u) dU^2 + \overbar{\mathcal{L}}(v) dV^2 - 2\rho_c \mathcal{L}(u) \overbar{\mathcal{L}}(v) dU \, dV  }{2 (1-\rho_c^2 \mathcal{L} (u) \overbar{\mathcal{L}}(v)  )} \, , 
\end{aligned}
\end{equation}
and trace-reversing to obtain the deformed stress tensor in the $(U, V)$ coordinates yields
\begin{equation}
\label{eq:TcomponentsTT}
T_{\alpha \beta}^{(\lambda)} dx^\alpha \, dx^\beta = \frac{\mathcal{L}(u) dU^2 + \overbar{\mathcal{L}}(v) dV^2 + 2\rho_c \mathcal{L}(u) \overbar{\mathcal{L}}(v) dU \, dV  }{2 (1-\rho_c^2 \mathcal{L}(u) \overbar{\mathcal{L}}(v)  )}\,.
\end{equation}
It is straightforward to show that \eqref{eq:TcomponentsTT} obeys the $\TT$ trace flow equation and is conserved
\begin{equation}
    \partial_V T^{(\lambda)}_{UU} + \partial_U T^{(\lambda)}_{VU} =  \partial_V T^{(\lambda)}_{UV} + \partial_U T^{(\lambda)}_{VV}   = 0\,.
\end{equation}
From this Fefferman-Graham analysis, we have therefore determined the deformed black hole solutions for constant $(\mathcal{L}(u), \overbar{\mathcal{L}}(v)) \equiv (\mathcal{L}_\lambda, \overbar{\mathcal{L}}_\lambda)$. In terms of the temporal and angular coordinates $\phi$ and $T$, \eqref{eq:TcomponentsTT} becomes 
\begin{equation}
T_{\alpha \beta}^{(\lambda)} dx^\alpha \, dx^\beta = \frac{(\mathcal{L}_\lambda + \overbar{\mathcal{L}}_\lambda -2 \rho_c \mathcal{L}_\lambda \overbar{\mathcal{L}}_\lambda )dT^2+(\mathcal{L}_\lambda + \overbar{\mathcal{L}}_\lambda +2 \rho_c \mathcal{L}_\lambda \overbar{\mathcal{L}}_\lambda ) d\phi^2 + 2 (\mathcal{L}_\lambda - \overbar{\mathcal{L}}_\lambda) d\phi \, dT}{2 (1-\rho_c^2 \mathcal{L}_\lambda \overbar{\mathcal{L}}_\lambda  )} \, , 
\end{equation}
where $(U, V) = (\phi + T, \phi - T)$. Therefore, in the $(T, \phi)$ coordinates and restoring factors of $4 \pi G \ell$, we find that the deformed energy and angular momentum are
\begin{equation}
\begin{aligned}
\label{eq:rohioe2}
E_\lambda &= \int^R_0 d\phi~ T^{(\lambda)}_{TT} = \frac{R (\mathcal{L}_\lambda+ \overbar{\mathcal{L}}_\lambda - 2 \rho_c \mathcal{L}_\lambda \overbar{\mathcal{L}}_\lambda )}{8 \pi G \ell (1-\rho_c^2 \mathcal{L}_\lambda \overbar{\mathcal{L}}_\lambda )}, \quad J_\lambda = \int^R_0 d\phi~ T^{(\lambda)}_{T\phi} =  \frac{R (\mathcal{L}_\lambda - 
 \overbar{\mathcal{L}}_\lambda)}{8 \pi G \ell (1-\rho_c^2 \mathcal{L}_\lambda \overbar{\mathcal{L}}_\lambda )}\,.
\end{aligned}
\end{equation}
The functions $(\mathcal{L}_\lambda, \overbar{\mathcal{L}}_\lambda)$ are fixed in terms of  $(\mathcal{L}_0, \overbar{\mathcal{L}}_0)$ by equating the undeformed and deformed angular momenta and event horizon areas \cite{Guica:2019nzm}. This is possible because the $\TT$ flow preserves the boundary theory's degeneracy of states, which implies that the horizon area of the black hole is unchanged by the deformation. The angular momentum is holographically dual to the momentum $P_n$ of the state in the field theory, which is quantized in units of $\frac{1}{R}$ and thus cannot flow with the deformation parameter because $\lambda$ is continuous. In fact, we expect that these two assumptions should hold for \emph{any} stress tensor deformation of the boundary field theory (including root-$\TT$), since any flow equation for the spectrum which is driven by a function of only energies and momenta will also preserve degeneracies.

We have already determined the angular momentum, so we now consider the horizon areas. The undeformed event horizon in the Fefferman-Graham gauge is at 
\begin{equation}
\label{eq:eventhorizonU}
\rho_h = \frac{1}{\sqrt{\mathcal{L}_0 \overbar{\mathcal{L}}_0  }} \, , 
\end{equation}
and \eqref{eq:BanadosMetric} evaluated at \eqref{eq:eventhorizonU} is
\begin{equation}
ds^2|_{\rho_h = (\mathcal{L}_0 \overbar{\mathcal{L}}_0 )^{-\frac{1}{2}} }   = \frac{\ell^2 \mathcal{L}_0 \overbar{\mathcal{L}}_0 }{4} d\rho^2 + \left( \sqrt{\mathcal{L}_0} - \sqrt{\overbar{\mathcal{L}}_0} \right)^2 dT^2 + \left( \sqrt{\mathcal{L}_0} + \sqrt{\overbar{\mathcal{L}}_0} \right)^2 d\phi^2 +2 \left( \mathcal{L}_0 - \overbar{\mathcal{L}}_0 \right) dTd\phi\,.
\end{equation}
For the deformed black hole metric, substituting \eqref{eq:g0TT} - \eqref{eq:g4TT} into \eqref{eq:BanadosMetric} evaluated at the event horizon 
\begin{equation}
\rho_h= \frac{1}{\sqrt{\mathcal{L}_\lambda \overbar{\mathcal{L}}_\lambda  }} \, , 
\end{equation}
we obtain
\begin{equation}
ds^2|_{\rho_h = (\mathcal{L}_\lambda \overbar{\mathcal{L}}_\lambda  )^{-\frac{1}{2}}} = \frac{\ell^2 \mathcal{L}_\lambda \overbar{\mathcal{L}}_\lambda}{4} d\rho^2 + \frac{\left( \sqrt{\mathcal{L}_\lambda} - \sqrt{\overbar{\mathcal{L}}_\lambda }   \right)^2 dT^2 + \left( \sqrt{\mathcal{L}_\lambda} + \sqrt{\overbar{\mathcal{L}}_\lambda } \right)^2 d\phi^2 + 2 (\mathcal{L}_\lambda - \overbar{\mathcal{L}}_\lambda  ) d\phi dT }{ \left(1+\rho_c \sqrt{\mathcal{L}_\lambda \overbar{\mathcal{L}}_\lambda }\right)^2} \, , 
\end{equation}
which has event horizon area
\begin{equation}
A^{(\lambda)} = \int^R_0 d\phi \sqrt{g_{\phi \phi}}|_{\rho_h = (\mathcal{L}_\lambda \overbar{\mathcal{L}}_\lambda )^{-\frac{1}{2}} }  = R \frac{\sqrt{\mathcal{L}_\lambda} + \sqrt{\overbar{\mathcal{L}}_\lambda}}{1+\rho_c \sqrt{\mathcal{L}_\lambda \overbar{\mathcal{L}}_\lambda }}\,.
\end{equation}
Equating the undeformed and deformed event horizon areas and angular momenta, we arrive at the constraints for $(\mathcal{L}_\lambda, \overbar{\mathcal{L}}_\lambda)$,
\begin{equation}
\label{eq:AreaAngularMomentumTT}
\sqrt{\mathcal{L}_0} + \sqrt{\overbar{\mathcal{L}}_0} = \frac{\sqrt{\mathcal{L}_\lambda} + \sqrt{\overbar{\mathcal{L}}_\lambda}   }{1+\rho_c \sqrt{\mathcal{L}_\lambda \overbar{\mathcal{L}}_\lambda }}, \quad \mathcal{L}_0 - \overbar{\mathcal{L}}_0 = \frac{ \mathcal{L}_\lambda - \overbar{\mathcal{L}}_\lambda     }{1 -  \rho_c^2 \mathcal{L}_\lambda \overbar{\mathcal{L}}_\lambda  }\,.
\end{equation}
The solution to \eqref{eq:AreaAngularMomentumTT} is
\begin{equation}
\begin{aligned}
\label{eq:solsTT1}
\mathcal{L}_\lambda &= \frac{- \left(1 + \rho_c (\mathcal{L}_0   -   \overbar{\mathcal{L}}_0  )  \right) \sqrt{\rho_c^2 \left( \mathcal{L}_0 - \overbar{\mathcal{L}}_0 \right)^2 - 2 \rho_c \left( \mathcal{L}_0 + \overbar{\mathcal{L}}_0 \right) + 1  }  + \rho_c^2 \left( \mathcal{L}_0 - \overbar{\mathcal{L}}_0  \right)^2 - 2\rho_c \overbar{\mathcal{L}}_0 + 1  }{2 \rho_c^2 \mathcal{L}_0},\\
\overbar{\mathcal{L}}_\lambda &= \frac{- \left(1 - \rho_c (\mathcal{L}_0   -   \overbar{\mathcal{L}}_0  )  \right) \sqrt{\rho_c^2 \left( \mathcal{L}_0 - \overbar{\mathcal{L}}_0 \right)^2 - 2 \rho_c \left( \mathcal{L}_0 + \overbar{\mathcal{L}}_0 \right) + 1  }  + \rho_c^2 \left( \mathcal{L}_0 - \overbar{\mathcal{L}}_0  \right)^2 - 2\rho_c \mathcal{L}_0 + 1  }{2 \rho_c^2 \overbar{\mathcal{L}}_0}\,.
\end{aligned}
\end{equation}
Substituting \eqref{eq:solsTT1} into the energy equation \eqref{eq:rohioe2}, we arrive at the well-established $\TT$-deformed energy expressed in terms of the field theory energy $E_0$ and momentum $P_0$,
\begin{equation}
\begin{aligned}
E_\lambda &= \frac{R}{8 \pi G \ell \rho_c} \left(1 - \sqrt{1 - 2 \rho_c \left(\mathcal{L}_0 + \overbar{\mathcal{L}}_0\right)  + \rho_c^2 \left( \mathcal{L}_0 - \overbar{\mathcal{L}}_0 \right)^2  } \right)
\\&= \frac{R}{2\lambda} \left( \sqrt{1 + \frac{4 \lambda E_0}{R} + \frac{4 \lambda^2 P_0^2}{R^2}  } -1\right)  \,,
\end{aligned}
\end{equation}
where the undeformed energy $E_0$, angular momentum $J_0$ (which corresponds to the momentum $P_0$ in the CFT), and deformation parameter with units restored are
\begin{equation}
    E_0 = \frac{R}{8 \pi G \ell} (\mathcal{L}_0 + \overbar{\mathcal{L}}_0), \quad J_0 = \frac{R}{8 \pi G \ell} \left( \mathcal{L}_0 - \overbar{\mathcal{L}}_0 \right) = P_0 , \quad \lambda =- 4 \pi G \ell \rho_c \, .
\end{equation}

\subsection{Chern-Simons Formalism}
\label{app:A2}
To obtain the $\TT$-deformed Chern-Simons connections, we use the coordinate transformation in \eqref{eq:statedepcoordTT} to obtain
\begin{equation}
\begin{aligned}
A(\rho_c)&= -\frac{1}{2\rho} L_0 d\rho + \frac{1}{\ell} \left( -\sqrt{\rho} \mathcal{L}_\lambda L_{-1} + \frac{1}{\sqrt{\rho}} L_1 \right) \left( \frac{dU - \rho_c \overbar{\mathcal{L}}_\lambda dV}{1-\rho_c^2 \mathcal{L}_\lambda \overbar{\mathcal{L}}_\lambda  } \right)\,,\\
\overbar{A}(\rho_c) &= \frac{1}{2\rho} L_0 d\rho + \frac{1}{\ell} \left(\frac{1}{\sqrt{\rho}} L_{-1} - \sqrt{\rho} \overbar{\mathcal{L}}_\lambda L_1 \right) \left( \frac{dV - \rho_c \mathcal{L}_\lambda dU  }{1-\rho_c^2 \mathcal{L}_\lambda \overbar{\mathcal{L}}_\lambda} \right)\,.
\end{aligned}
\end{equation}
We can see that the deformed gauge fields obey a mixed boundary condition 
\begin{equation}
\rho_c \overbar{\mathcal{L}}_\lambda A_U(\rho_c) + A_V(\rho_c) = 0, \quad \overbar{A}_U(\rho_c) + \rho_c \mathcal{L}_\lambda \overbar{A}_V(\rho_c) = 0\,.
\end{equation}
To convert the connections from the $(U, V)$ coordinates to the $(T, \phi)$ coordinates, we recall that
\begin{equation}
    \begin{aligned}
        A &= A_\alpha dx^\alpha 
        \\&= A_U dU + A_V dV
        \\&= \left( A_U + A_V \right) d\phi+\left( A_U - A_V \right) dT \, ,  
    \end{aligned}
\end{equation}
yielding 
\begin{equation}
\begin{aligned}
A_\phi = A_U + A_V, \quad A_T = A_U - A_V\,, \quad \overbar{A}_\phi = \overbar{A}_U + \overbar{A}_V\,, \quad \overbar{A}_T = \overbar{A}_U - \overbar{A}_V\,.
\end{aligned}
\end{equation}
Hence
\begin{equation}
A_\phi (\rho_c) = \frac{1}{\ell} \frac{1-\rho_c \overbar{\mathcal{L}}_\lambda}{1- \rho_c^2 \mathcal{L}_\lambda  \overbar{\mathcal{L}}_\lambda } \left( -\sqrt{\rho} \mathcal{L}_\lambda  L_{-1} + \frac{1}{\sqrt{\rho}}L_1 \right) \,, \quad A_T (\rho_c) =  \frac{1+\rho_c \overbar{\mathcal{L}}_\lambda}{1-\rho_c \overbar{\mathcal{L}}_\lambda} A_\phi(\rho_c) \, , 
\end{equation}
and 
\begin{equation}
\overbar{A}_\phi(\rho_c) = \frac{1}{\ell} \frac{1 - \rho_c \mathcal{L}_\lambda}{1-\rho_c^2 \mathcal{L}_\lambda \overbar{\mathcal{L}}_\lambda}  \left(\frac{1}{\sqrt{\rho}} L_{-1} - \sqrt{\rho} \overbar{\mathcal{L}}_\lambda L_1 \right) \,, \quad   \overbar{A}_T(\rho_c) = - \frac{1+\rho_c \mathcal{L}_\lambda}{1- \rho_c \mathcal{L}_\lambda}  \overbar{A}_\phi (\rho_c)\,.
\end{equation}
The boundary connections obey a similar relation as the bulk connections
\begin{equation}
\begin{aligned}
\label{eq:TTdeformedboundaryA}
a_\phi (\rho_c) =  \frac{1}{\ell} \frac{1-\rho_c\overbar{\mathcal{L}}_\lambda}{1- \rho_c^2 \mathcal{L}_\lambda  \overbar{\mathcal{L}}_\lambda } \left( -\mathcal{L}_\lambda L_{-1} + L_1\right)\,, \quad a_T (\rho_c) =  \frac{1+\rho_c \overbar{\mathcal{L}}_\lambda}{1-\rho_c \overbar{\mathcal{L}}_\lambda} a_\phi(\rho_c) \, , 
\end{aligned}
\end{equation}
and
\begin{equation}
\label{eq:TTdeformedboundaryAb}
\overbar{a}_\phi (\rho_c) =  \frac{1}{\ell} \frac{1 - \rho_c \mathcal{L}_\lambda}{1-\rho_c^2 \mathcal{L}_\lambda \overbar{\mathcal{L}}_\lambda} \left(  L_{-1} -  \overbar{\mathcal{L}}_\lambda L_1   \right)\,, \quad \overbar{a}_T(\rho_c) = - \frac{1 + \rho_c \mathcal{L}_\lambda}{1-\rho_c \mathcal{L}_\lambda} \overbar{a}_\phi(\rho_c)\,.
\end{equation}
We may also study the black hole entropy and horizon areas using these deformed connections in the same way as we did in the root-$\TT$ deformed case around equation \eqref{eq:Entropy}. The analogues of the matrices $\lambda_\phi$ and $\overbar{\lambda}_\phi$ in equation (\ref{root_TT_lambda_matrices}), which are simply the diagonalized versions of $a_\phi$ and $\overbar{a}_\phi$, for the $\TT$-deformed connections \eqref{eq:TTdeformedboundaryAb}, are
\begin{equation}
    \begin{aligned}
        \lambda_\phi &= \frac{1}{\ell} \left( \begin{array}{cc}
           \frac{\left( 1- \rho_c \overbar{\mathcal{L}}_\lambda \right) \sqrt{\mathcal{L}_\lambda} }{1-\rho_c^2 \mathcal{L}_\lambda \overbar{\mathcal{L}}_\lambda   }  &\quad 0  \\
          0   & \quad -  \frac{\left( 1- \rho_c \overbar{\mathcal{L}}_\lambda \right) \sqrt{\mathcal{L}_\lambda} }{1-\rho_c^2 \mathcal{L}_\lambda \overbar{\mathcal{L}}_\lambda   }
        \end{array} \right)   \,, \\
        \overbar{\lambda}_\phi &= \frac{1}{\ell} \left( \begin{array}{cc}
         -  \frac{\left( 1- \rho_c \mathcal{L}_\lambda \right) \sqrt{\overbar{\mathcal{L}}_\lambda} }{1-\rho_c^2 \mathcal{L}_\lambda \overbar{\mathcal{L}}_\lambda   }  &\quad 0  \\
          0   & \quad   \frac{\left( 1- \rho_c \mathcal{L}_\lambda \right) \sqrt{\overbar{\mathcal{L}}_\lambda} }{1-\rho_c^2 \mathcal{L}_\lambda \overbar{\mathcal{L}}_\lambda   }
        \end{array} \right) \, .
    \end{aligned}
\end{equation}
Using the equation $S = C \operatorname{Tr} \left( (\lambda_\phi - \overbar{\lambda}_\phi)L_0 \right)$ for the entropy, which we quoted in \eqref{eq:Entropy}, we find an expression for the deformed entropy $S^{(\lambda)}$:
\begin{equation}
\label{eq:TTentropies}
     S^{(\lambda)} = \frac{C}{\ell} \left(  \frac{\sqrt{\mathcal{L}_\lambda} + \sqrt{\overbar{\mathcal{L}}_\lambda} }{1+\rho_c \sqrt{\mathcal{L}_\lambda \overbar{\mathcal{L}}_\lambda} }  \right)\,.
\end{equation}
Setting (\ref{eq:TTentropies}) equal to the undeformed entropy
\begin{align}
    S^{(0)} = \frac{C}{\ell} \left( \sqrt{\mathcal{L}_0} + \sqrt{\overbar{\mathcal{L}}_0} \right) \, , 
\end{align}
then reproduces the area equation \eqref{eq:AreaAngularMomentumTT}.

Following \cite{He:2020hhm}, we can now read off the variation of the boundary action which is compatible with the relations (\ref{eq:TTdeformedboundaryA}) and (\ref{eq:TTdeformedboundaryAb}) for the deformed boundary connections:
\begin{align}\label{app_CS_deformed_delta_S}
    \delta S = -\frac{\ell}{8 \pi G} \int_{\partial M} dT \, d\phi \, &\Bigg( \operatorname{Tr} \left[ \left( a_T(\rho_c) - \frac{1+\rho_c \overbar{\mathcal{L}}_\lambda   }{1-  \rho_c \overbar{\mathcal{L}}_\lambda    } a_\phi (\rho_c) \right) \delta a_\phi (\rho_c) \right] \nonumber \\
    &\qquad - \Tr \left[ \left( \overbar{a}_T (\rho_c) + \frac{1 + \rho_c \mathcal{L}_\lambda  }{1 - \rho_c \mathcal{L}_\lambda} \overbar{a}_\phi (\rho_c)  \right) \delta \overbar{a}_\phi (\rho_c)   \right] \Bigg) \, .
\end{align}
We see that, when the constraints (\ref{eq:TTdeformedboundaryA}) and (\ref{eq:TTdeformedboundaryAb}) are satisfied, the variation (\ref{app_CS_deformed_delta_S}) collapses to $\delta S_{\text{bdry}} = 0$. This guarantees a well-defined variational principle.

To determine this boundary action in terms of $\mathcal{L}_\lambda$, $\overbar{\mathcal{L}}_\lambda$, and $\rho_c$, we must first evaluate the variations of the boundary connections. The variations of \eqref{eq:TTdeformedboundaryA} and \eqref{eq:TTdeformedboundaryAb}  are
\begin{equation}
\begin{aligned}
\label{eq:variationsTT}
\delta a_\phi (\rho_c) &= \frac{(1 -  \rho_c \overbar{\mathcal{L}}_\lambda ) \left( L_{-1} - \rho_c^2 \overbar{\mathcal{L}}_\lambda L_1 \right) \delta \mathcal{L}_\lambda  -\rho_c \left(  \mathcal{L}_\mu L_{-1} - L_1   \right) \left( 1-\rho_c \mathcal{L}_\lambda \right) \delta \overbar{\mathcal{L}}_\lambda}{\ell \left(1 - \rho_c^2 \mathcal{L}_\lambda \overbar{\mathcal{L}}_\lambda \right)^2}\,,\\
\delta \overbar{a}_\phi (\rho_c) &= \frac{- \rho_c \left(L_{-1} - \overbar{\mathcal{L}}_\mu L_1 \right) \left( 1 -\rho_c \overbar{\mathcal{L}}_\lambda \right) \delta \mathcal{L}_\lambda -\left( \rho_c^2 \mathcal{L}_\mu L_{-1} - L_1 \right)  \left( 1 - \rho_c \mathcal{L}_\lambda \right)  \delta \overbar{\mathcal{L}}_\lambda }{\ell \left( 1 - \rho_c^2 \mathcal{L}_\lambda \overbar{\mathcal{L}}_\lambda \right)^2}\,.
\end{aligned}
\end{equation}
The variation of the boundary piece is
\begin{equation}
\begin{aligned}
\label{eq:boundarypieceTT}
\delta S_{\text{bdry}}=  \frac{\ell}{8 \pi G} \int_{\partial M} dT \, d\phi \left(    \frac{1+\rho_c\overbar{\mathcal{L}}_\lambda   }{1-   \rho_c \overbar{\mathcal{L}}_\lambda    } \operatorname{Tr} \left(a_\phi (\rho_c)  \delta a_\phi (\rho_c)\right)  + \frac{1 + \rho_c  \mathcal{L}_\lambda  }{1 - \rho_c \mathcal{L}_\lambda} \operatorname{Tr} \left( \overbar{a}_\phi (\rho_c)   \delta \overbar{a}_\phi (\rho_c)  \right)  \right) \, , 
\end{aligned}
\end{equation}
 and using \eqref{eq:variationsTT}, the traces evaluate to
\begin{equation}
    \begin{aligned}
    \label{eq:TTtraces}
 \operatorname{Tr} \left(a_\phi (\rho_c)  \delta a_\phi (\rho_c)\right) &=  \frac{(1- \rho_c \overbar{\mathcal{L}}_\lambda )^2 (1 + \rho_c^2 \mathcal{L}_\lambda \overbar{\mathcal{L}}_\lambda)  \delta \mathcal{L}_\lambda - 2\rho_c \mathcal{L}_\lambda (1-\rho_c \mathcal{L}_\lambda)(1-\rho_c \overbar{\mathcal{L}}_\lambda )   \delta \overbar{\mathcal{L}}_\lambda }{\ell^2(1-\rho_c^2 \mathcal{L}_\lambda \overbar{\mathcal{L}}_\lambda  )^3} \,, \\
            \operatorname{Tr} \left(\overbar{a}_\phi (\rho_c)  \delta \overbar{a}_\phi (\rho_c)\right) &= \frac{ -2\rho_c \overbar{\mathcal{L}}_\lambda (1-\rho_c \mathcal{L}_\lambda)(1-\rho_c \overbar{\mathcal{L}}_\lambda)   \delta \mathcal{L}_\lambda + (1- \rho_c \mathcal{L}_\lambda)^2 (1+\rho_c^2 \mathcal{L}_\lambda \overbar{\mathcal{L}}_\lambda) \delta \overbar{\mathcal{L}}_\lambda }{\ell^2 (1-\rho_c^2 \mathcal{L}_\lambda \overbar{\mathcal{L}}_\lambda  )^3} \,.
    \end{aligned}
\end{equation}
 Substituting \eqref{eq:TTtraces} into \eqref{eq:boundarypieceTT}, the varied boundary action in terms of $\delta \mathcal{L}_\lambda$ and $\delta \overbar{\mathcal{L}}_\lambda$ is
\begin{equation}
    \delta S_{\text{bdry}} = \frac{1}{8 \pi G \ell} \int_{\partial M} dT \, d\phi \left( \left(\frac{1- \rho_c \overbar{\mathcal{L}}_\lambda  }{1-\rho_c^2 \mathcal{L}_\lambda \overbar{\mathcal{L}}_\lambda  }\right)^2 \delta \mathcal{L}_\lambda   + \left(\frac{1- \rho_c\mathcal{L}_\lambda  }{1-\rho_c^2 \mathcal{L}_\lambda \overbar{\mathcal{L}}_\lambda  }\right)^2 \delta \overbar{\mathcal{L}}_\lambda \right) \, , 
\end{equation}
from which $S_{\text{bdry}}$ can be read off as
\begin{equation}
\label{eq:boundaryactionTT}
    S_{\text{bdry}} = \frac{1}{8 \pi G \ell} \int_{\partial M} dT \, d\phi \, \frac{\mathcal{L}_\lambda  + \overbar{\mathcal{L}}_\lambda - 2 \rho_c \mathcal{L}_\lambda  \overbar{\mathcal{L}}_\lambda }{1 - \rho_c^2 \mathcal{L}_\lambda \overbar{\mathcal{L}}_\lambda}\,.
\end{equation}
After integration over $\phi$ in \eqref{eq:boundaryactionTT}, we find that
\begin{align}
    S_{\text{bdry}} &=  \int \, d T \, \frac{R \left( \mathcal{L}_\lambda  + \overbar{\mathcal{L}}_\lambda - 2 \rho_c \mathcal{L}_\lambda  \overbar{\mathcal{L}}_\lambda  \right) }{8\pi G \ell (1 - \rho_c^2 \mathcal{L}_\lambda \overbar{\mathcal{L}}_\lambda)} \, \nonumber \\
    &=  \int \, d T \, E_\lambda \, , 
\end{align}
where we have used the expression for $E_\lambda$ in \eqref{eq:rohioe2}. Therefore the boundary Lagrangian density in the Chern-Simons formalism agrees with the deformed mass (or energy) of the bulk spacetime as computed in the metric formalism.

We emphasize again that it was not clear \emph{a priori} that the boundary Chern-Simons action would necessarily reproduce the mass of the deformed spacetime. Although this is true in the undeformed theory, after adding a boundary deformation which implements mixed boundary conditions in the bulk, one would need to compute the Hamiltonian in order to argue that the Chern-Simons boundary action will agree with the deformed spacetime mass in general. However, in this case, we have seen by explicit computation that the two agree, at least for the class of Ba\~{n}ados-type solutions we are considering.

\bibliographystyle{utphys}
\bibliography{ref}

\end{document}